\documentclass[twocolumn]{aastex631}
\usepackage{amsmath}
\usepackage{svg}
\usepackage{subfloat}
\usepackage{rotating}

\shorttitle{Testing Chemodynamical Tagging with Open Clusters in GALAH DR4}
\shortauthors{Barth et al.}
\graphicspath{{./}{figures/}}

\makeatletter
\let\frontmatter@title@above=\relax
\makeatother
\usepackage{lipsum}

\usepackage{hyperref}

\begin{document}

\title{Trials and Tribulations of Chemodynamical Tagging: \\ Investigating the Recovery of Open Clusters in the \textit{Gaia} DR3-GALAH DR4 Catalog}

\author[0000-0003-0602-5287]{Nicholas A. Barth}
\affiliation{Department of Astronomy \\ University of Florida, 
1772 Stadium Rd, 
Gainesville, FL, 32611, USA.}

\author[0000-0002-0057-3699]{David Mendez}
\affiliation{Department of Physics\\ University of Florida, P.O. Box 118440, Gainesville, FL 32611, USA.}

\author[0000-0002-8504-8470]{Rana Ezzeddine}
\affiliation{Department of Astronomy \\ University of Florida, 
1772 Stadium Rd, 
Gainesville, FL, 32611, USA.}

\author[0000-0003-4769-3273]{Yuxi(Lucy) Lu}
\affiliation{Department of Astronomy\\ The Ohio State University, Columbus, 140 W 18th Ave, OH 43210, USA}
\affiliation{Center for Cosmology and Astroparticle Physics (CCAPP)\\ The Ohio State University, 191 W. Woodruff Ave., Columbus, OH 43210, USA}

\author[0000-0002-1333-8866]{Leslie M. Morales}
\affiliation{Department of Astronomy \\ University of Florida, 
1772 Stadium Rd, 
Gainesville, FL, 32611, USA.}

\author[0000-0002-9879-3904]{Zachary R. Claytor}
\affiliation{Space Telescope Science Institute \\ 3700 San Martin Drive, 
Baltimore, MD, 21218, USA.}
\affiliation{Department of Astronomy \\ University of Florida, 
1772 Stadium Rd, 
Gainesville, FL, 32611, USA.}

\author[0000-0002-4818-7885]{Jamie Tayar}
\affiliation{Department of Astronomy \\ University of Florida, 
1772 Stadium Rd, 
Gainesville, FL, 32611, USA.}



\begin{abstract}

Chemodynamical tagging has been suggested as a powerful tool to trace stars back to their birth clusters. However, the efficacy of chemodynamical tagging as a means to recover individual stellar clusters is still under debate. In this study, we present a detailed investigation of chemodynamical tagging of open clusters using both dynamical and chemical data from the \textit{Gaia} DR3 and GALAH DR4 surveys, respectively. Using a sample of open clusters and mock field stars, we conduct a bootstrap analysis to evaluate every unique combination of orbital components ($E, J_R, J_\phi, J_Z$) and chemical abundances ([X/Fe] for O, Na, Mg, Al, Si, K, Ca, Sc, Ti, Cr, Mn, Ni, Cu, Y, and Ba) on how well they recover open clusters when used as parameters in the clustering algorithm, HBDSCAN. We find that using primarily dynamical orbital parameters leads to the highest recovery rate of open cluster stars. Nevertheless, even employing the best performing parameter combinations leads to low open cluster recovery rates. We find that, in most cases, chemodynamical tagging of open clusters using blind clustering algorithms is not efficient, which is in line with previous theoretical and observational work. However, we show that the addition of cuts based on metallicity, age, and birth radii in order to reduce the size of the clustering catalog can marginally improve the recovery rate of open clusters. 

\end{abstract}



\section{Introduction} \label{sec:intro}

The era of large photometric, spectroscopic, and astrometric surveys has significantly improved our understanding of how the Milky Way (MW) was formed and evolved to what we observe today. Using positions and velocities from photometric and astrometric surveys (e.g. \citealt{2021A&A...649A...1G}), we have access to the dynamical information of millions of stars, while spectroscopic surveys (e.g. \citealt{2012RAA....12..723Z}, \citealt{2017AJ....154...94M}, \citealt{2021MNRAS.506..150B}) provide detailed chemical abundances for hundreds of thousands of stars. Astronomers have long used chemodynamical properties, including the chemical abundances, as well as the Galactic orbital information, to partition the Milky Way into major overdensities, including thin and thick disc components, as well as substructures in the bulge and the inner and outer halo (\citealt{2015MNRAS.453..758H}, \citealt{2016ARA&A..54..529B}). Using these major overdensities is a form of \textit{weak} chemodynamical tagging, where stars are categorized based on the Galactic environment in which they were formed (\citealt{2002ARA&A..40..487F}, \citealt{2021A&A...654A.151C}). As new surveys came online and the number of stars with chemodynamical measurements began to grow, substructures within the major subdivisions of the Milky Way are now being recovered, including new globular clusters (e.g. \citealt{2011A&A...533A..69C}, \citealt{2017MNRAS.466.1010S}, \citealt{2018ApJ...860...70C}), open clusters (OCs) (e.g. \citealt{2006AJ....131..455D}, \citealt{2007AJ....133.1161D}, \citealt{2012MNRAS.421.1231T}, \citealt{Hunt2023}), stellar streams (e.g. \citealt{2011MNRAS.415.1138P}, \citealt{2020ApJ...901...48N}, \citealt{2020ApJ...901...23H}, \citealt{2021ApJ...907...10L}), and other unique stellar populations formed or accreted during the Milky Way's dynamic history. This process of identifying stars that were born in the same formation events in the same systems is known as \textit{strong} chemodynamical tagging (\citealt{2002ARA&A..40..487F}, \citealt{2021A&A...654A.151C}).

Strong chemical tagging has been shown to be a strong tool in identifying presently bound, tidal associations, however, using strong chemical tagging to recover structures that have been dispersed throughout the Galaxy since their initial formation has been shown to be challenging (e.g. \citealt{2012MNRAS.421.1231T}, \citealt{2013MNRAS.428.2321M}, \citealt{2015ApJ...807..104T}, \citealt{2021A&A...654A.151C}). The strength of strong chemical tagging is the assumption that star clusters are chemically homogeneous at birth, and preserve this chemical information even after the cluster has long since been physically associated (e.g. \citealt{2006AJ....131..455D}, \citealt{2016ApJ...817...49B}, \citealt{2018MNRAS.478..425B}). We thus expect to be able to disentangle dispersed star clusters from large all-sky surveys based on their unique chemical signature.

The use of OCs has been instrumental in testing the different methods of chemical tagging. Studies done by \citet{2013MNRAS.428.2321M} and \citet{2015A&A...577A..47B} used high-precision abundances to perform robust, blind clustering methods, empirical cluster probability functions and K-means++, respectively, which were applied to entire star catalogs, such as GALAH, APOGEE, or compiled observations from high-resolution instruments like NARVAL, HARPS, or UVES, without any prior restrictions based on the properties of known OCs (\citealt{2003EAS.....9..105A}, \citealt{2003Msngr.114...20M}, \citealt{2000SPIE.4008..534D}). Blind clustering methods can assess how well known OCs can be recovered using chemical parameters and indicate if unknown OCs can be found using these methods. Other works have shown some success in recovering OCs using hierarchical clustering algorithms \citep{2018IAUS..334..128S}, t-distributed stochastic neighbour embedding (t-SNE) visual identification \citep{2018MNRAS.473.4612K}, and non-hierarchical clustering algorithms \citep{2019A&A...629A..34G}. Using the APOGEE chemical abundances, \citet{2021A&A...654A.151C} showed that the overlap in the chemical signatures between OCs is large and hinders accurate grouping of real star clusters using unsupervised clustering techniques. Each of the aforementioned studies focused solely on identifying OCs in chemical space alone. 

In terms of their dynamics, star clusters that form in the MW disc tend to disperse over a few $\sim100$ Myr, and it has been suggested that it is possible to trace back their formation sites using dynamics alone for young clusters \citep{2019ARA&A..57..227K}. Orbital actions can be calculated utilizing the kinematic data of a star, are conserved quantities under adiabatic changes to the underlying Galactic potential (\citealt{2008gady.book.....B}, \citealt{2014MNRAS.441.3284S}, \citealt{2016MNRAS.457.2107S}). Since OC stars are born in the same location, their initial orbits are similar and therefore, for a potential which varies slowly enough the actions should remain similar long after the dissolution of the cluster.

The goal of chemodynamical tagging is to combine both the chemical and dynamical information in order to classify stars to a unique formation site or event. One common way to test chemodynamical tagging is to attempt to recover stars in well-known OCs using clustering algorithms to distill large catalogs of stars into statistical groups using their dynamical and chemical properties, as OCs are still dynamically bound associations with homogeneous chemical abundances up to ~0.02-0.03 dex (\citealt{2006AJ....131..455D}, \citealt{2016MNRAS.463..696L}, \citealt{2016ApJ...817...49B}). By comparing the statistical group labels to the true labels of the OC stars, we can assess how well the clustering algorithm performed.

\textit{Gaia} kinematic data has paved the way for searching for OCs in the kinematic space, using positional and dynamical properties to identify OCs. \citet{2018A&A...615A..49C}, \citet{2020A&A...633A..99C} use unsupervised membership assignment based on photometric and astrometric measurements from \textit{Gaia}. \citet{2021ApJ...923..129J} employs Gaussian mixture models using the parallax and proper motions of stars, while \citet{Hunt2023} uses a hierarchical density clustering algorithm to identify OCs based on kinematic information provided by \textit{Gaia}. 

While these previous studies have demonstrated some of the challenges encountered in recovering OCs using different clustering algorithms, no comprehensive study has yet been conducted which evaluates both the chemical \textit{and} dynamical parameters of OC stars on how well they perform when used as clustering parameters. In this work we aim to thoroughly test the recovery rates of known OCs from the entire \textit{Gaia} DR3-GALAH DR4 catalog (hereafter, GGC), from optimal chemodynamical parameters using blind clustering algorithms. Furthermore, we test if we can improve the recovery rates based on cuts to the GGC using prior knowledge of the cluster properties including their metallicities, ages and birth radii. 


We discuss our choice of surveys and OC catalogs in Section \ref{sec:data}. Investigations into the choice of chemodynamical parameters is presented in Section \ref{sec:methods}, along with our high-dimensional clustering methodology. We report our results for optimal chemodynamical parameters and efficacy of using prior knowledge to retrieve the OCs in Section \ref{sec:Results}. The impact of these results on the field of chemodynamical tagging and our conclusions are found in Section \ref{sec:Discussion} and \ref{sec:Conclusions}, respectively.

\section{Data} \label{sec:data}

The aim of our study is to test optimal chemodynamical clustering methods that can recover open star cluster members from large catalogs, and their efficacy. We thus require two things: a large selection of stars that have both astrometric information and chemical abundances, and a sample of OCs that have verified member stars. We aim to measure how well parameters from these two components can be implemented to disentangle and retrieve OCs from field stars in the GALAH field. 

We employ data from the \textit{Gaia} DR3 survey for astrometric data \citep{2021A&A...649A...1G} and the GALAH-DR4 survey for chemical abundances \citep{2024arXiv240919858B}. With the release of Data Release 3, the \textit{Gaia} survey has measured precise positions, proper motions, and radial velocities of 1.8 billion sources \citep{2021A&A...649A...1G}. The GALAH-DR4 survey has obtained spectra for over $900,000$ nearby stars using the HERMES spectrograph at the Anglo-Australian Telescope \citep{2021MNRAS.506..150B}. Each star in the GALAH survey can have up to 27 elemental abundances determined, which include Li, C, O, Na, Al, K, Mg, Si, Ca, Ti, Sc, V, Cr, Mn, Co, Ni, Cu, Zn, Y, Ba, La, Rb, Mo, Ru, Nd, Sm, and Eu. However, the GALAH survey does not have measured abundances for all 27 elements for every star in its catalog. Thus, we restrict the elements we use as chemical parameters in our clustering tests to O, Na, Mg, Al, Si, K, Ca, Sc, Ti, Cr, Mn, Ni, Cu, Y, and Ba. Over 60 percent of the stars in the GCC have abundance measurements for each of these elements. The elements that have abundances for less than 60 percent of stars in the GGC (Li, C, V, Co, Zn, La, Rb, Mo, Ru, Nd, Sm, and Eu), we do not consider in our evaluation of chemodynamical parameters and their effectiveness at recovering OCs. At the intersection of these two surveys, we have a cross-matched catalog of $\sim917,571$ stars that have astrometric measurements from \textit{Gaia} and abundance measurements from GALAH.

In addition, we employ OC data from a catalog compiled by \citet{2020A&A...633A..99C}, which contains cluster members for $1,481$ Milky Way OCs. There are $1,205$ stars that are spread among 54 clusters from the \citet{cg2020} catalog that have measurements in the GGC. Following work done by \citet{2022AJ....164...85M}, we impose an additional constraint to the OC members that the [Fe/H] measurements of the members stars must fall within 3$\sigma$ of the OC's [Fe/H] mean abundance. We decided to only use star clusters that have 10 or more member stars as not to bias our results to parameter combinations that are only able to recover OCs with low membership counts. This leaves us with 29 OCs with 1,082 total stars, (hereafter, OCC) which are listed in Table \ref{tab:ocs}, along with their median metallicities taken from the GALAH [Fe/H] abundances. 

We also take derived stellar parameters from the GGC in order to determine intrinsic properties of each OC in the OCC. In Sections \ref{sec:dyn}, \ref{sec:ages}, and \ref{sec:br} we outline how we derive orbital parameters, stellar ages, and stellar birth radii for each star in the OCC, respectively. The final OCC has OCs that are varied in their intrinsic properties, such as their metallicities, ages, birth radii, and membership counts, which allows us to probe any possible dependencies on clustering recovery rates to the types of OC.


\begin{deluxetable*}{l|ccccccc}




\tablecaption{Properties of the OCC sample used in this work. $\sigma_{\rm [Fe/H]}$ corresponds to the standard deviation of GALAH DR4 [Fe/H] measurements for member stars in that cluster. Similarly, $\sigma_{\rm Age}$ is the median age dispersion from the MIST isochrone ages determined in Section \ref{sec:ages}. Finally, $\sigma_{R_b}$ is the standard deviation of $R_b$ values determined for each OC in Section \ref{sec:br}.}
\label{tab:ocs}

\tablenum{1}

\tablehead{\colhead{Cluster} & \colhead{$N_{\rm members}$} & \colhead{[Fe/H]} & \colhead{$\sigma_{\rm [Fe/H]}$} & \colhead{Age} & \colhead{$\sigma_{\rm Age}$} & \colhead{$R_b$} & \colhead{$\sigma_{\rm R_b}$}\\
\colhead{} & \colhead{} & \colhead{dex} & \colhead{dex} & \colhead{Gyr} & \colhead{Gyr} & \colhead{kpc} & \colhead{kpc}} 

\startdata
Alessi 24  & 13 & -0.28 & 0.31 & 3.8 & 2.7 & 8.0 & 4.0\\
Berkeley 32  & 11 & -0.27 & 0.17 & 5.9 & 3.1 & 6.0 & 2.0\\
Blanco 1  & 35 & -0.11 & 0.14 & 3.2 & 1.9 & 7.0 & 2.0\\
Collinder 135  & 27 & -0.05 & 0.08 & 6.9 & 3.2 & 3.0 & 3.0\\
Collinder 140  & 13 & -0.13 & 0.23 & 5.5 & 3.1 & 5.0 & 3.0\\
Collinder 261  & 167 & 0.03 & 0.12 & 5.5 & 2.8 & 4.0 & 2.0\\
IC 4651  & 18 & 0.0 & 0.07 & 2.8 & 0.6 & 6.0 & 1.0\\
IC 4665  & 22 & -0.11 & 0.2 & 8.2 & 5.1 & 3.0 & 5.0\\
Mamajek 4  & 13 & 0.07 & 0.08 & 3.5 & 3.5 & 5.0 & 3.0\\
Melotte 22  & 52 & -0.05 & 0.14 & 4.6 & 3.5 & 5.0 & 4.0\\
Melotte 25  & 36 & 0.05 & 0.14 & 4.8 & 3.4 & 4.0 & 3.0\\
NGC 1901  & 17 & -0.22 & 0.19 & 3.5 & 2.7 & 8.0 & 3.0\\
NGC 2112  & 28 & -0.1 & 0.15 & 2.7 & 1.1 & 7.0 & 2.0\\
NGC 2204  & 35 & 0.15 & 0.21 & 1.5 & 1.8 & 5.0 & 2.0\\
NGC 2232  & 14 & -0.27 & 0.26 & 4.4 & 3.9 & 8.0 & 6.0\\
NGC 2360  & 12 & -0.35 & 0.21 & 2.9 & 1.4 & 10.0 & 3.0\\
NGC 2451B  & 27 & -0.09 & 0.16 & 7.4 & 3.1 & 4.0 & 3.0\\
NGC 2516  & 47 & -0.18 & 0.21 & 4.2 & 3.6 & 7.0 & 4.0\\
NGC 2548  & 26 & -0.37 & 0.2 & 1.2 & 0.3 & 12.0 & 2.0\\
NGC 2632  & 70 & 0.12 & 0.09 & 4.4 & 3.2 & 4.0 & 2.0\\
NGC 2682  & 158 & -0.0 & 0.08 & 5.5 & 3.0 & 4.0 & 2.0\\
NGC 3114  & 16 & -0.33 & 0.24 & 2.2 & 1.1 & 10.0 & 3.0\\
NGC 5822  & 17 & -0.1 & 0.28 & 2.1 & 2.3 & 8.0 & 3.0\\
NGC 6124  & 36 & -0.16 & 0.24 & 2.0 & 2.1 & 9.0 & 3.0\\
NGC 6253  & 17 & 0.27 & 0.1 & 4.7 & 2.1 & 2.0 & 1.0\\
Ruprecht 145  & 10 & -0.19 & 0.13 & 1.4 & 0.5 & 9.0 & 1.0\\
Ruprecht 147  & 65 & 0.06 & 0.07 & 5.1 & 2.8 & 4.0 & 2.0\\
Trumpler 10  & 59 & -0.04 & 0.1 & 6.6 & 4.1 & 4.0 & 3.0\\
UPK 612  & 31 & -0.18 & 0.25 & 4.1 & 3.0 & 7.0 & 4.0\\
\enddata



\tablerefs{\citet{cg2020}}
\end{deluxetable*}

\subsection{Dynamical Parameters} \label{sec:dyn}

OCs begin to dissipate due to dynamical interactions after about 100 Myr \citep{1988AJ.....95..771J}, which would hinder our ability to recover clusters using their current kinematic positions (right ascension, declination, proper motions, and radial velocities) as a clustering tool. To circumvent this issue, we can transform a star's positional coordinates ($\textbf{x,v}$) to action-angle coordinates (\textbf{$\theta$,J}), which remain relatively more static than positional coordinates in a slow varying Milky Way potential (\citealt{2008gady.book.....B}, \citealt{2014MNRAS.441.3284S}, \citealt{2016MNRAS.457.2107S}). Using the positions, proper motions, and radial velocities from the \textit{Gaia} survey, we derive action coordinates ($J_R$, $J_\phi$, $J_Z$) and orbital energy ($E$) for all of the stars in the OCC as well as the stars in the GGC. We derive the parameters using the \texttt{galpy} Python package, which takes the positional data and computes the dynamical information in a given Milky Way potential \citep{2015ApJS..216...29B}. We use the {\fontfamily{qcr}\selectfont McMillan17} potential derived in \citet{2017MNRAS.465...76M} in order to calculate orbital actions. The {\fontfamily{qcr}\selectfont McMillan17} potential is a simplistic axisymmetric potential which allows one to calculate actions using the Stackel Fudge technique \citep{2014MNRAS.441.3284S}. Multiple observational constraints of the Milky Way were used to fit the potential, which makes it representative of the true Milky Way potential. In the {\fontfamily{qcr}\selectfont McMillan17} potential, the Sun's distance from the Galactic center is defined as $R_0=8.20$kpc, and its tangential velocity is defined as $v_0=233$km s$^{-1}$. 

\begin{figure}
    \centering
    \includegraphics[width=\linewidth]{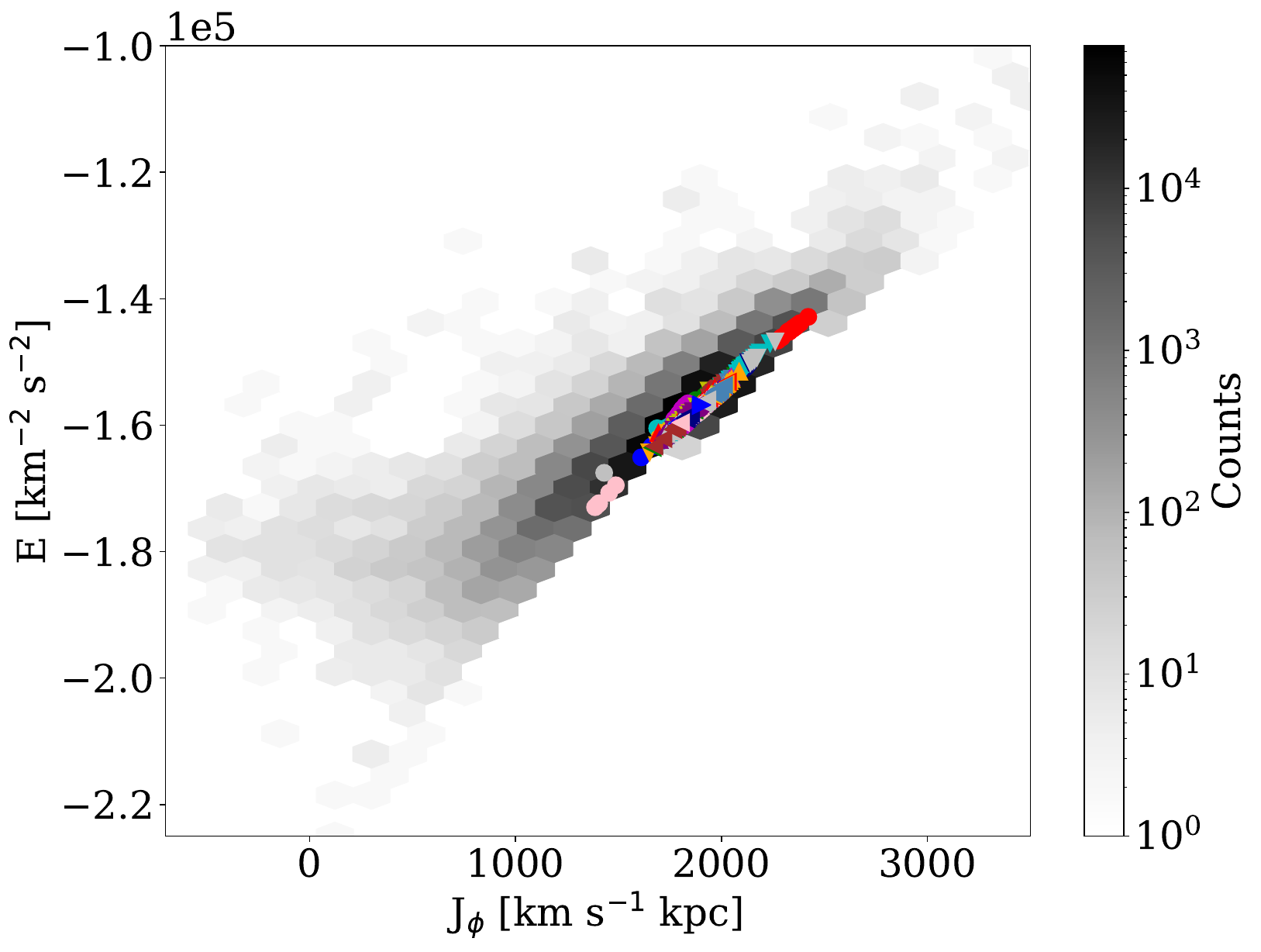}
    \includegraphics[width=\linewidth]{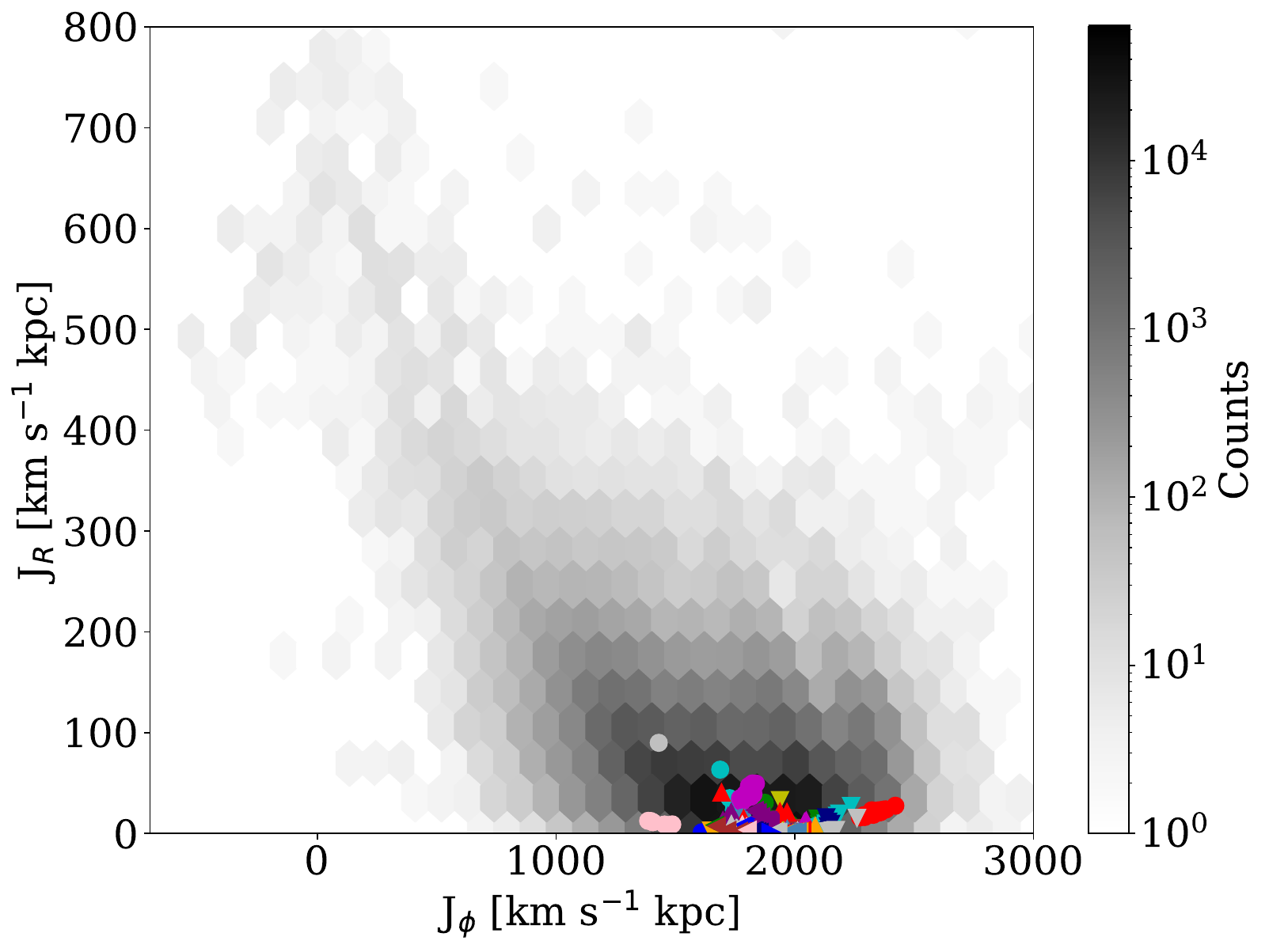}
    \caption{\textit{Top:} Distribution of stars in the $E-J_\phi$ space for stars in the GGC. The greyscale, 2D-histogram represents the entire GGC in the $E-J_\phi$ space. Overlaid in solid colors are the members stars of OCs from the OCC, colored by cluster. \textit{Bottom:} Same as the top panel, but for the $J_R - J_\phi$ space.}
    \label{fig:actions}
\end{figure}

In a slow varying potential, a stellar cluster should remain grouped in action space even if the cluster dissolves spatially across its host galaxy \citep{2008gady.book.....B}. As can be seen in Figure \ref{fig:actions}, stars in the same OCs are grouped together in action space. OCs that lie outside of the MW disk (low $J_\phi$, high $J_R$), are more separated from other OCs in action space, while the OCs within the MW disk overlap heavily in action space. It is unsurprising that OCs are grouped together in action space, since they are still physically associated structures. It should be noted however, that there are many components of the Milky Way that can lead to the dissolution of star clusters and affect the distribution of stellar actions as they orbit the galactic plane, including giant molecular clouds (GMCs), the Milky Way spiral arms, and the Milky Way bar (\citealt{2014MNRAS.441.3284S}, \citealt{2016MNRAS.457.2107S}). 

\subsection{Stellar Ages} \label{sec:ages}

When compared to the formation timescale of the Milky Way, OCs have a small spread in ages among their member stars, which could drastically help our clustering efforts if we could limit the GGC to similarly aged stars to each OC. However, the determination of stellar ages is a long-standing issue in astronomy (e.g. \citealt{2008ApJ...687.1264M}). With the stellar parameters available in the GGC, we can roughly estimate the ages of the stars in the GGC using the open-source package {\fontfamily{qcr}\selectfont kiauhoku}\footnote{\tt https://github.com/zclaytor/kiauhoku} \citep{2020ApJ...888...43C} to fit isochrones to stellar parameters. 
Stellar isochrones and evolutionary tracks are common tools used for inferring the ages of star clusters. Various stellar models exist, each with different choices in model physics that impact the timescale of the stellar evolution process (e.g. MESA Isochrones and Stellar Tracks, MIST, \citealt{2016ApJ...823..102C}; \citealt{2011Paxton} PAdova and tRieste Stellar Evolutionary Code, PARSECv2.0, \citealt{2022A&A...665A.126N}; Dartmouth Stellar Evolution Program, DSEP, \citealt{2008Ap&SS.316...31D}; \citealt{2006Bjork&Chaboyer}; Yonsei-Yale, Y$^2$/YaPSI \citealt{2017ApJ...838..161S}; A Bag of Stellar Tracks and Isochrones, BaSTI, \citealt{2024MNRAS.527.2065P}). We use {\fontfamily{qcr}\selectfont kiauhoku} to estimate the GGC ages using the surface gravity, metallicity, and effective temperature across four model grids; Yale Rotation Evolution Code (YREC; \citealt{2022ApJ...927...31T}; \citealt{1989ApJ...338..424P}), MIST, DSEP, and Garching Stellar Evolution Code (GARSTEC; \citealt{2013MNRAS.429.3645S}; \citealt{2008Ap&SS.316...99W}). The models used here are the same model grids as those in \citealt{2022ApJ...927...31T}, computed with the same input physics and starting conditions. Differences include choices in atmospheric boundary conditions, implementation of mixing length theory, and diffusion, among others. For a detailed overview of the input physics, we refer the reader to Table 1 of \citealt{2022ApJ...927...31T}. We search within a mass range of 0.6 to 2.0 $\mathrm{M}_\odot$, metallicity range [Fe/H] = –1.0 to +0.5, and from the zero-age main sequence to the tip of the red giant branch. However, as the MIST models extend into later evolutionary phases, we allow these tracks to reach the asymptotic giant branch. Stars outside of the -1.0 to +0.5 [Fe/H] range are unable to be adequately fit by the model grids, therefore we exclude them from the rest of this work, which brings the total number of stars in the GGC to 483,433. 

\begin{figure}
    \centering
    \includegraphics[width=\linewidth]{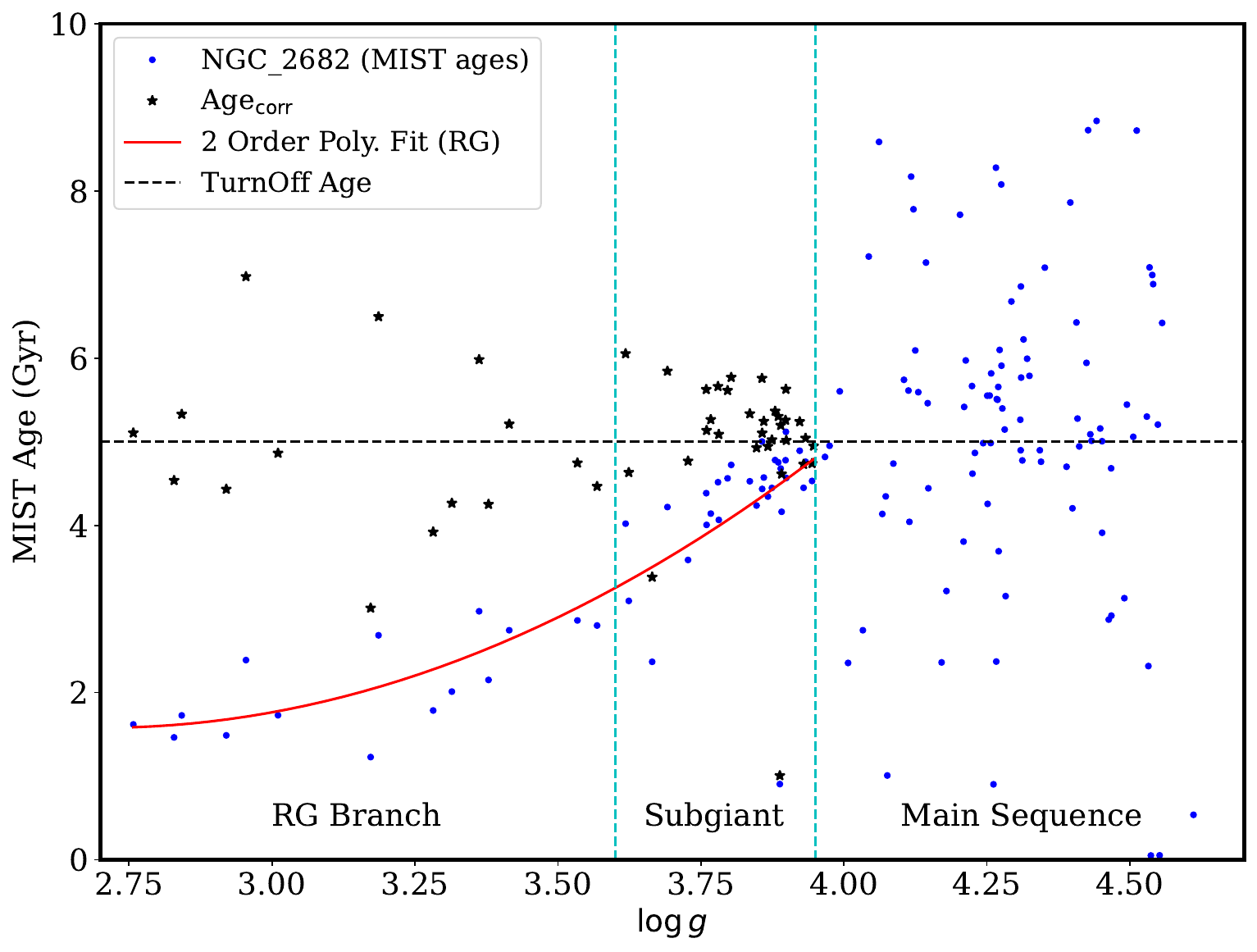}
    \caption{Relative age correction based on the OC NGC 2682. We show the separation between the main sequence, subgiant, and red giant branch with the dashed, cyan vertical lines. The horizontal dashed, black line indicates the turn-off age for NGC 2682. The MIST derived ages are plotted as a function of $\log g$, shown as blue points. The red line shows a 2$^{\rm nd}$ order polynomial fit to the RGB stars, defined as having a $\log g  < 3.95$. The black stars show the ages of the RGB stars after applying the relative age correction.}
    \label{fig:age_corr}
\end{figure}

{\fontfamily{qcr}\selectfont kiauhoku} operates by resampling evolutionary tracks to equivalent evolutionary phases (EEPs; \citealt{2016Dotter}) and then interpolating stellar parameters based on initial mass, metallicity, and EEP \citep{2022ApJ...927...31T}. The combination of initial mass, metallicity, and EEP serves as a starting point for Nelder-Mead optimization \citep{Nelder&Mead} with a mean squared error loss function, achieving convergence to the solution setting the tolerance to $10^{-6}$ for a successful fit. While we opt to use the MIST ages for the rest of this work, we computed ages using four different model grids to estimate the systematic errors arising from different input physics and modeling techniques. Recent studies have shown that mass offsets between model grids are typically around 5\% during the main sequence and subgiant phases \citep{2022ApJ...927...31T}, yet can exceed 20\% as stars evolve onto the red giant branch (Morales et al., submitted). These mass differences lead to age offsets between models that can approach 100\%. Using the OC M67 (NGC 2682), \citet{2024RNAAS...8..201B} demonstrates that there is a systematic offset in ages for red giant branch (RGB) stars. By fitting a polynomial to the ages as a function of $\log g$ for the RGB stars in NGC 2682, as shown in Figure \ref{fig:age_corr}, we derive a relative function needed to correct the ages to the main sequence turn-off age. We apply this age correction on RGB stars in the GGC, defined as stars with a $\log g < 3.95$. Although this study focuses on relative ages, we advise caution when relying on evolutionary models for precise age measurements and recommend referring to \citealt{2022ApJ...927...31T} to identify areas most affected by model choice uncertainties.

\subsection{Birth Radii} \label{sec:br}

OCs consist of co-evolved stars that are born in the same region of space in the Galactic disk. However, over time, the angular momenta of the stars can be permanently changed by interactions with the spiral arms and the central bar of the MW (\citealt{2002MNRAS.336..785S}, \citealt{2008ApJ...675L..65R}, \citealt{2010ApJ...722..112M}, \citealt{2013A&A...553A.102D}, \citealt{2020A&A...638A.144K}), causing them to migrate away from their birthplace. 
Work done by \citet{2018MNRAS.481.1645M} has shown that there is a correlation between a star's metallicity and age with the radius at which it is born in the MW. 
Building upon this method, \citet{2022arXiv221204515L} developed a relation to obtain birth radii for stars in the MW disk empirically.
As birth radii can only be obtained for stars born after the MW disk has started to form \citep{2022MNRAS.515L..34L}, the method is only calibrated for stars with [Fe/H] $>-1.0$.
Since this method also did not take into account the growth of the Galactic disk \citep{2024arXiv241017326R}, the birth radii estimated should be treated as relative birth radii instead of absolute.
Using this relationship, we derive the relative birth radius for each star in the GGC. 
The hope is that clustering stars of a similar birth radii can potentially improve our recovery rates as stars in each cluster should have very similar birth radii. 
While this is a powerful tool when searching for dispersed star clusters, there are some limitations to the relationship. Namely, the method was calibrated on stars that were older than 2 Gyr and with [Fe/H] $>-1.0$. As a result, some of the younger or metal-poor stars in the OCC have abnormally large birth radii, indicating their ex-situ nature or deviation from the field population.

The birth radii are estimated using the equation taken from \citet{2022arXiv221204515L},
\begin{equation}\label{eq4}
    \mathrm{R_b(age, [Fe/H])} = \frac{\mathrm{[Fe/H]} - \mathrm{[Fe/H]}(0,\tau)}{\mathrm{\nabla [Fe/H](\tau)}}.
\end{equation}
In which $\mathrm{[Fe/H]}(0,\tau)$ and $\mathrm{\nabla [Fe/H](\tau)}$ are the derived relation of the metallicity evolution in the Galactic center, and the temporal evolution of the metallicity gradient for the MW interstellar medium, respectively.
The values for $\mathrm{[Fe/H]}(0,\tau)$ and $\mathrm{\nabla [Fe/H](\tau)}$ are taken from Table.~1 in \citet{2022arXiv221204515L}.

\section{Methods} \label{sec:methods}

To test different chemical and dynamical parameters to use for the optimal recovery of open star clusters in N-dimensional space, we use the Hierarchical Density-Based Spatial Clustering of Applications with Noise (HDBSCAN) algorithm \citep{McInnes2017} to group Milky Way OCs and compare predicted group labels to the true OC labels from \citet{Hunt2023}. While there are many clustering algorithms, HDBSCAN was chosen as it has been shown to be the most sensitive and effective clustering algorithm for recovering OCs in \textit{Gaia} data (\citealt{2020MNRAS.496.5101P}, \citealt{2021A&A...646A.104H}). We use OCs from the OCC  as a benchmark to test the performance of each chemodynamical parameter combination. Our clustering results are evaluated using the V-measure score, 

\begin{equation}
    v = \frac{(1+\beta) \times h \times c}{\beta \times h + c}
\end{equation}

which is a weighted average of the completeness, $c$, and homogeneity, $h$ of the recovered stars of an individual cluster. The completeness is defined as a measure of all members of a given class (in our case, an OC) which are assigned to the same statistical cluster. If one OC is spread over many statistical clusters, then the completeness is low ($c\sim0$), however if most of the stars in an OC are grouped in the same statistical cluster, then the completeness is high ($c\sim1$). The homogeneity measures that each statistical cluster contains mainly members of a single class. A statistical cluster with stars from many different OCs has a low homogeneity ($h\sim0$), while a statistical cluster with stars mostly from one OC will have a high homogenesity ($h\sim1$). We use a $\beta$ value of 1.0, which prescribes an equal weight to the completeness and homogeneity. V-measure scores closer to 1 indicate a better match of input cluster labels to HBDSCAN labels, while scores closer to 0 indicate a worse match \citep{rosenberg-hirschberg-2007-v}. We compute the V-measure score by comparing the true labels of the verified OC members from the OCC to the generated and predicted labels from HBDSCAN.

\subsection{Deriving Optimal Chemodynamical Clustering Parameters} \label{sec:optimal}

The challenge with chemodynamical clustering comes from the large dimensionality associated with a high number of parameters (\citealt{2015ApJ...807..104T}, \citealt{2021A&A...654A.151C}). As the number of dimensions increases, the sparser a dataset becomes in the high-dimensional space, which limits density based methods such as HBDSCAN. In this work, we aim to reduce number of parameters to cluster on by measuring whether parameter usefulness parameter patterns and combinations emerge and their success in groupings and recovering OCs from the overall GGC. 

\begin{figure}
    \centering
    \includegraphics[width=\linewidth]{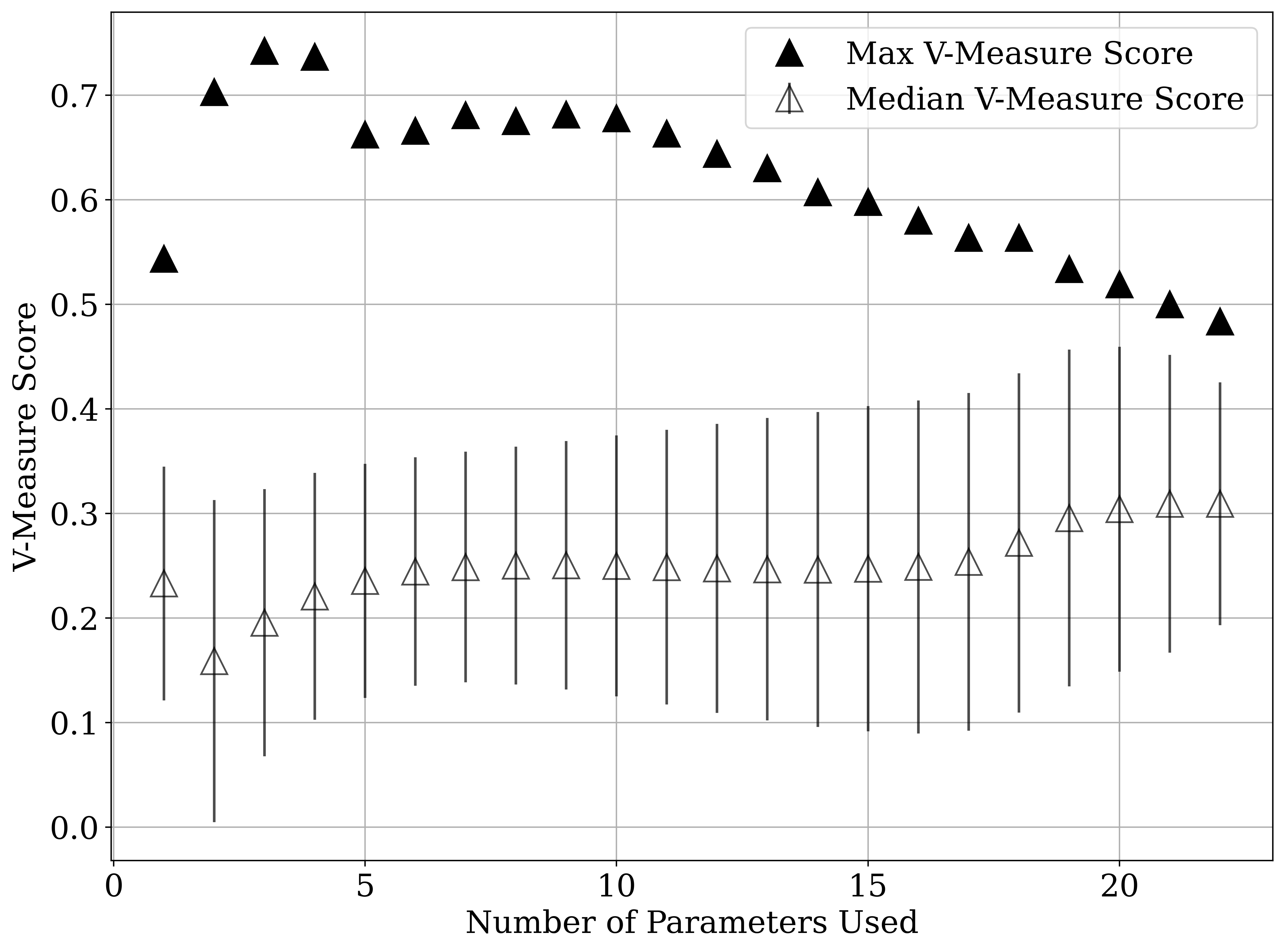}
    \caption{V-measure score determined from recovering OCs from the OCC as a function of number of parameters used in the clustering algorithm. The filled triangles represent the maximum V-measure score achieved with a specific number of parameters, while the unfilled triangles represent the median V-measure score. The error bars on the median V-measure score indicate the standard deviation of the V-measure score distribution.}
    \label{fig:dim_falloff}
\end{figure}

The first step that we take is meant to decrease the so-called \textit{curse of dimensionality} \citep{bellman1957dynamic}, which is commonly encountered in the field of machine learning. In general, as the number of parameters used in a clustering algorithm increases, the performance the algorithm decreases, because the intrinsic dimension of the data set is relatively large compared to number of data points. With the metallicity, orbital energy ($E$), action potentials ($J_\phi$, $J_Z$, $J_R$), and 15 chemical elements, there are 20 parameters to consider as clustering parameters, which amount to over 1 million unique parameter combinations. For an initial test, we apply the HBDSCAN on just the OCs in the OCC, where we run all different possible parameter combinations and measure the maximum V-measure score achieved as a function of number of parameters used in the clustering algorithm. As we increase the number of parameters, we determine the median and maximum V-measure score for all parameter combinations at each set number of parameters. We show in Figure \ref{fig:dim_falloff}, the median V-measure score achieved with any number of parameters used is between 0.2-0.3, with a spread in V-measure scores of $\sim0.1$. We find that a maximum V-measure score of $\sim0.7$ is achievable using 3-4 parameters and decreases as the number of parameters increases. Thus, in order to search for the best parameter set to maximize our recovery, we limit our future tests to only using parameter combinations of 2 to 5 parameters, which brings our total number of unique parameter combinations down to 9,108.

Before we attempt to recover OCs within the GGC, we wish to narrow down the parameters combinations we consider by testing which combinations achieve the highest V-measure score when only clustering OCs. We evaluate each parameter combination on how well they can recover OCs with mock field stars. We use the OCC and emulate field stars by generating twice the amount of stars than OC members, with abundances and dynamical values randomly sampled from the ranges found in the OCC. In order to get the average performance for each parameter combination, we perform a bootstrap analysis of randomly sampling from the OCC and synthesized field stars.

\begin{figure}
    \centering
    \includegraphics[width=\linewidth]{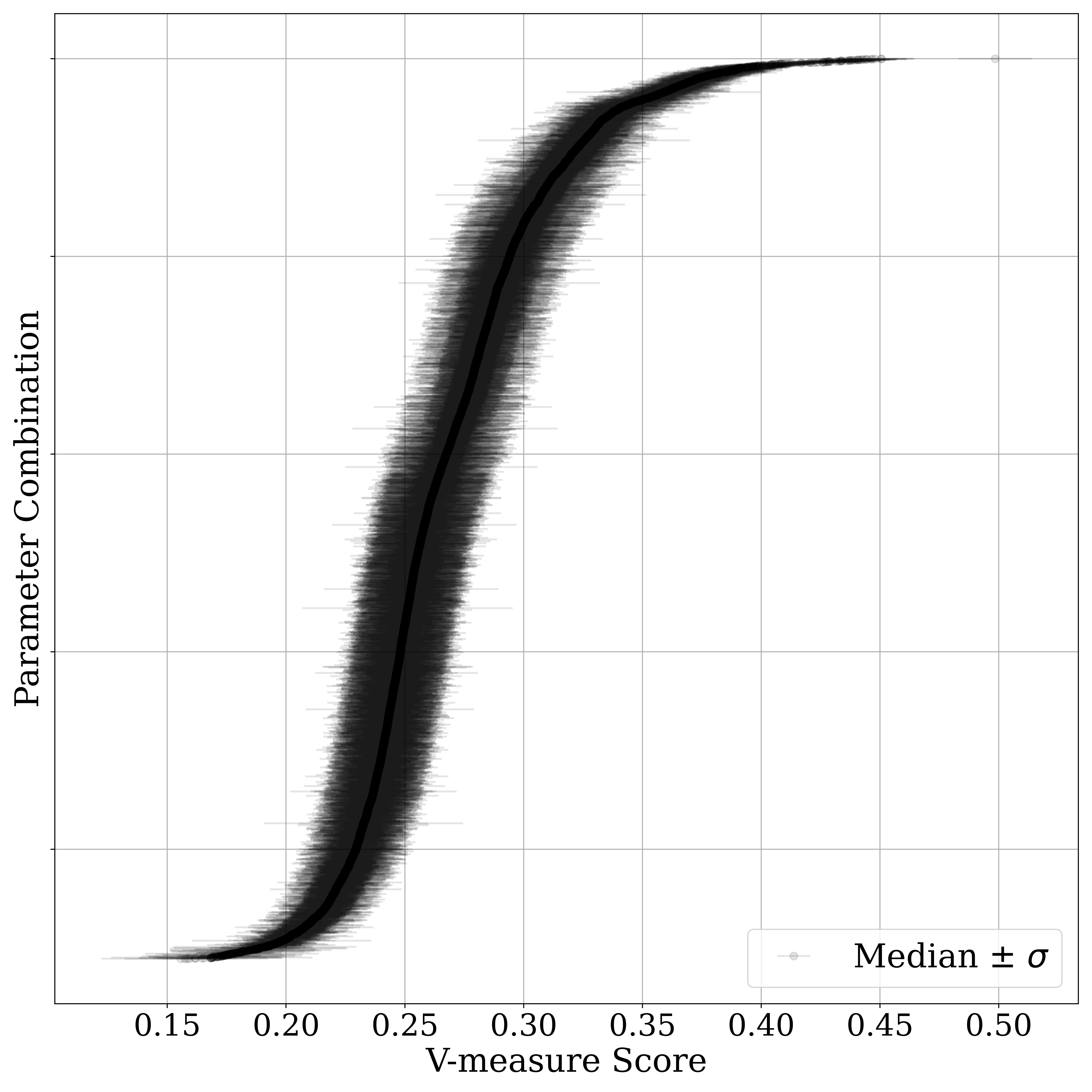}
    \includegraphics[width=\linewidth]{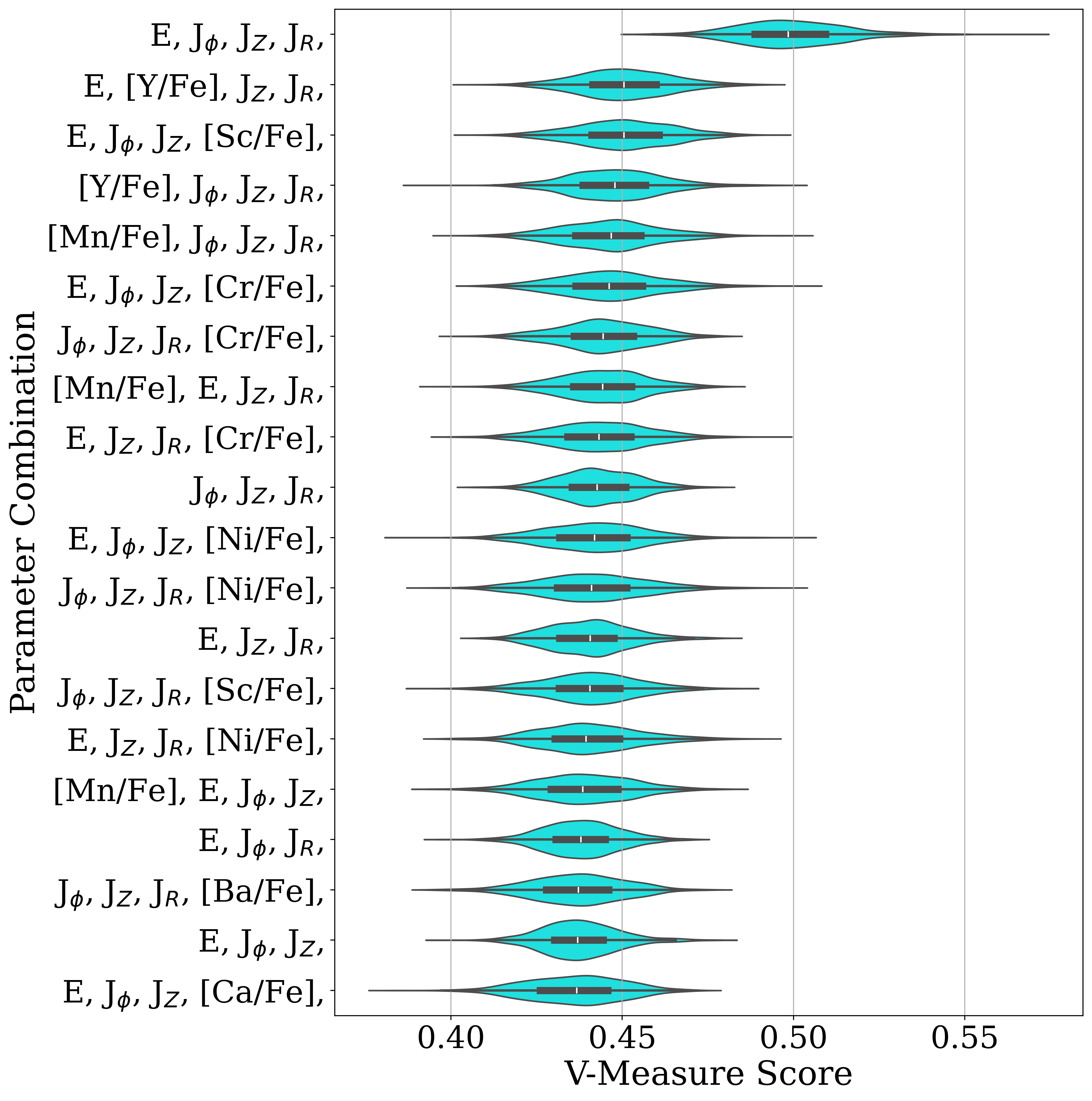}
    \caption{\textit{Top: }Median V-measure score for all 9,108 unique combinations of 2-5 parameters from 1000 random samplings of OCs and mock field stars. The list of parameters used in the clustering include $E$, $J_R$, $J_\phi$, $J_Z$, [Fe/H], and [X/Fe] abundances for O, Na, Al, K, Mg, Si, Ca, Ti, Sc, Cr, Mn, Ni, Cu, Y, and Ba. \textit{Bottom: }A zoom in of the top 20 parameter combinations with the highest median V-measure scores from out bootstrap analysis. The central point and horizontal bar for each combination show the median V-measure score and its associated standard deviation. The violin plots represent the distribution of V-measure scores for all 1,000 samplings of the bootstrap analysis.}
    \label{fig:param_all}
\end{figure}

For 1000 iterations, we pick a new set of 80\% OC and field star dataset stars each time and run HBDSCAN with all 9,108 unique parameter combinations, obtaining a distribution of V-measure scores for each unique parameter combination. The top panel of Figure \ref{fig:param_all} shows the median V-measure score and its associated standard deviation for all 9,108 different parameter combinations. We see that a majority of the parameter combinations fail to achieve a V-measure score of larger than 0.3, as applied to our OCC plus field stars. The bottom panel of Figure \ref{fig:param_all} shows the median V-measure score for the top 20 parameter combinations. The highest V-measure score we find is 0.5 for the dynamical parameter combination $E$, $J_\phi$, $J_Z$, and $J_R$. We note that among the top 20 V-measure parameter combinations, the dynamical parameters dominate, however for some cases, some chemical parameters including Y, Sc, Mn, Cr, Ni, Ba, and Ca also yield V-measures $\geq 0.4$ along with other dynamical properties. While on average a combination of $E$, $J_\phi$, $J_Z$, and $J_R$ provides the best V-measure score, our analysis indicates that there can be specific cases where adding various abundances as a clustering parameter work better at recovering OCs. 

\subsection{Recovering Open Clusters in the GGC}

Once we identified the optimal clustering chemodynamical parameters, we then proceed to evaluate how these parameter combinations are able to recover the OCs from the entire GGC. Recovering OCs in the GGC presents a tougher challenge than recovering OCs in the OCC, as our goal now is to recover a star cluster within a sample of hundred of thousands of field stars. For each of the top 20 parameter combinations found in Section \ref{sec:optimal}, we use HBDSCAN to split the GGC into statistical clusters, with numerical group labels assigned by HBDSCAN. By comparing the group labels to the true cluster labels from the OCC, we can evaluate how well each OC is recovered. For each OC, we compute a V-measure score for all of the statistical clusters that contains member stars of that OC. Statistical clusters that do not contain a member star for the targeted OC are not included when calculating the V-measure score. This method provides us with a V-measure score that represents how well each OC is recovered using each parameter combination. 

As a result of our training set being entirely made up of undissolved OCs, the method explored here will most likely only work best for identifying undissolved or perhaps partially dissolved star clusters. The ability to recover dissolved star clusters relies on the conservation of action-angles. In an ideal and slowly varying Milky Way potential the actions would remain nearly static, however perturbations due to the Milky Way's spiral arms, giant molecular clouds, or minor merger events, could alter the actions enough to make recovering dissolved clusters impossible (\citealt{2011MNRAS.413.2509G}, \citealt{2014MNRAS.441.3284S}, \citealt{2016MNRAS.457.2107S}). 


\section{Results} \label{sec:Results}

\subsection{Chemodynamical Tagging using HBDSCAN} \label{sec:blind}

\begin{figure*}
    \centering
    \includegraphics[width=\linewidth]{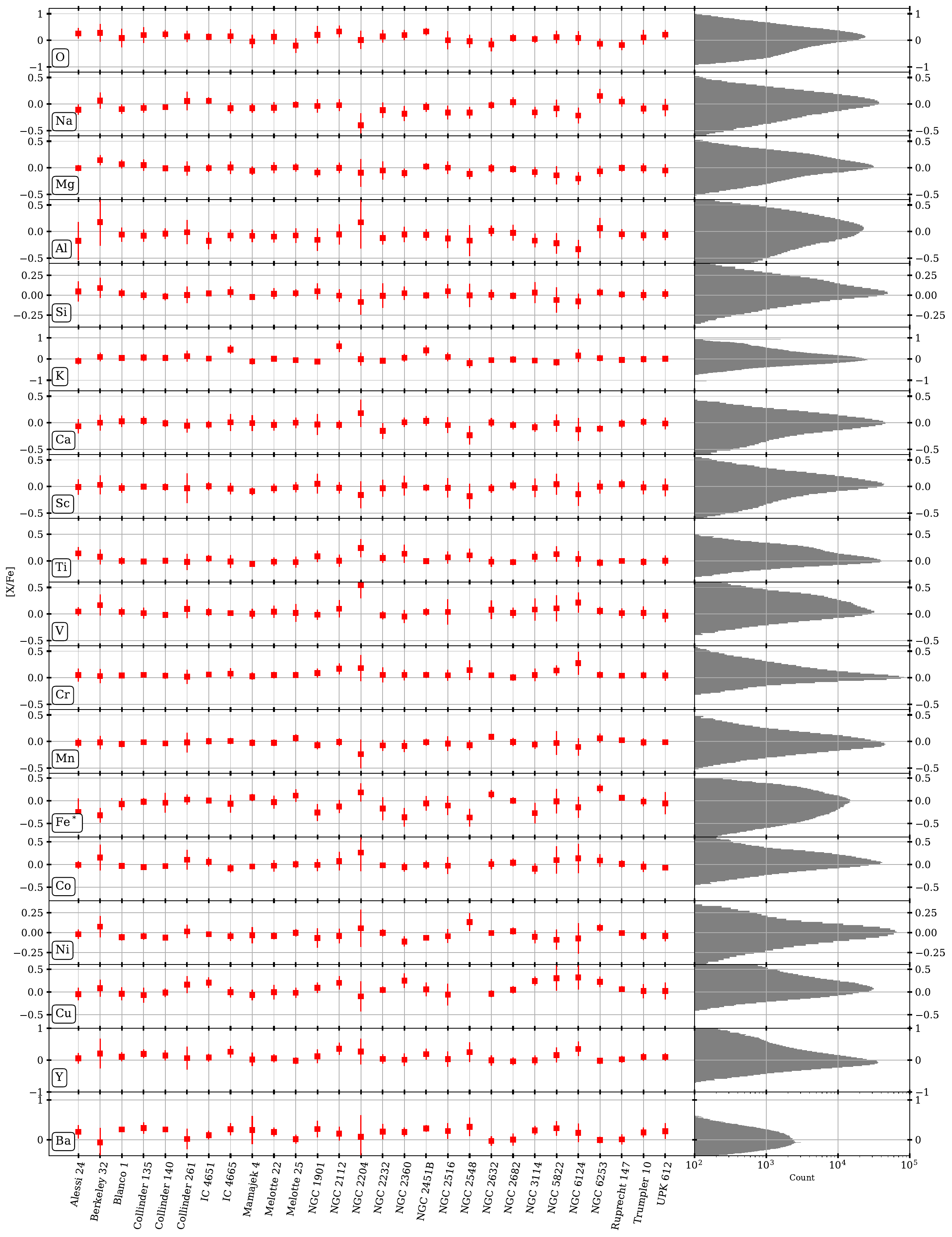}
    \caption{Distribution of GALAH element abundances for the OCC sample. The left panels show the [X/Fe] abundance for each element for each of the 28 OC, with error bars representing the standard deviation of abundance measurements within each cluster. The histograms in the right panels show the distribution of [X/Fe] abundances for the entire GGC. \textit{*Note:} For the element Fe, we plot the [Fe/H] abundance.}
    \label{fig:el_dist}
\end{figure*}

In Section \ref{sec:methods} we established a method to evaluate how specific parameter combinations perform in recovering OCs in the GGC with HDBScan clustering. For the 29 OCs in the OCC, using the top 20 parameter combinations based on our tests in Section \ref{sec:optimal}, we find an almost zero V-measure value score for most OCs. This result indicates that while the top 20 parameter combination derived from recovering the OCs from the OCC performed relatively better with V-measure scores $>0.4$ on average, it does more poorly when applied to the entire GGC using the same method and same parameters. 
Our results are similar to other studies that have attempted to apply automated chemodynamical tagging methods to OCs (e.g. \citealt{2013MNRAS.428.2321M}, \citealt{2015A&A...577A..47B}, and \citealt{2021A&A...654A.151C}). These previous studies, as well as the work presented here highlight the difficulty in automating the recovery of OCs using blind clustering algorithms. This can be mainly attributed to the fact that methods utilizing primarily chemical tagging to identify and recover star cluster members are faced with the large overlap in chemical abundance space across most elements among OCs, as well as with the chemical distribution in the GGC (see Figure \ref{fig:el_dist}). The overlap prevents blind methods from separating the OCs from the bulk of the GGC stars. In terms of the dynamical properties, which are more distinguishable among OCs as can be seen in Figure \ref{fig:actions}, we find slightly better success in recovering some of the OCs. However, as the OCs are still physically associated in the Galaxy and in their dynamical parameters with the large number of GGC stars, we find that recovering a small number of stars from the entire GGC is not feasible. We thus look into the possibility of increasing the recovery rates, by decreasing the size of the GGC by performing strategic cuts on the data.


\subsection{Applying [Fe/H], Age, and Birth Radii Restrictions} \label{sec:cuts}

In Section \ref{sec:blind}, we demonstrated that blind chemodynamical clustering techniques yield recovery rates near zero for most OCs . In some cases, prior information about the OCs, such as metallicity, age, and $R_b$, may be known of the star clusters in which there is interest to recover from a larger sample. These cases include pre-existing clusters for which we are interested to find additional member stars, or where there's interest to find the star cluster of one or more stars. In these cases, it is possible to reduce the large GGC by selecting stars that fall in a narrow range of prior parameters such as [Fe/H], age, and $R_b$. Due to the uncertainty of determining ages through spectroscopic stellar parameters (see Section \ref{sec:ages}), which also permeates to our determination of $R_b$, we decide to use these parameters to limit the GGC as opposed to using them as clustering parameters. We first try to establish a static cut size from the entire GGC, for each of these parameters chosen by the typical distribution of [Fe/H], age, and $R_b$ among our sample of OCs. We then apply our clustering method for each of these cuts and evaluate the V-measure scores for each OC as before.

The first static cut we apply is based on the [Fe/H] of each OC. For each OC, we limit the GGC to stars with a [Fe/H]$\pm0.3$ dex of the mean [Fe/H] value for that OC. We then apply our clustering method on the reduced GGC dataset, which is repeated for each OC, obtaining a new V-measure score based on a [Fe/H] cut specific to that OC. The next cut uses the ages determined in Section \ref{sec:ages}. The stellar ages determined for OCs in the OCC are sensitive to the evolutionary stage of the member stars, along with the uncertainty of stellar parameters provided from the GCC. For OCs that are dominated by giant stars, we have a narrow distribution of ages, however, in OCs that have dwarf member stars, the determined ages have a much wider distribution. The distribution of ages for each OC can be seen in the top middle panel of the figures in Appendix \ref{sec:app}. To account for the varying size of the age distributions, we are forced to take a large age cut of $\pm3$ Gyr to successfully include the member stars in our reduction of the GGC. We then apply a cut using the $R_b$ determined in Section \ref{sec:br}. Since the $R_b$ is derived from a relationship between a star's [Fe/H] and age, the large spread in ages also causes a large spread in the birth radii for a given OC. As a result, we take a conservative $\pm3$ kpc cut to the birth radii of the GCC in order to capture most of the targeted OC. 


Finally, we apply all three of the cuts simultaneously to the GGC for each OC, in order to focus solely on stars that have similar [Fe/H], age, and birth radii values. We find that even with the size of the GGC greatly reduced, the OCs did not show any significant improvement compared to the clustering results with no cuts applied to the GGC. It should be noted that each static cut is intended to reduce the size of the GGC, however, in some cases, the inter-cluster distribution of ages and birth radii is larger than the size of the cut, leading to some member stars being cut from the clustering group as well. This particularly affects clusters with low membership counts, such as Berkley 32, NGC 2232, and IC 4665.



\subsection{Effects of Different Nucleosynthesis Groups} \label{sec:elGroups}

We have so far demonstrated that OCs are incredibly difficult to recover among the large all-sky surveys such as \textit{Gaia} and GALAH, despite decreasing the size of the GGC with [Fe/H], age and $R_b$ cuts. While we find that the best overall parameter combinations include primarily different combinations of the dynamical parameters $E$, $J_\phi$, $J_Z$, and $J_R$, we also aim to test whether the inclusion of various nucleosynthesis groups , such as $\alpha$-elements, Fe-peak elements, and neutron capture elements, could improve the recovery of certain OCs \citep{2024ApJ...972...69M}. We thus repeat our clustering method using the top performing parameter combination from Section \ref{sec:optimal}, the dynamical parameters ($D = E, J_\phi, J_Z, J_R$), and add different representative elements from different nucleosynthesis groups. We use [O/Fe] and [Mg/Fe] for the $\alpha$-element group, [Mn/Fe] and [Fe/H] for the Fe-peak elements, and [Y/Fe] and [Ba/Fe] for the neutron capture elements. Using this procedure, we can compare the V-measure scores to properties of the OCs to see if there are specific circumstances where a particular nucleosynthesis group might be useful in recovering star clusters. The V-measure score for these combinations can be found in the solid areas of the bar graphs in Figure \ref{fig:ngc2632},  \ref{fig:ngc2682}, Figure \ref{fig:alessi24} as examples, and the plots in Appendix \ref{sec:app} for all other clusters.

Additionally, we test the recovery rates for applying [Fe/H], age, and $R_b$ cuts for each of these nucleosynthesis groups. From Section \ref{sec:cuts}, we see that a static cut size does not improve the V-measure score for most OCs. Thus, we explore how different flexible cut sizes perform by adopting a range of values for the [Fe/H], age, and $R_b$ and look for the maximum V-measure score we can achieve for each OC using a combination of different cut sizes for each parameter. For [Fe/H], we use cut sizes of 0.1, 0.2, 0.3, 0.4, 0.5, 1.0, 1.5, 2.0, 2.5, and $\infty$ (which indicates no cut was taken in the [Fe/H] values) and take all of the stars that in the GGC that have [Fe/H] values that fall within the median of each cluster's [Fe/H]$\pm$ the cut size. Similarly we test age cut sizes of 0.5, 1.0, 1.5, 2.0, 2.5, 3.0, 3.5, 4.0, 4.5, 5.0, and $\infty$ Gyr, and $R_b$ cut sizes of 2, 4, 6, 8, 10, 12, 14, 16, and $\infty$ kpc. We then apply our clustering method for each scenario using the chemical group combinations ($D+$ [O/Fe], [Mg/Fe], [Mn/Fe], [Fe/H], [Y/Fe], and [Ba/Fe]). The hatched shaded areas on the bottom right bar graphs of the plots in Figure \ref{fig:ngc2632} (NGC 2632), Figure \ref{fig:ngc2682} (NGC 2682), Figure \ref{fig:alessi24} (Alessi 24), along with the plots in Appendix \ref{sec:app}, show the max V-measure score found using a combination of cuts for each OC. 

\begin{figure*}
    \centering
    \includegraphics[width=\linewidth]{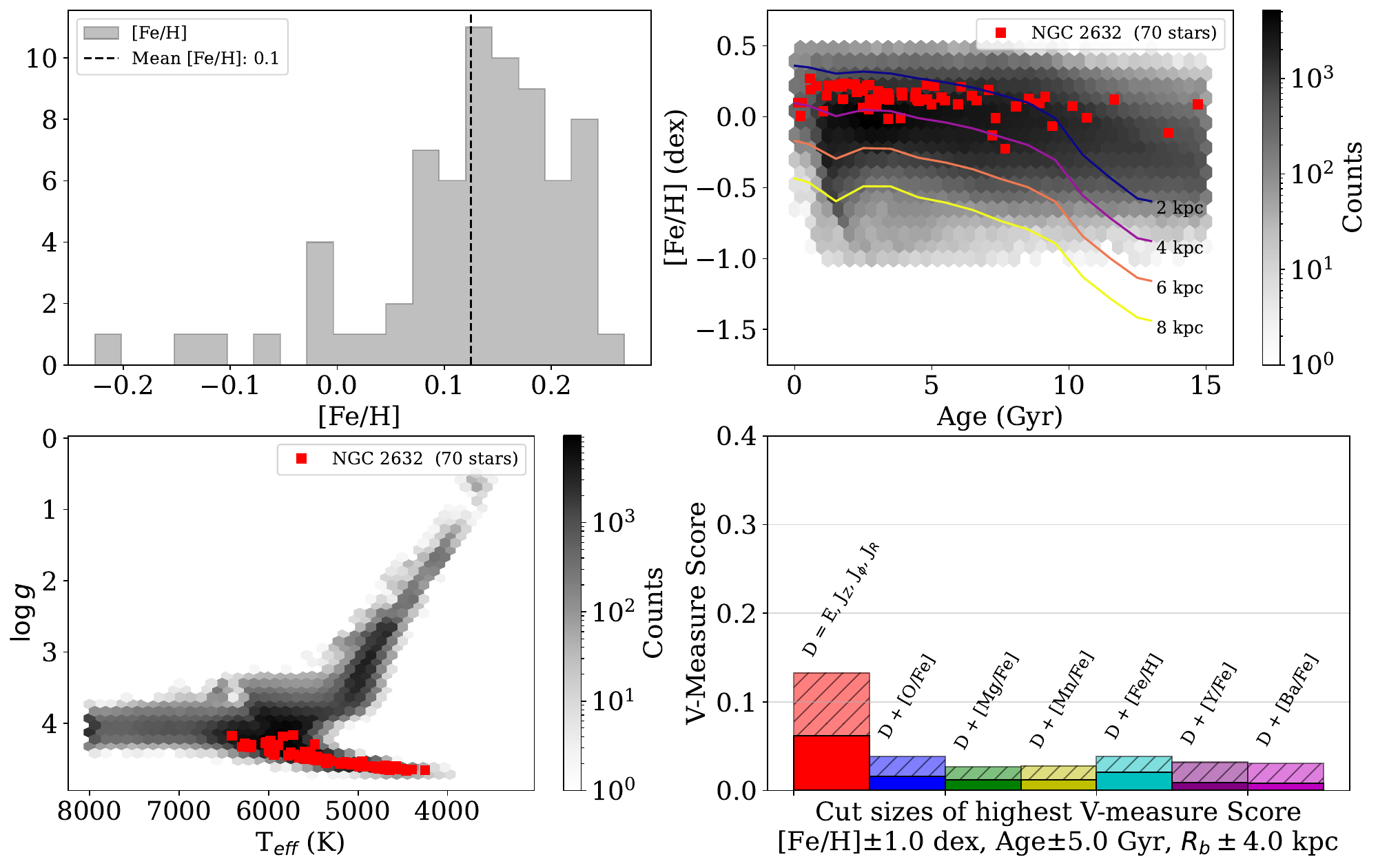}
    \caption{Characteristics of the OC NGC 2632. \textit{Top left}: Distribution of [Fe/H] abundances for member stars in NGC 2632. The mean metallicity from the \citet{2020A&A...633A..99C} sample is shown as the vertical, dashed line. \textit{Top right}: 2D histogram of GALAH [Fe/H] abundances and ages derived in this work using stellar evolutionary track fitting, using the MIST isochrone models. Overlaid as red squares are the member stars of NGC 2632. The horizontal lines correspond to mono-values of $R_b$ calculated from the GGC.  \textit{Bottom left}: Kiel diagram showing evolutionary stages of member stars of NGC 2632, compared to the GGC. \textit{Bottom right}: Our clustering results showing how well NGC 2632 can be recovered using different parameter combinations. The solid colored areas  represent the V-measure score of clustering on the entire GGC for different parameter combinations, while the hatched colored bars represents the best possible  V-measure scores achieved using a combination of [Fe/H], age, and birth radii parameter cuts. The plots for each OC from our sample are found in Appendix \ref{sec:app}.}
    \label{fig:ngc2632}
\end{figure*}

\begin{figure*}
    \centering
    \includegraphics[width=0.9\linewidth]{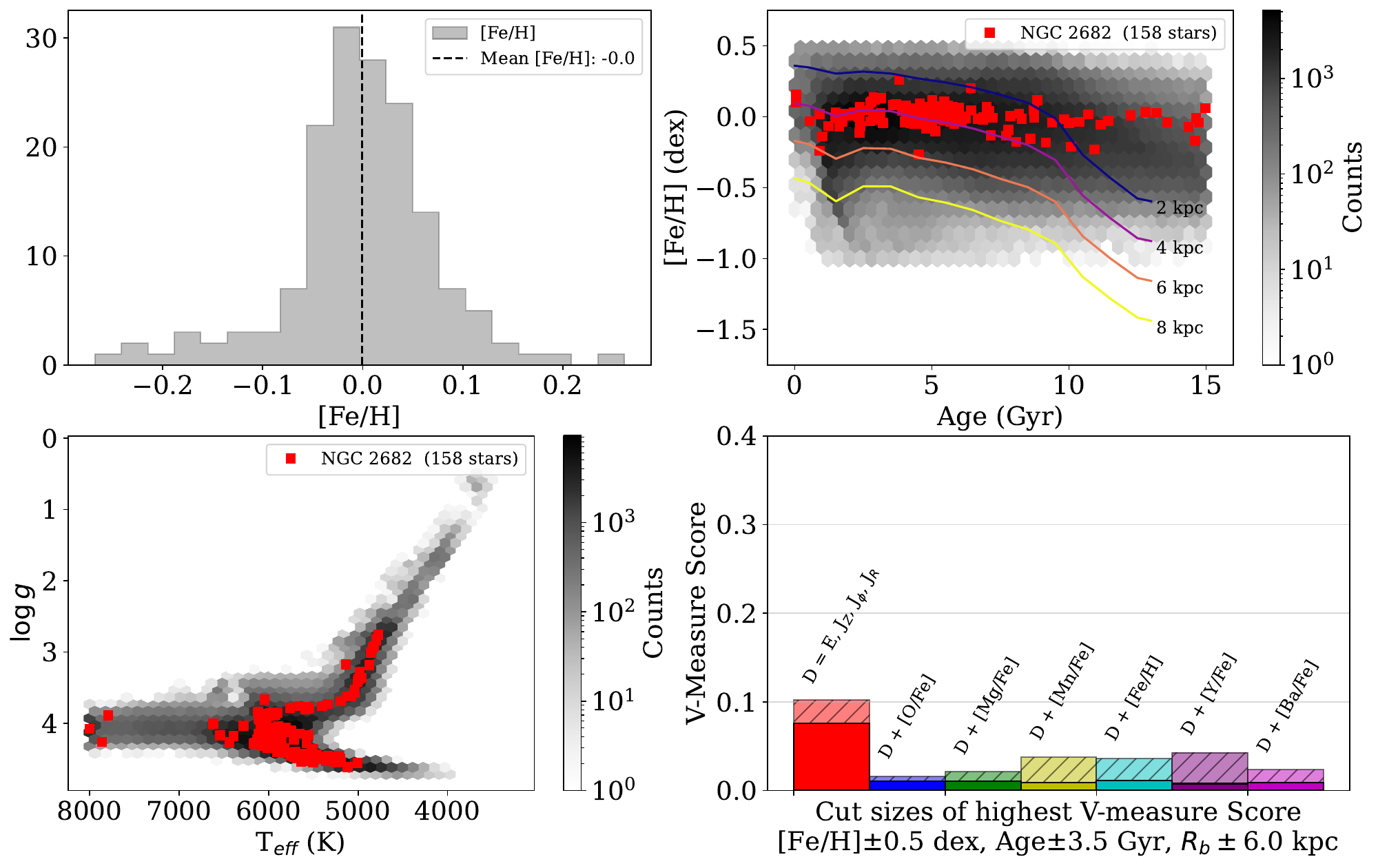}
    \caption{Same as Figure \ref{fig:ngc2632}, but for the OC, NGC 2682 (M67).}
    \label{fig:ngc2682}
\end{figure*}

\begin{figure*}
    \centering
    \includegraphics[width=\linewidth]{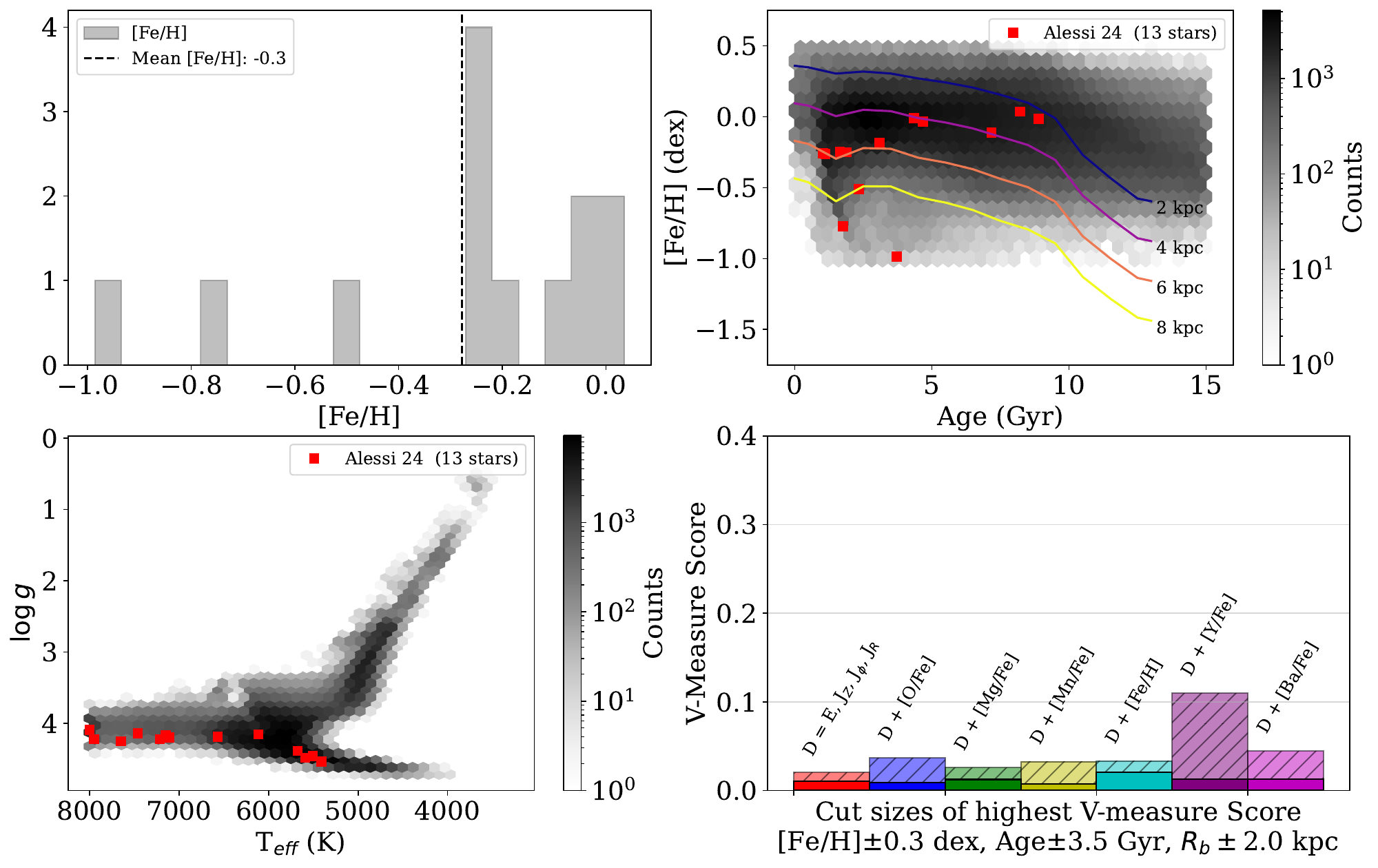}
    \caption{Same as Figure \ref{fig:ngc2632}, but for the OC, Alessi 24.}
    \label{fig:alessi24}
\end{figure*}

We find that the addition of an element abundance from different nucleosynthesis groups to the dynamical parameters ($E, J_Z, J_\phi, J_R$) can sometimes achieve a higher V-measure score than only using the dynamical parameters as clustering parameters. One example is for Alessi 24, which has a V-measure increase from near 0 to a V-measure score of 0.1 when Y is used as a clustering parameter, as can be seen in the bar graph in Figure \ref{fig:alessi24}. However, and as expected for OCs that have abundances similar to the bulk of the GGC, we do not find significant improvement to the V-measure scores with the inclusion of abundances from various nucleosynthesis groups.

While we find that by varying the size of [Fe/H], age, and $R_b$ cuts for each OC, we can marginally increase the V-measure scores from the near 0 values found with the static cuts performed in Section \ref{sec:cuts}, we fail to see any recovery rate above 0.1 for any of the OCs.



\section{Discussion} \label{sec:Discussion}

It has been demonstrated that chemical abundances alone struggle to accurately tag and recover star clusters (\citealt{2013MNRAS.428.2321M}, \citealt{2015A&A...577A..47B}, \citealt{2018MNRAS.473.4612K}, \citealt{2018ApJ...853..198N}, \citealt{2020MNRAS.496.5101P}, \citealt{2021A&A...654A.151C}). The use of dynamical parameters of stars has improved the recovery rate of star clusters, however dynamics alone struggles to identify clusters that are close to the Milky Way disc (\citealt{2018A&A...615A..49C}, \citealt{2020A&A...633A..99C}, \citealt{2021ApJ...923..129J}). The purpose of this work is to evaluate the combination of both chemical and dynamical parameters on their ability to recover OCs. Additionally, we explore improving the recovery rate by reducing the ratio between field stars to member stars by applying cuts to our catalogs based on prior knowledge of the OCs. We discuss the results of our work by addressing the following questions.

\textit{What are the chemodynamical parameters that return the highest recovery rate when used as clustering parameters in the HBDSCAN algorithm?} We performed a bootstrap analysis of the parameters available from the GGC in order to determine the best parameter combinations for recovering OCs. In our first test, we clustered only the OC members and a comparable amount of mock field stars, and found a maximum V-measure score of $\sim0.5$, while a majority of parameter combinations returned a V-measure score of $<0.3$. A maximum V-measure score of $\sim0.5$ indicates that on average, around half of the stars in each OC are recovered into the same statistical cluster. Similarly, the fact that most parameter combinations do not achieve a V-measure score of 0.3 indicates that a majority of parameter combinations do not recover even a third of most OCs into the same statistical clusters. Due to OCs still being physically associated structures, it is unsurprising that all of the top parameter combinations include at least one orbital dynamic parameter. From our bootstrap analysis, we find that the abundances of Ca, Sc, Mn, Cr, Ni, Ba, and Y, show up in the top parameter combinations. Interestingly, Mn and Cr have a low spread in average abundances among the OCs, however the range of Cr abundances in the entire GGC is quite narrow, which could make it easier to tag stars with offset Cr abundances. As can be seen in Figure \ref{fig:el_dist}, Ba and Y have large dispersions in the average OC abundances, ranging up to 0.5 dex, which could explain their usefulness as a clustering parameter. The usefulness of Ba and Y as a parameter to disentangle open clusters has been noted previously (\citealt{2009A&A...501..973D}, \citealt{2012ApJ...747...53M}, \citealt{2018A&A...617A.106M}, \citealt{2021A&A...654A.151C}). It should be noted however, that these abundances only perform well when paired with 3 dynamic parameters. 



The lack of elements in our top parameter combinations demonstrates the difficulty of chemical tagging star clusters. The challenge with chemical tagging is that the majority of OCs have overlapping chemical signatures (\citealt{2015A&A...577A..47B}, \citealt{2020MNRAS.496.5101P}, \citealt{2021A&A...654A.151C}). Similarly, \citet{2024arXiv240802228B} uses Milky Way-like galaxy simulations to show that the main issue with using the chemical abundances of OCs as clustering parameters is that the inter-cluster spread of abundances is so small that it is difficult to distinguish stars from different OCs, which can be seen in Figure \ref{fig:el_dist}.

\textit{Can we improve the recovery rates of OCs by using prior information of OC targets to trim the catalogs we are clustering?} The difficulty of chemodynamical tagging increases drastically when we apply our top parameter combinations in order to recover OCs when field stars are included from the GGC. When clustering, we must not only consider the overlapping chemical signatures between the OCs, but also the overlapping signatures from the field stars that largely outnumber the OC member stars. Work done by \citet{2018ApJ...853..198N} shows that $\sim30\%$ of random pairs of MW disk stars are chemically indistinguishable, which has the potential to add false-positives to identified cluster members. Other attempts at blindly clustering OCs using abundance surveys and automated clustering algorithms have also reported difficulty in accurately recovering the OCs (\citealt{2013MNRAS.428.2321M}, \citealt{2015A&A...577A..47B}). Particularly, \citet{2024A&A...688A.165S}, who performs a similar study to ours, but with APOGEE data, finds that chemodynamical tagging can be effective at recovering OCs with a 75\% recovery rate with no field stars, but when they add in mock field stars with chemical abundance ranges similar to their open cluster catalog they see a decrease in the performance of clustering algorithms. Similarly, we find that clustering the entire \textit{Gaia}-GALAH catalog leads to near 0 recovery rates for most OCs in our sample.





In an attempt to improve the clustering rates, we apply parameter cuts to the GGC based on mean metallicity, age, and $R_b$ of each OC and their respective spreads in our determined values (0.3 dex for metallicity, 3 Gyr for age, and 3 kpc in $R_b$) in order to reduce the ratio of field stars to member stars for a particular OC.
We find that despite these cuts, no significant improvement in V-measure scores can be obtained for a majority of OCs using static cut sizes, which indicates that there is no optimal cut size which can be used for all OCs. We find, however, more success in a variable cut size, but do not find any trend between the properties of the OCs and the size of the cuts. The lack of trend in the size of the cuts indicates that the ability to recover an OC depends moreso on where the cluster lies in chemodynamical space, and less on the spread in its member stars properties.

\begin{figure*}
    \centering
    \includegraphics[width=\linewidth]{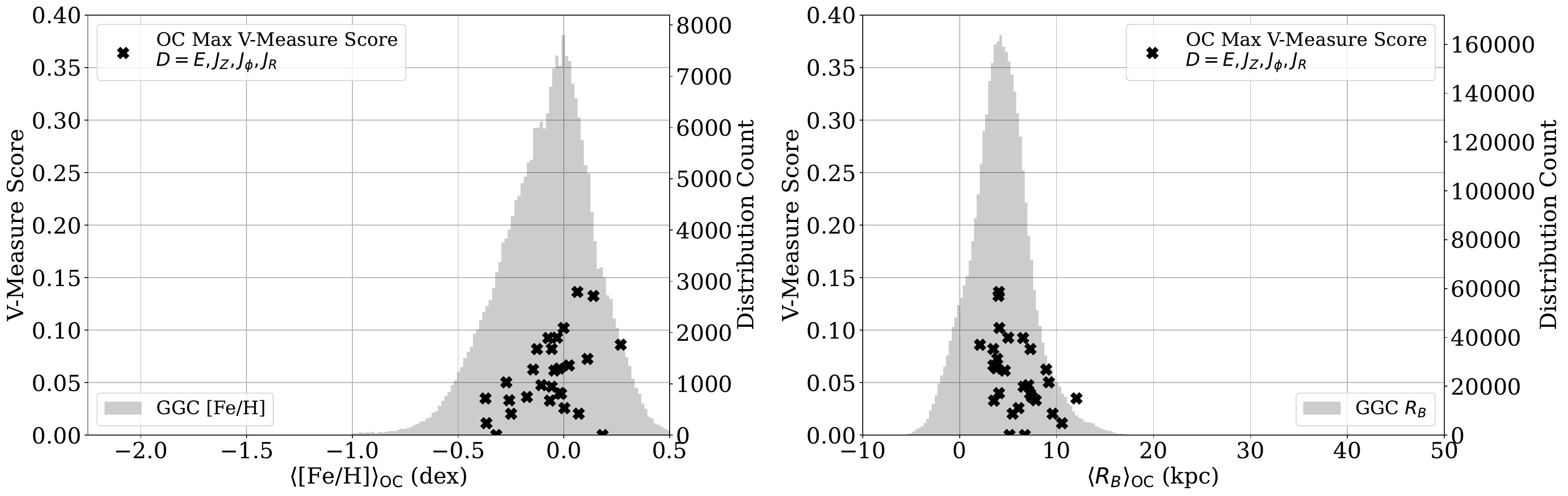}
    \caption{Maximum V-measure scores achieved for each OC in this work's sample, as a function of its mean [Fe/H] (\textit{left}) and $R_b$ (\textit{right}), are shown with filled Xs. The histogram represents the overall distribution of [Fe/H] and $R_b$ measurements in the GGC.}
    \label{fig:oc_trends}
\end{figure*}

\textit{What properties of OCs make them easier to be recovered?} When an OC's mean metallicity, age, or $R_b$ lies close to the bulk of the values of field stars from the GGC, even small cuts centered around the mean parameters will still have thousands of field stars compared to member stars. In Figure \ref{fig:oc_trends}, we plot the highest V-measure achieved for each OC as a function of the OCs' mean [Fe/H] and $R_b$. We see that all of our OCs lie within the overall [Fe/H] and $R_b$ distributions of the GGC. Broad cuts of [Fe/H] and $R_b$ still leave hundreds of thousands of stars in the clustering catalog, making the recovery of individual OCs infeasible. While spectroscopic surveys have vastly improved our ability to determine [Fe/H] for many stars, the process to determine stellar ages, and subsequently $R_b$, is much more uncertain (see Section \ref{sec:ages}).


\textit{What are the current limitations of chemodynamical tagging?}

There are unavoidable complications concerning chemodynamical tagging of OCs, including the overlap in chemical space and the overlap in dynamical space when in the disk of the galaxy. A majority of OC stars are young, and have yet to be perturbed by interactions within the disk, which leads to narrow ranges of $J_Z$ and $J_R$ in which OCs reside. Furthermore, the orbital values of a star is heavily dependent on the choice of Milky Way potential (\citealt{2008gady.book.....B}, \citealt{2014MNRAS.441.3284S}, \citealt{2016MNRAS.457.2107S}). Current work on improving our models of the Milky Way potential, including calculating non-axisymmetric potentials (\citealt{2018MNRAS.474.2706B}, \citealt{2019A&A...627A.123B}) and implementing time-dependent potentials to trace back stars to their birth cluster (\citealt{2015MNRAS.450.2812H}, \citealt{2019ApJ...879L..15H}, \citealt{2023A&A...673A.152I}), could further improve the precision of our calculated stellar actions. In terms of the chemical parameters, the GALAH survey has brought an unprecedented number of chemical abundances for hundreds of thousands of stars, however, the abundance uncertainties can reach up to 0.2-0.3 dex for even easily measurable elements, such as Fe and Mg \citep{2021MNRAS.506..150B}. 

Additionally, we mentioned that some elements were excluded from consideration because of the lack of abundances for many stars in the GGC. These elements, especially the neutron capture elements like Nd and Eu, have already been shown to be powerful tools in separating stellar populations (e.g. \citealt{2008ARA&A..46..241S}, \citealt{2011MNRAS.412.1203N}, \citealt{2014ApJ...797..123H}, \citealt{2018A&A...609A..13K}, \citealt{2024ApJ...972...69M}). As more spectroscopic studies map out more of the neutron capture abundances of stars in the Milky Way, we can evaluate how heavy element abundances perform at tagging stars to specific star clusters \citep{2024ApJ...972...69M}. The addition of more elements will also allow for different ``chemical" clocks to be used. Abundances are typically reported with respect to Fe (i.e. [X/Fe]), however, these ratios are sensitive to the timescale at which Type Ia supernovae produce Fe (\citealt{2009ARA&A..47..371T}, \citealt{2015A&A...580A.129S}, \citealt{2019A&A...626A..15H}). We could instead compare abundance ratios, such as [Eu/Mg] which trace rare \textit{r}-process events or [Y,Ba/Mg] which trace stellar ages, in order to better tag OCs \citep{2019A&A...631A.171S}.

\section{Conclusions} \label{sec:Conclusions}

The era of large all-sky surveys has brought rich and detailed properties of hundreds of thousands of stars, broadening our understanding of the Milky Way and the stellar populations that inhabit it. One group of stellar populations, open star clusters, has seen an explosion of discovery rates through clustering of overdensities in the kinematic space. However, as OCs diffuse across the Milky Way, we lose the ability to recover them using physical association. Thus we explore clustering methods using chemodynamical information that is preserved long after a cluster disperses.

\begin{itemize}
    \item We systematically evaluate chemodynamical parameters available from the \textit{Gaia} DR3 and, for the first time, the GALAH DR4 catalog on their ability to successfully recover known OCs. We report the optimal parameter combinations for obtaining the highest possible recovery rates for OCs in the Open Cluster Catalogue. Including orbital and kinematic properties significantly enhances the effectiveness of tagging star clusters, particularly those that have dispersed over time. The orbital properties provide a stable and distinct identifier than positional data, such as sky coordinates and proper motions, alone. Combining dynamical information with carefully selected chemical abundances further improves recovery rates, especially when coupled with prior knowledge about target clusters, such as their ages and birth radii.
    

    \item When evaluating the OC recovery in the entire $Gaia$-GALAH DR4 Catalogue (GGC), we show that blind clustering methods fail to recover a majority of clusters. We assess whether applying restrictions to the GGC based on known metallicities, ages, and birth radii improve the recovery rates of the OCs, and find that only in a few select cases (e.g. NGC 2632 and Alessi 24) do we find a marginal improvement ($\sim 10\%$) in the recovery rate of our clustering process.
    
    \item Finally, we find the maximum recovery rate achievable through a grid of restrictions to the \textit{Gaia}-GALAH catalog based on an individual OC's metallicity, ages, and birth radii for each OC in the OCC. The maximum recovery rate found out of all of the OCs in the OCC is less than 15\%. From these tests, we conclude that the use of stellar ages and birth radii is not enough to disentangle OCs from the field.
    
\end{itemize}

While the concept of strong chemical tagging holds promise for uncovering the history of star formation in the Milky Way, achieving its full potential requires overcoming significant challenges related to data quality, the complexity of chemical evolution, and the limitations of current analytical methods. As spectroscopic surveys obtain higher precision in their abundance and stellar parameter determination, it is necessary to re-access our ability to recover stars to their birth clusters.





\vspace{5mm}
N.B. and R.E. acknowledge support from the NASA Future Investigators in NASA Earth and Space Science and Technology (FINESST) Grant 21-ASTRO21-0075.

%

\vspace{5mm}
\facilities{\href{http://www.rc.ufl.edu}{Hipergator}}


\software{\href{http://github.com/jobovy/galpy}{Galpy}}




\clearpage
\bibliography{bib}{}
\bibliographystyle{aasjournal}

\newpage

\appendix

\section{Individual OC Properties} \label{sec:app}

The common assumption for OCs is that they are chemically homogeneous, similarly aged, and born in the same position, however, we show that there exists an intrinsic spread in each OC. For each OC in the OCC, we plot the GALAH metallicity distributions of all of the member stars. We also plot the distribution of the derived ages and birth radii (see Sections \ref{sec:ages} and \ref{sec:br}, respectively). Additionally, we show the Kiel diagram for each OC, which shows if an OC is dominated by dwarf stars, giants, or a combination of both. 

Finally, we plot the V-measure scores obtained for each OC using the dynamic parameters, $D = E, J_\phi, J_Z, J_R$, and the dynamic parameters with an additional element parameter. The solid areas of each bar plot show the V-measure score associated with our clustering algorithm on the entire GGC. The hatched regions show the maximum V-measure score obtained using a combination of metallicity, age, and birth radius restrictions to the GGC. By comparing the maximum V-measure score obtained for each OC to the properties of the OCs, we can determine what properties are important for increasing recovery rates when clustering.

\begin{subfigures}
    \begin{figure}
        \centering
        \includegraphics[width=0.9\linewidth]{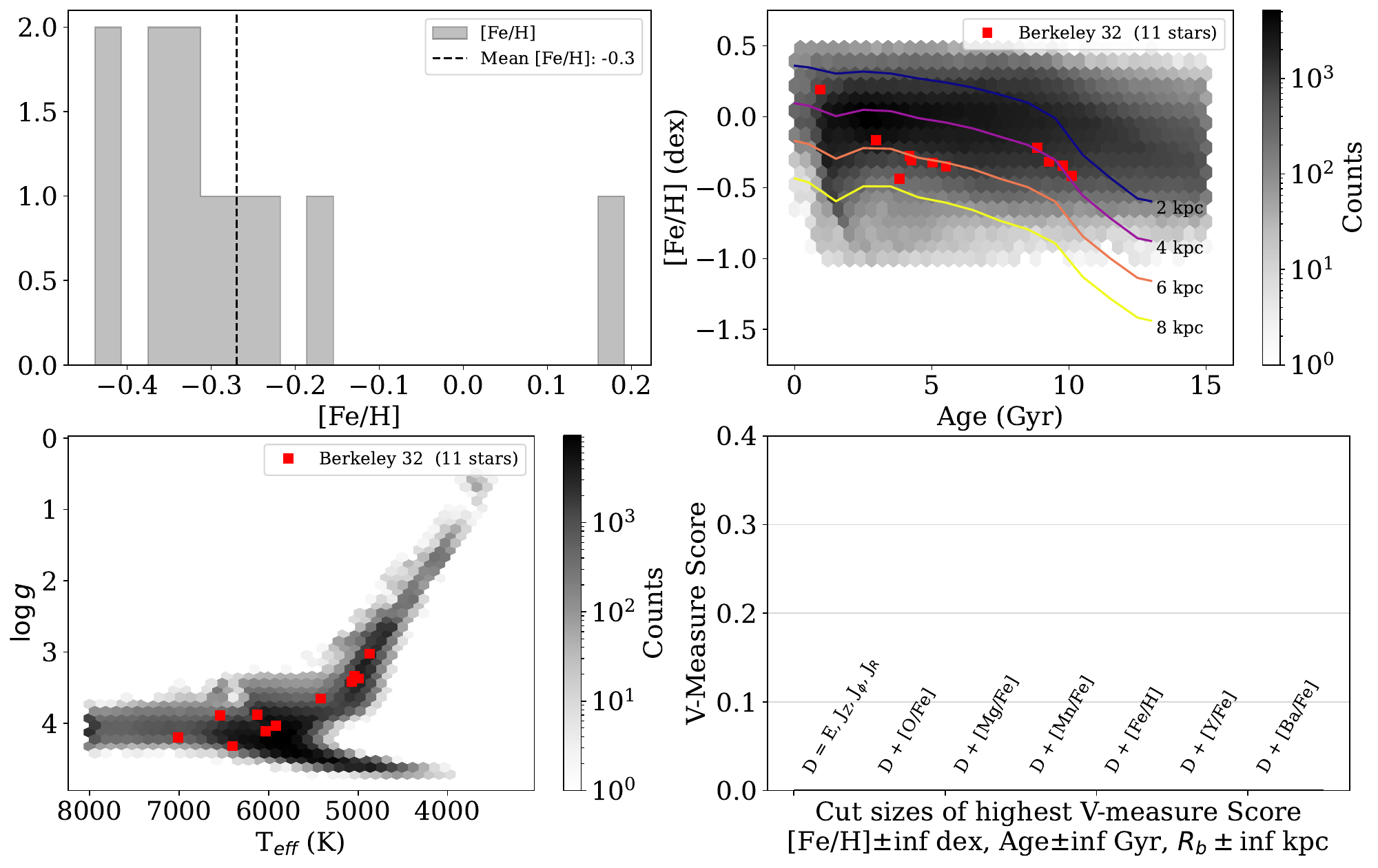}
        \caption{Characteristics of the open cluster Berkeley 32. \textit{Top left}: Distribution of [Fe/H] abundances for member stars in Berkeley 32. The cluster's mean metallicity is shown as the vertical, dashed line. \textit{Top right}: 2D histogram of GALAH [Fe/H] abundances and ages derived in this work using stellar evolutionary track fitting, using the MIST isochrone models. Overlaid as red squares are the member stars of Berkeley 32. The horizontal lines correspond to mono-values of $R_b$ calculated from the GGC.  \textit{Bottom left}: Kiel diagram showing evolutionary stages of member stars of Berkeley 32, compared to the GGC. \textit{Bottom right}: Results of how well Berkeley 32 is recovered using different parameter combinations. The solid area of each bar represent the V-measure score of clustering the GGC, while the hatched section of the plot represents the best V-measure score achieved using a combination of [Fe/H], age, and birth radii cuts to reduce the size of the GGC.}
        \label{fig:sum_plot_Berkeley_32}
    \end{figure}
    
    \begin{figure}
        \centering
        \includegraphics[width=0.9\linewidth]{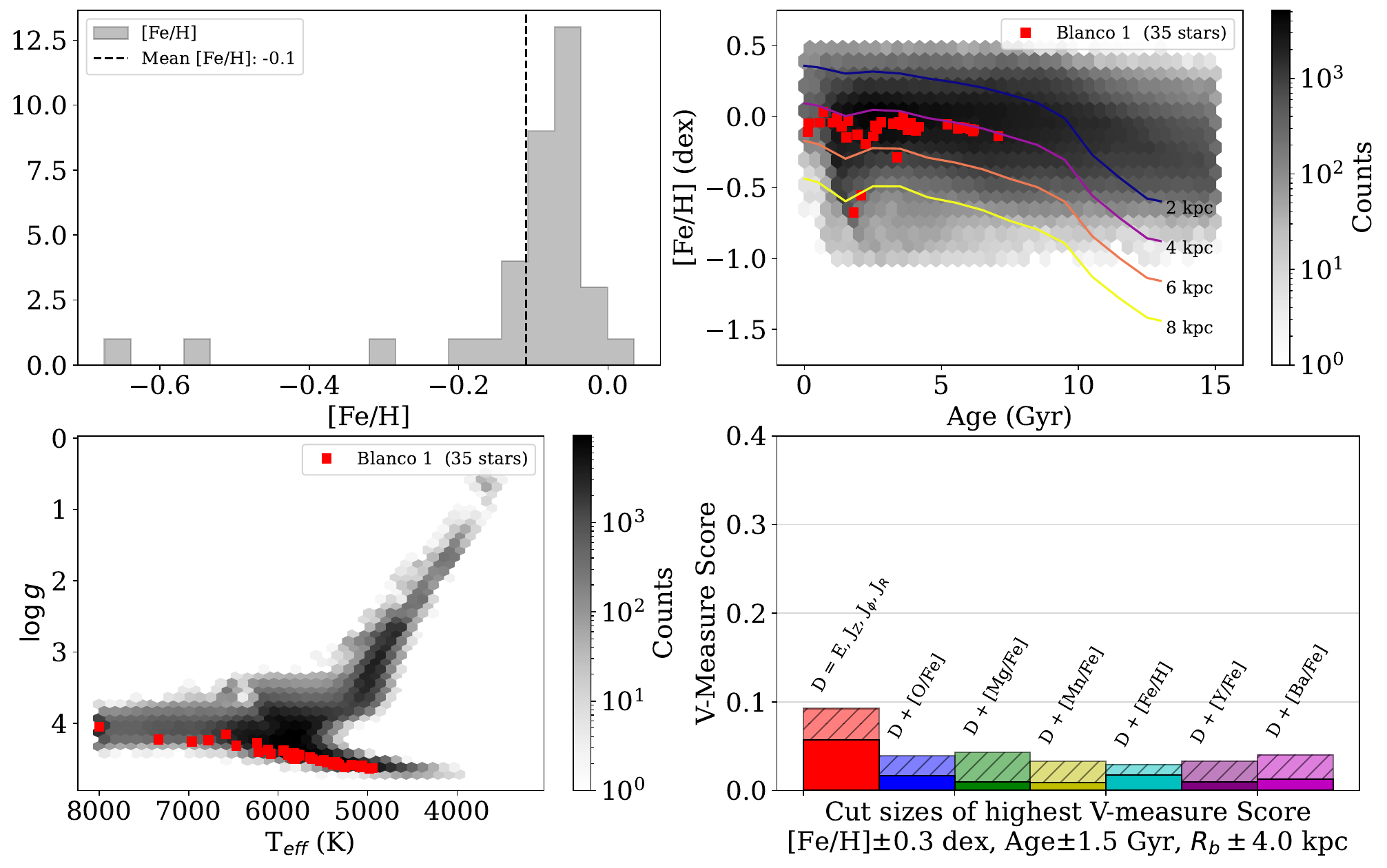}
        \caption{Same as Figure \ref{fig:sum_plot_Berkeley_32}, but for Blanco 1.}
    \end{figure}
    
    \begin{figure}
        \centering
        \includegraphics[width=0.9\linewidth]{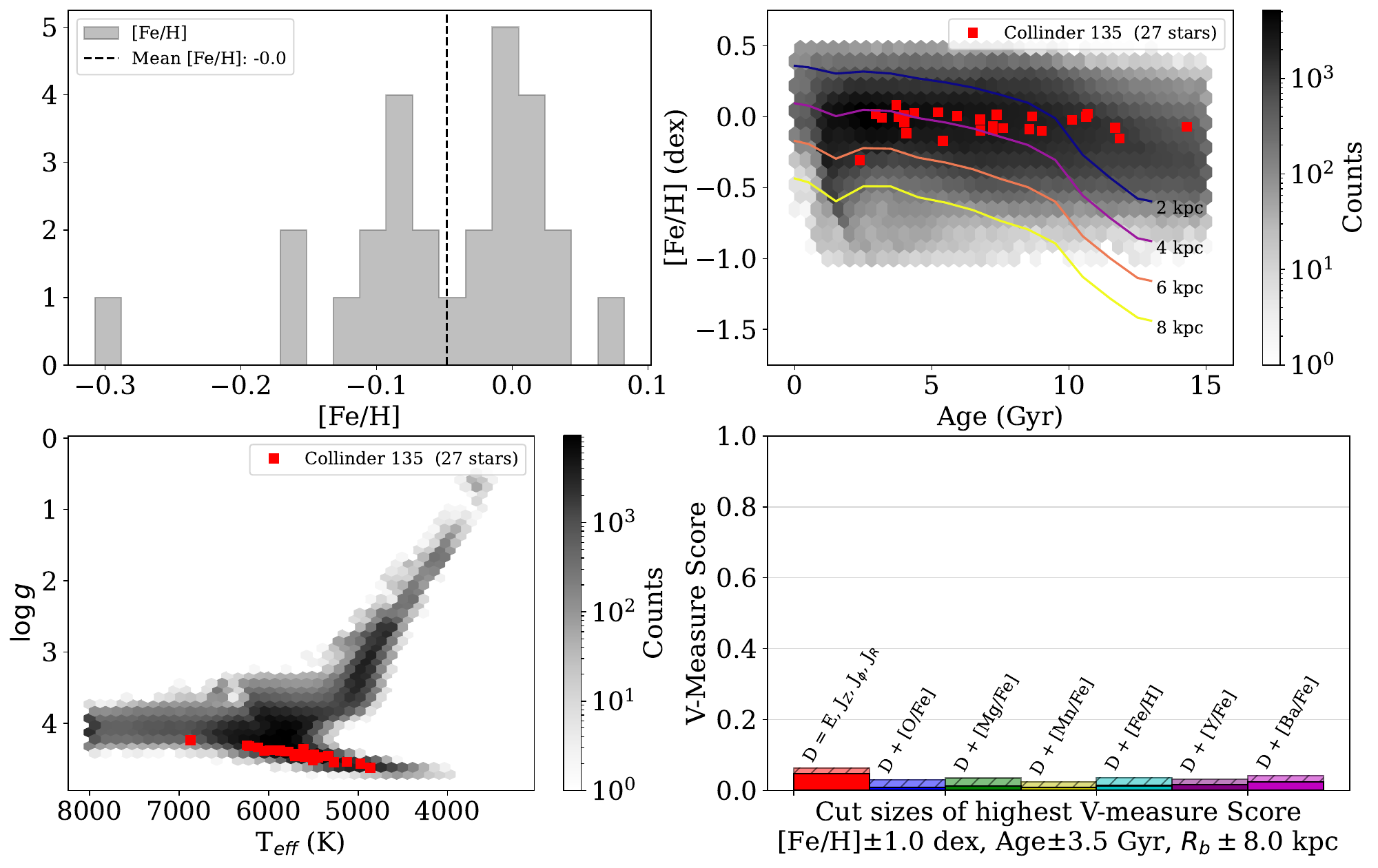}
        \caption{Same as Figure \ref{fig:sum_plot_Berkeley_32}, but for Collinder 135.}
    \end{figure}
    
    \begin{figure}
        \centering
        \includegraphics[width=0.9\linewidth]{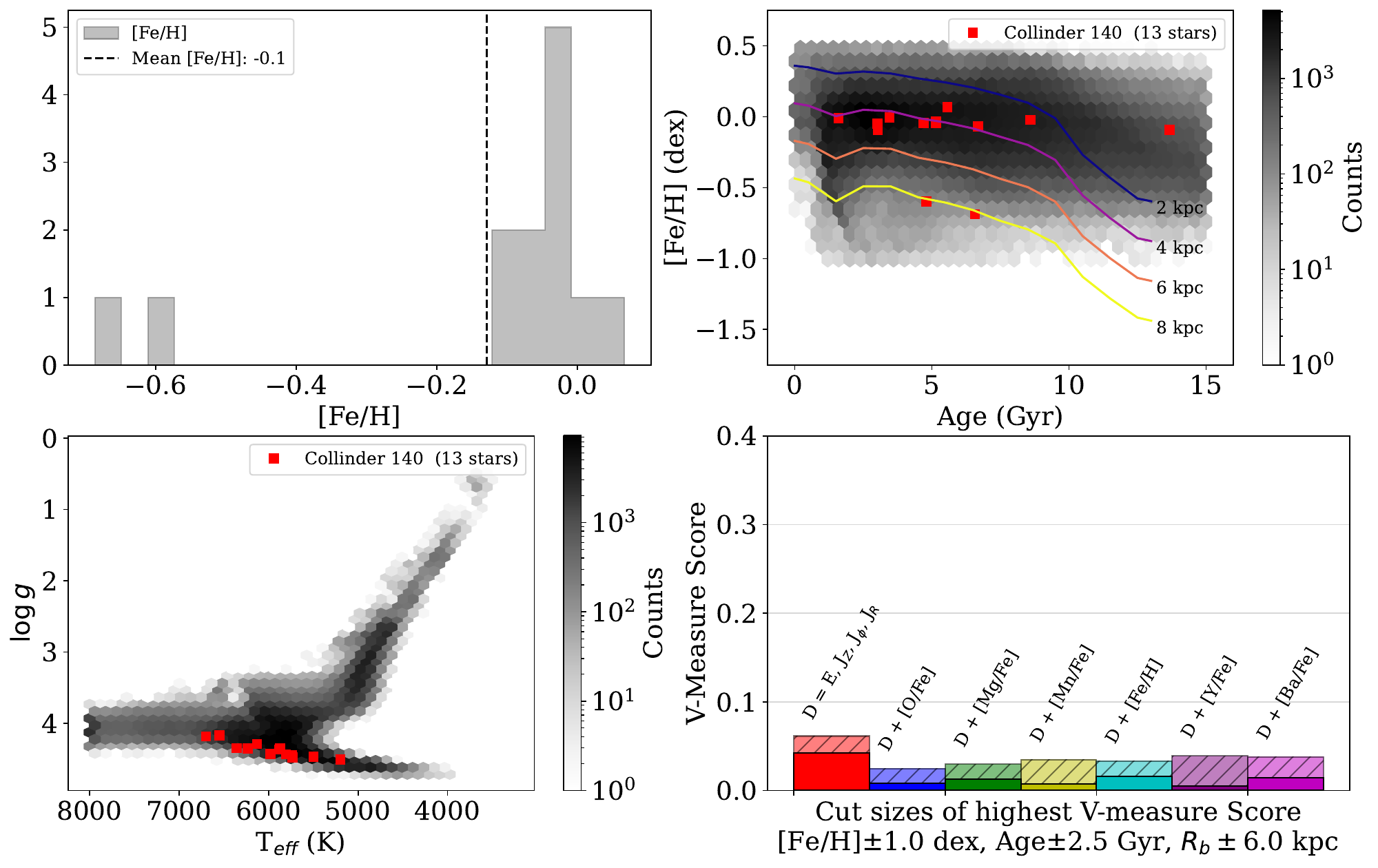}
        \caption{Same as Figure \ref{fig:sum_plot_Berkeley_32}, but for Collinder 140.}
    \end{figure}
    
    \begin{figure}
        \centering
        \includegraphics[width=0.9\linewidth]{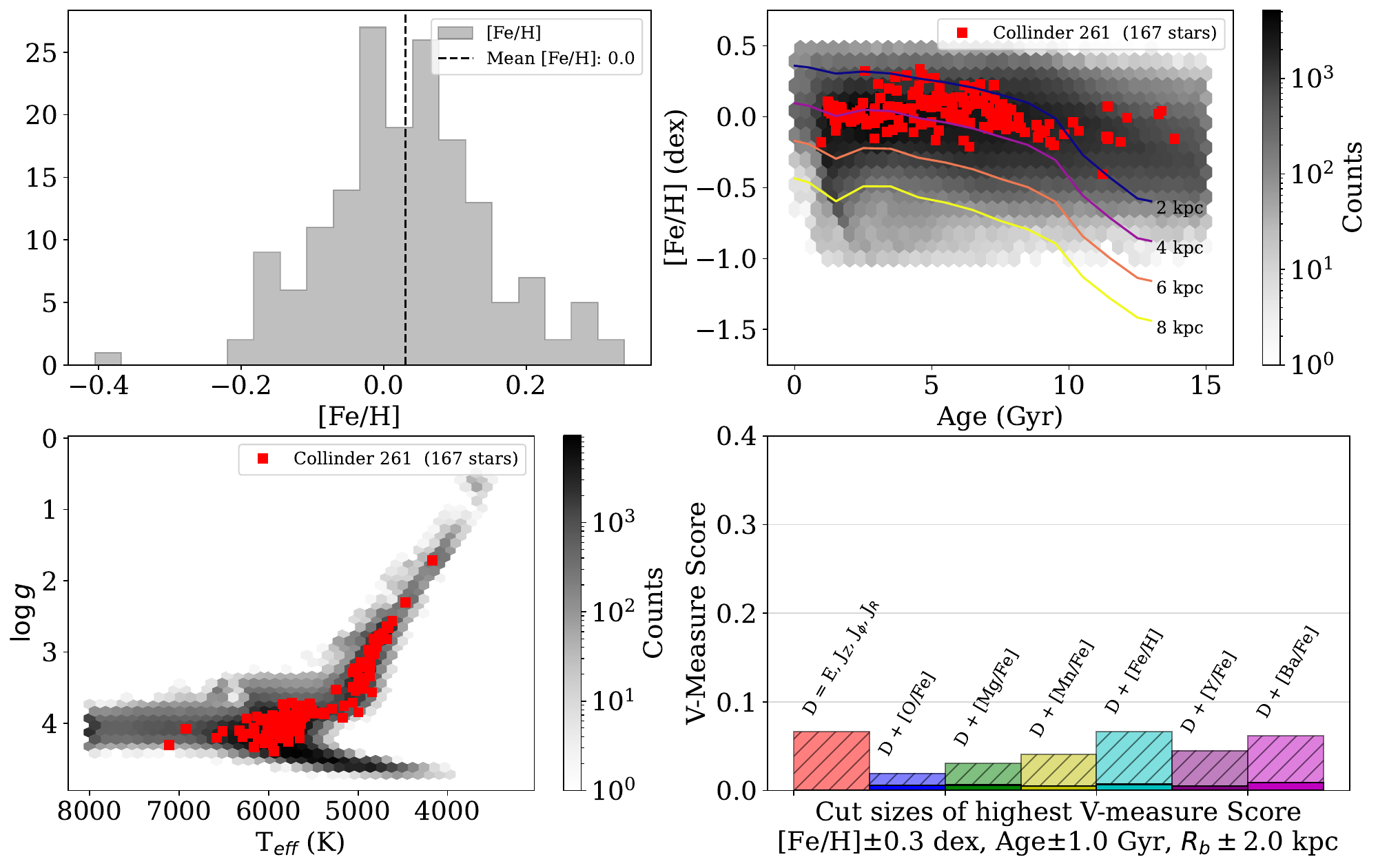}
        \caption{Same as Figure \ref{fig:sum_plot_Berkeley_32}, but for Collinder 261.}
    \end{figure}
    
    \begin{figure}
        \centering
        \includegraphics[width=0.9\linewidth]{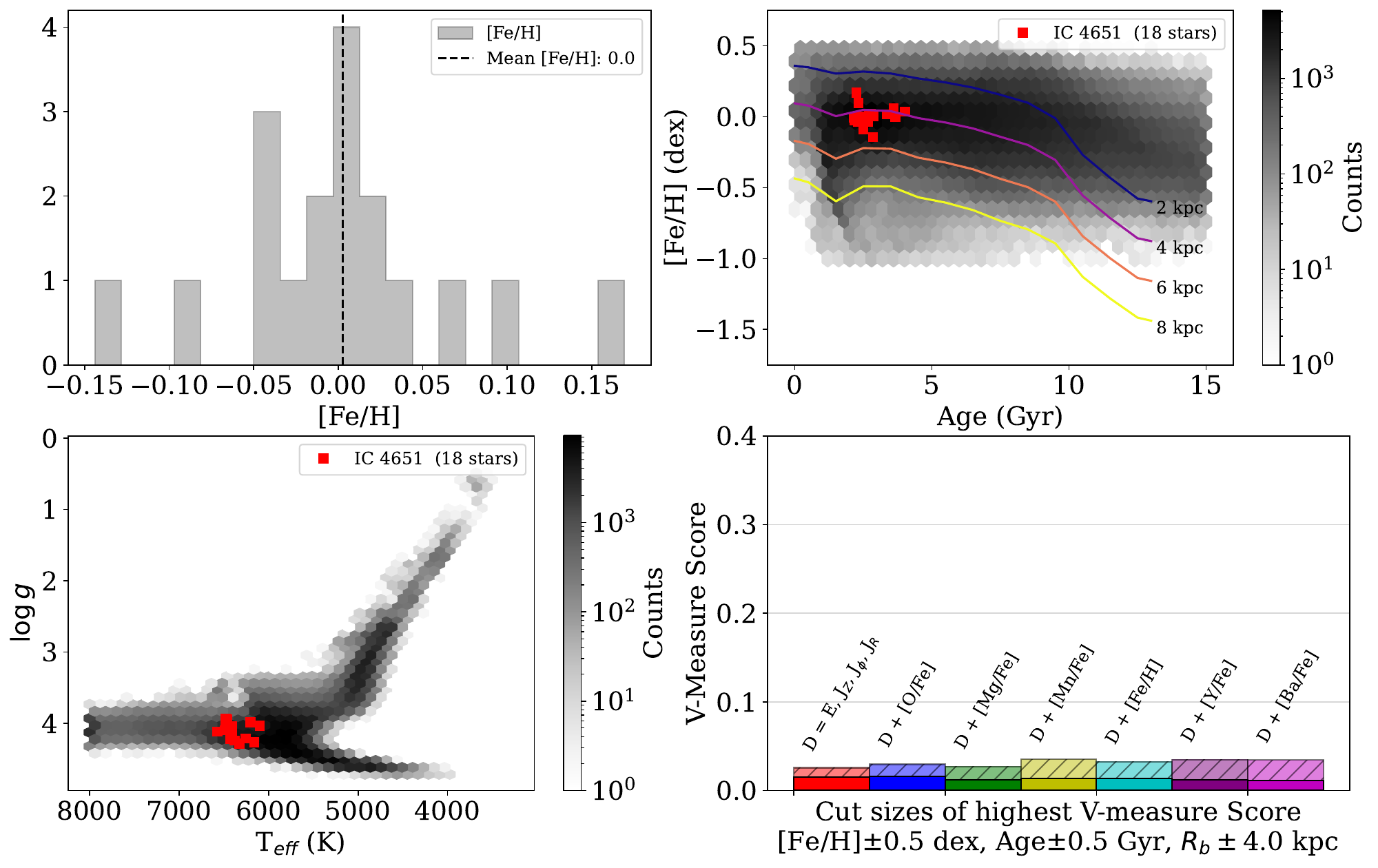}
        \caption{Same as Figure \ref{fig:sum_plot_Berkeley_32}, but for IC 4651.}
    \end{figure}
    
    \begin{figure}
        \centering
        \includegraphics[width=0.9\linewidth]{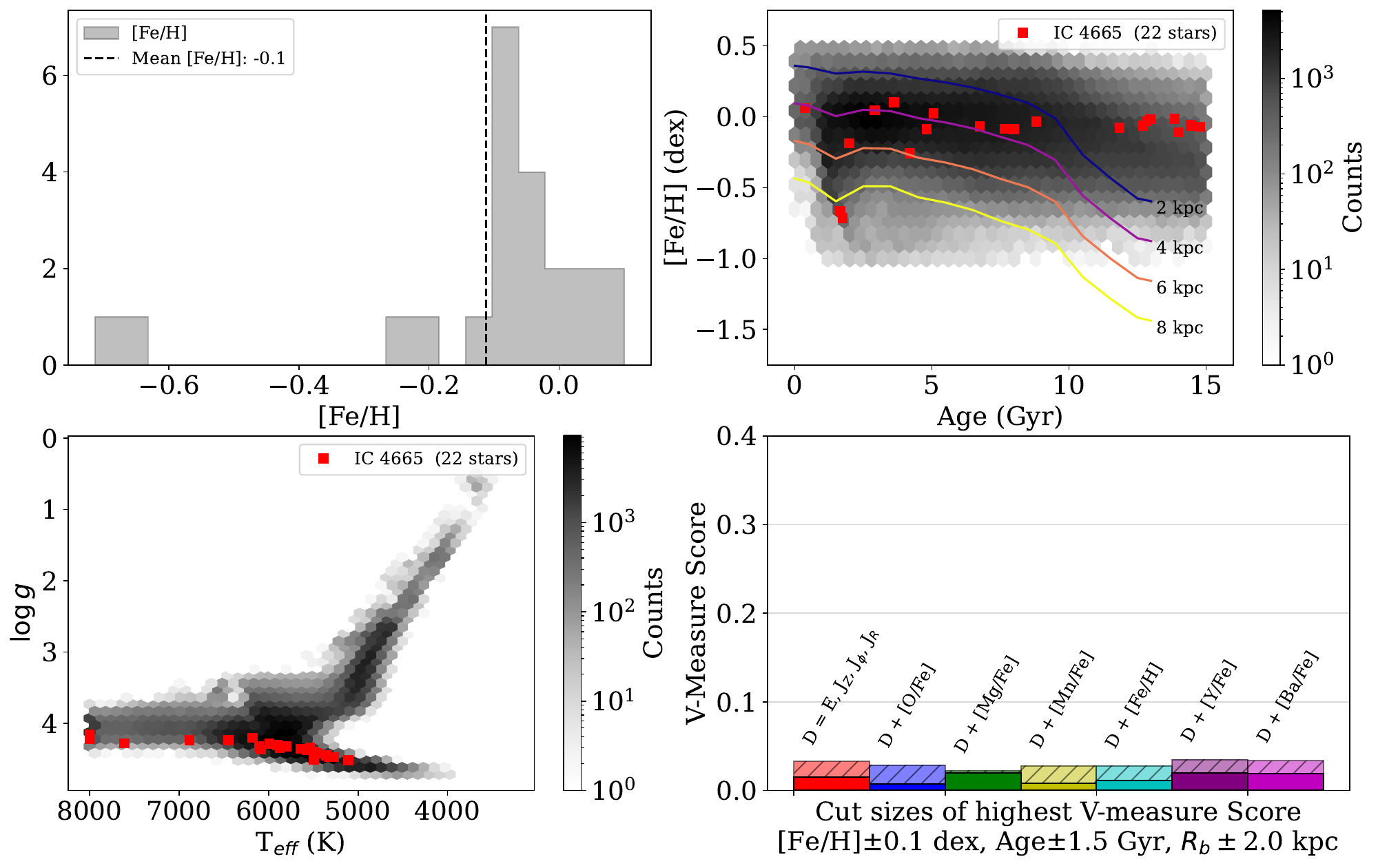}
        \caption{Same as Figure \ref{fig:sum_plot_Berkeley_32}, but for IC 4665.}
    \end{figure}
    
    \begin{figure}
        \centering
        \includegraphics[width=0.9\linewidth]{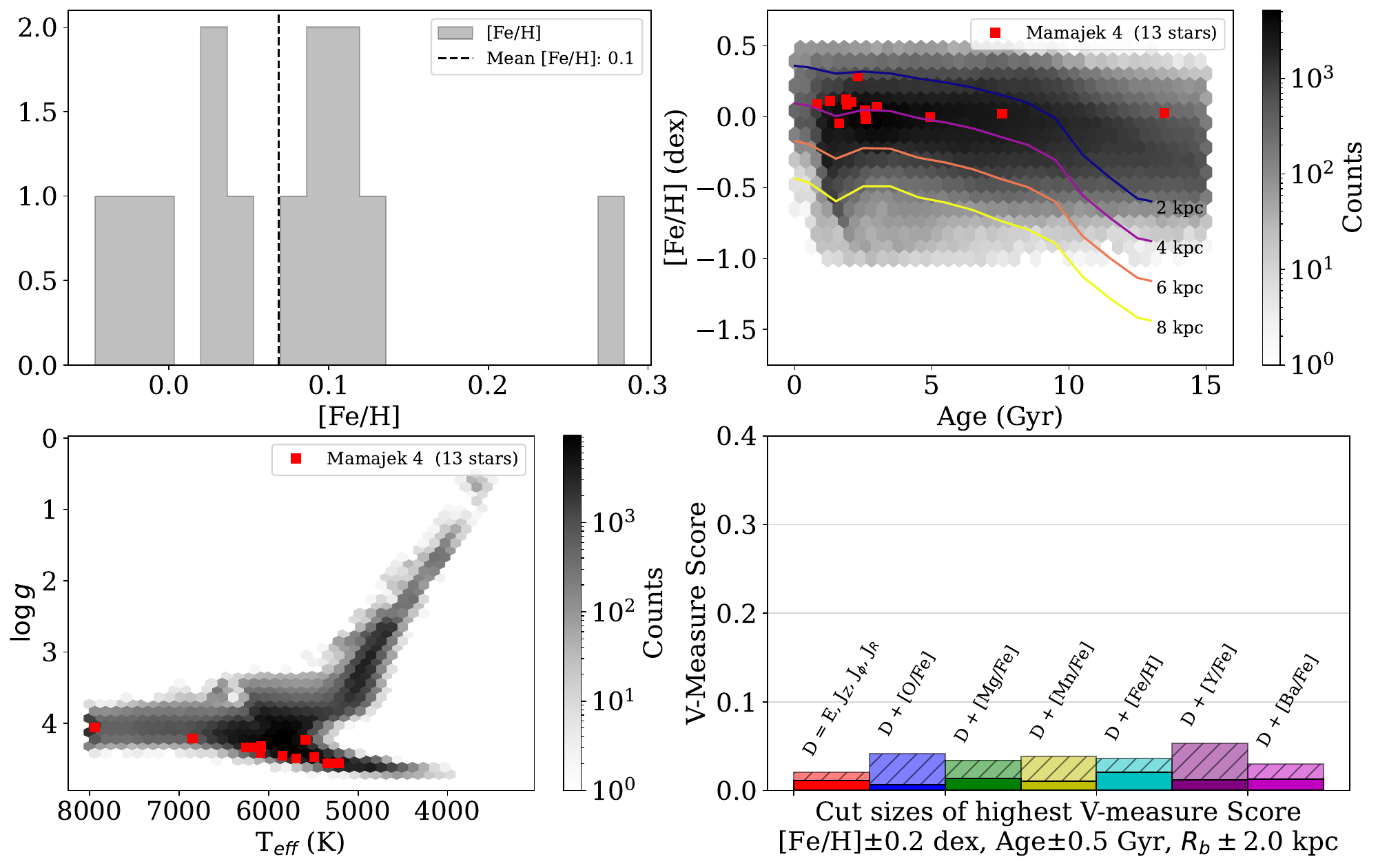}
        \caption{Same as Figure \ref{fig:sum_plot_Berkeley_32}, but for Mamajek 4.}
    \end{figure}
    
    \begin{figure}
        \centering
        \includegraphics[width=0.9\linewidth]{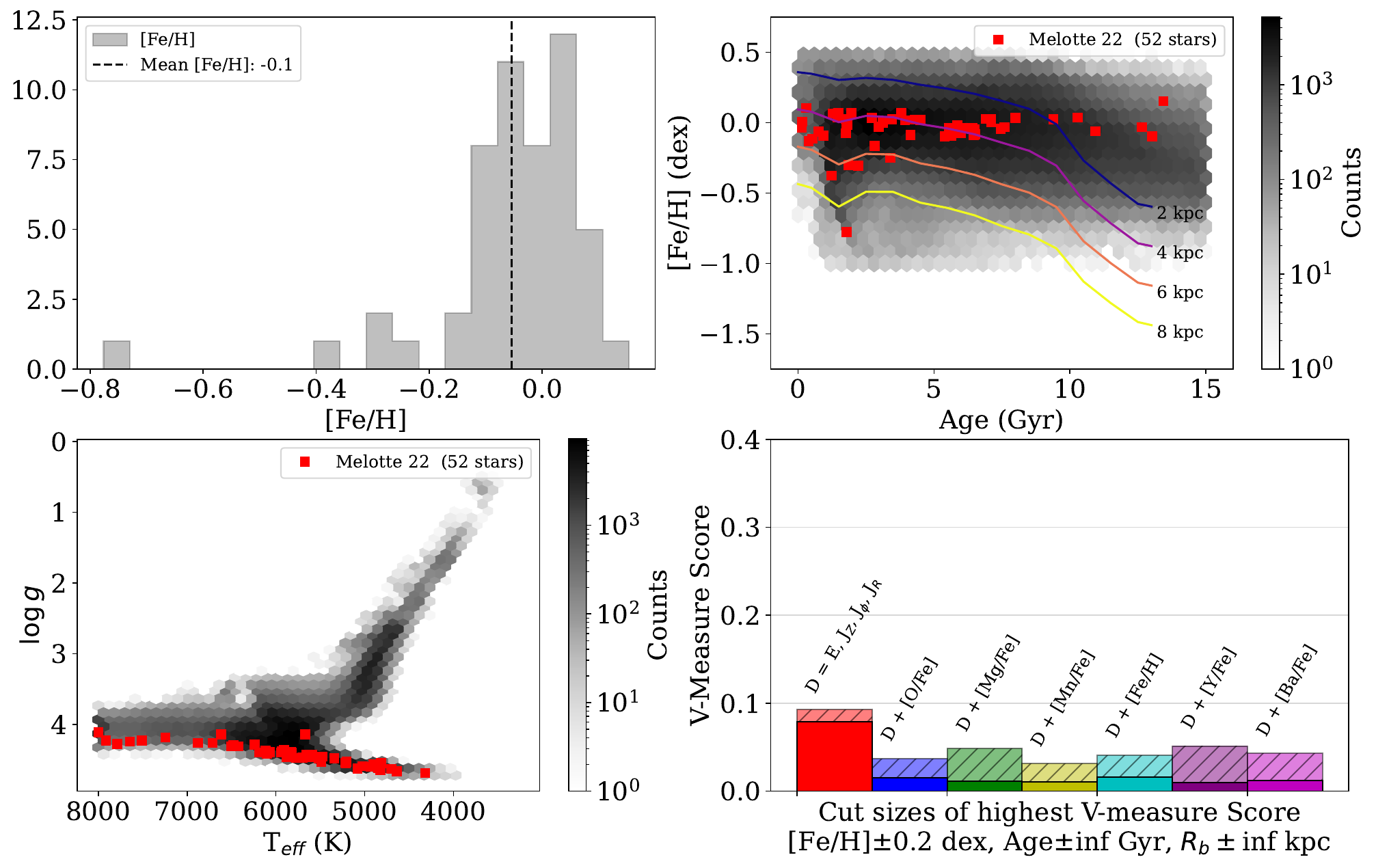}
        \caption{Same as Figure \ref{fig:sum_plot_Berkeley_32}, but for Melotte 22.}
    \end{figure}
    
    \begin{figure}
        \centering
        \includegraphics[width=0.9\linewidth]{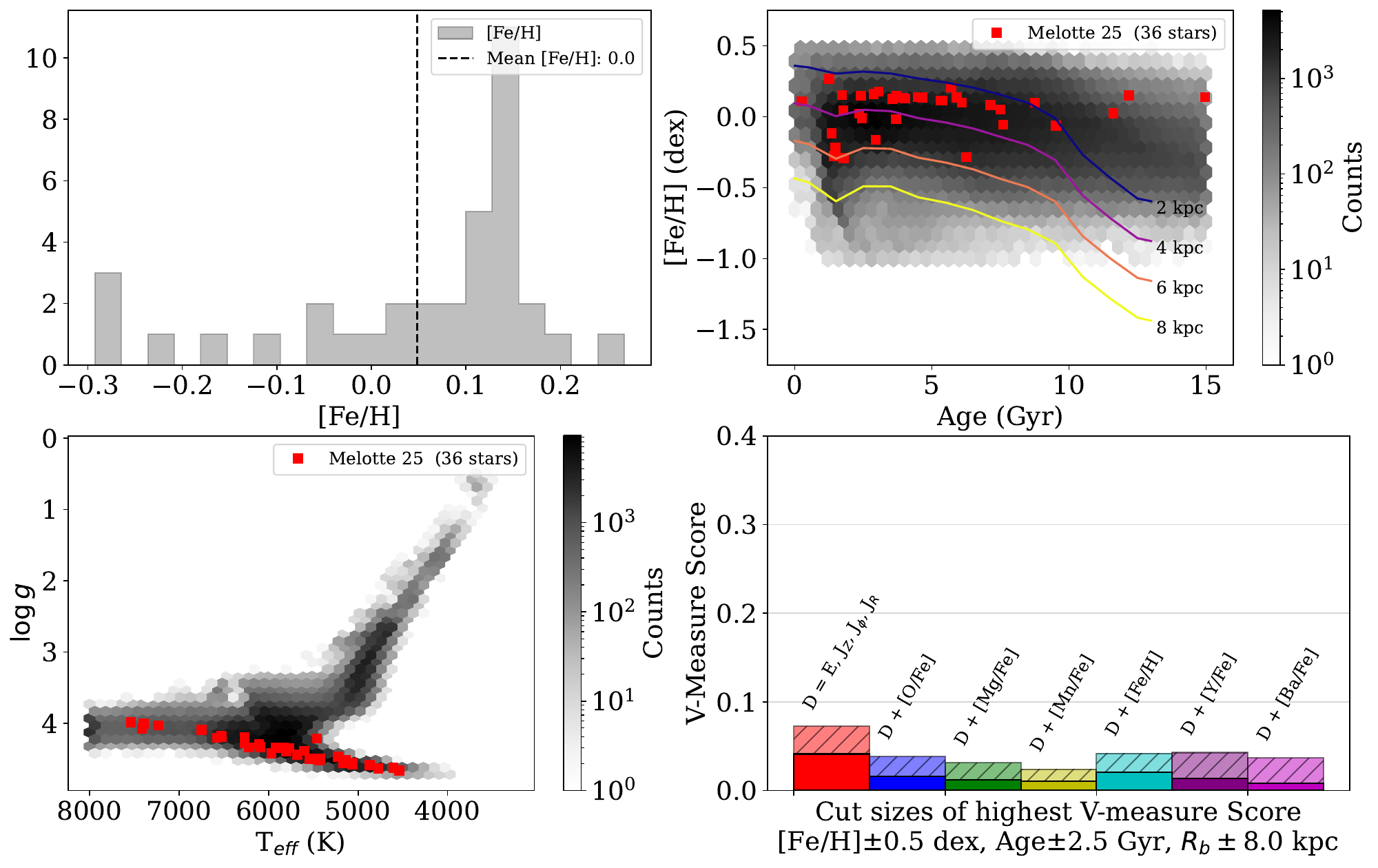}
        \caption{Same as Figure \ref{fig:sum_plot_Berkeley_32}, but for Melotte 25.}
    \end{figure}
    
    \begin{figure}
        \centering
        \includegraphics[width=0.9\linewidth]{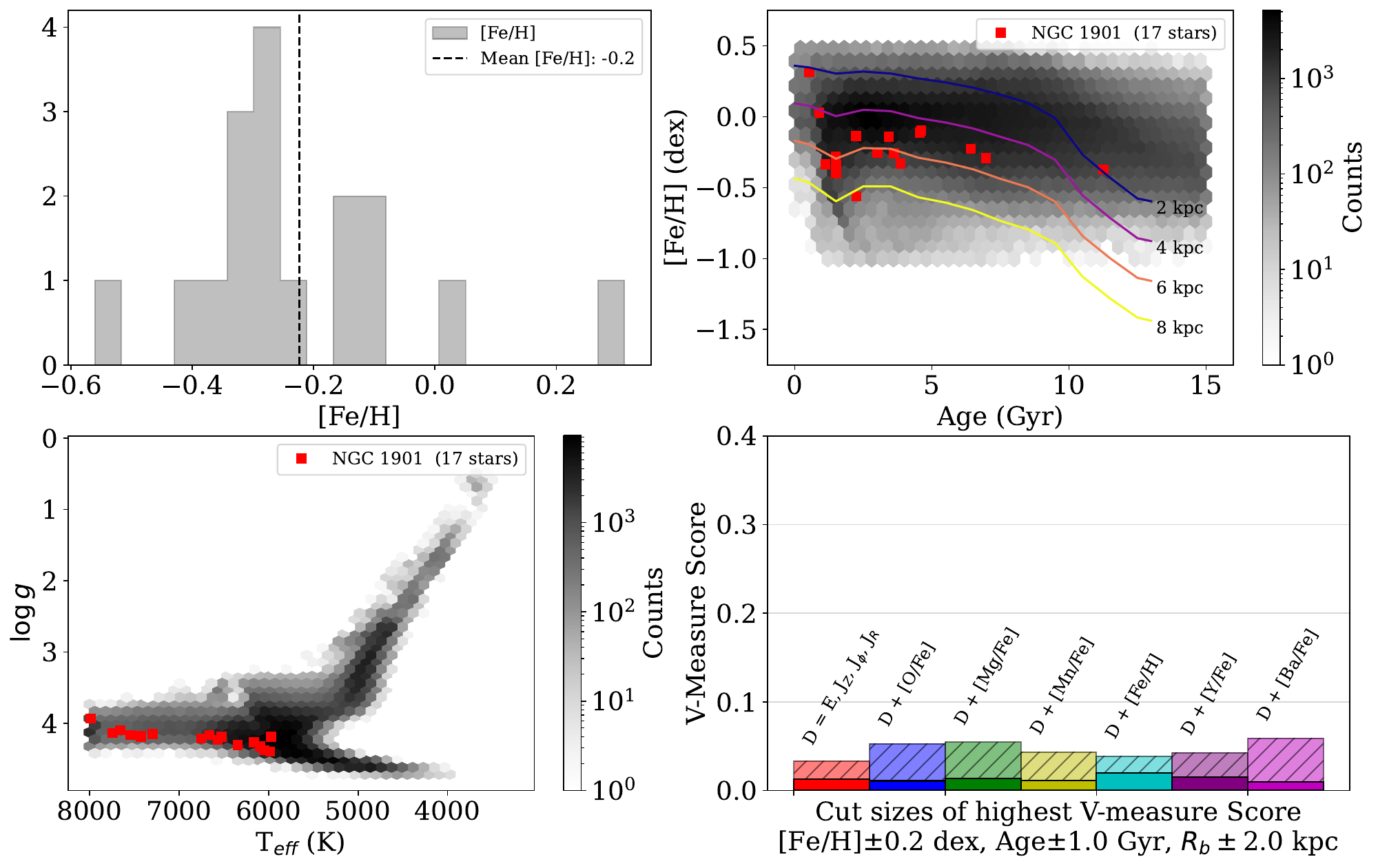}
        \caption{Same as Figure \ref{fig:sum_plot_Berkeley_32}, but for NGC 1901.}
    \end{figure}
    
    \begin{figure}
        \centering
        \includegraphics[width=0.9\linewidth]{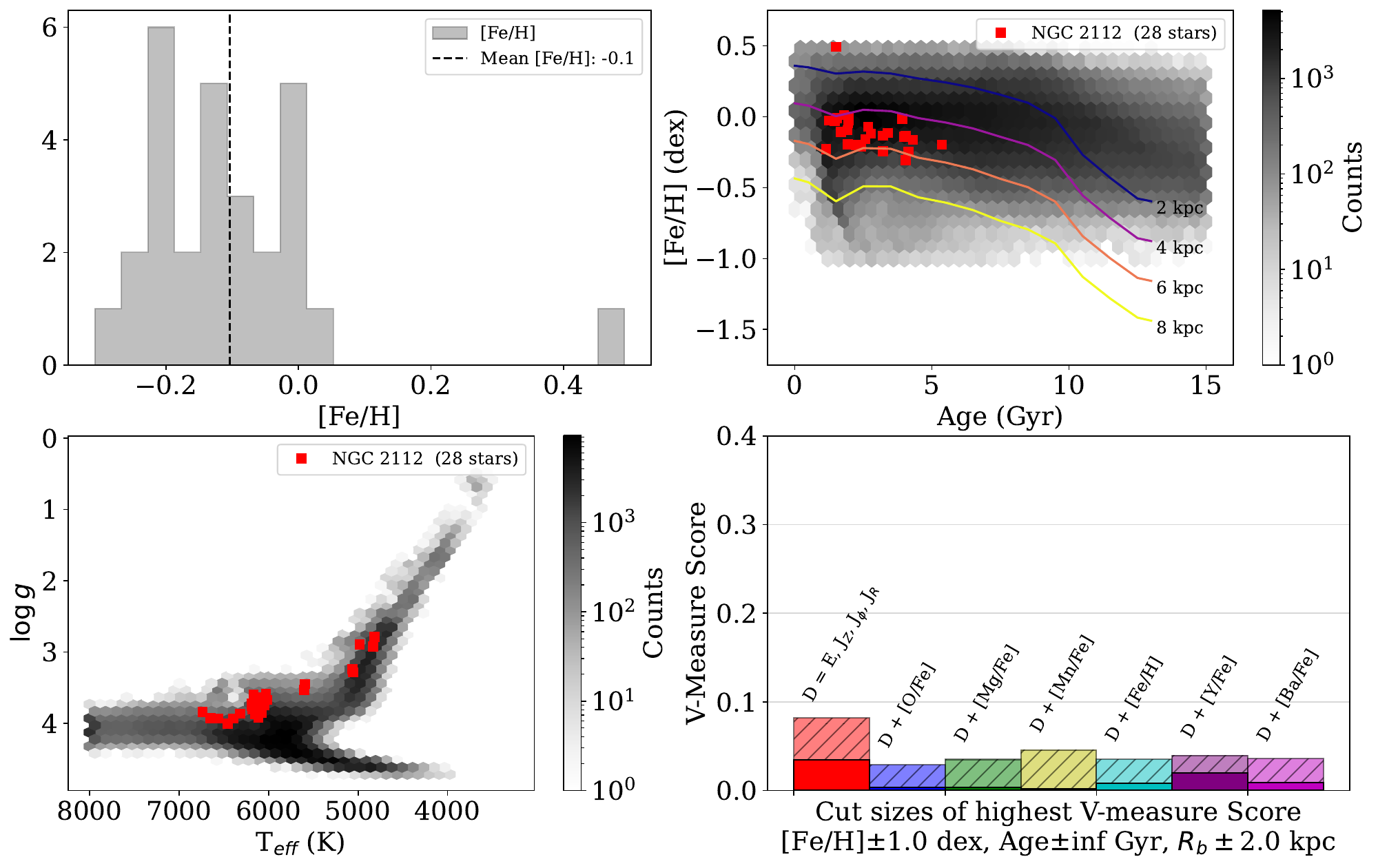}
        \caption{Same as Figure \ref{fig:sum_plot_Berkeley_32}, but for NGC 2112.}
    \end{figure}
    
    \begin{figure}
        \centering
        \includegraphics[width=0.9\linewidth]{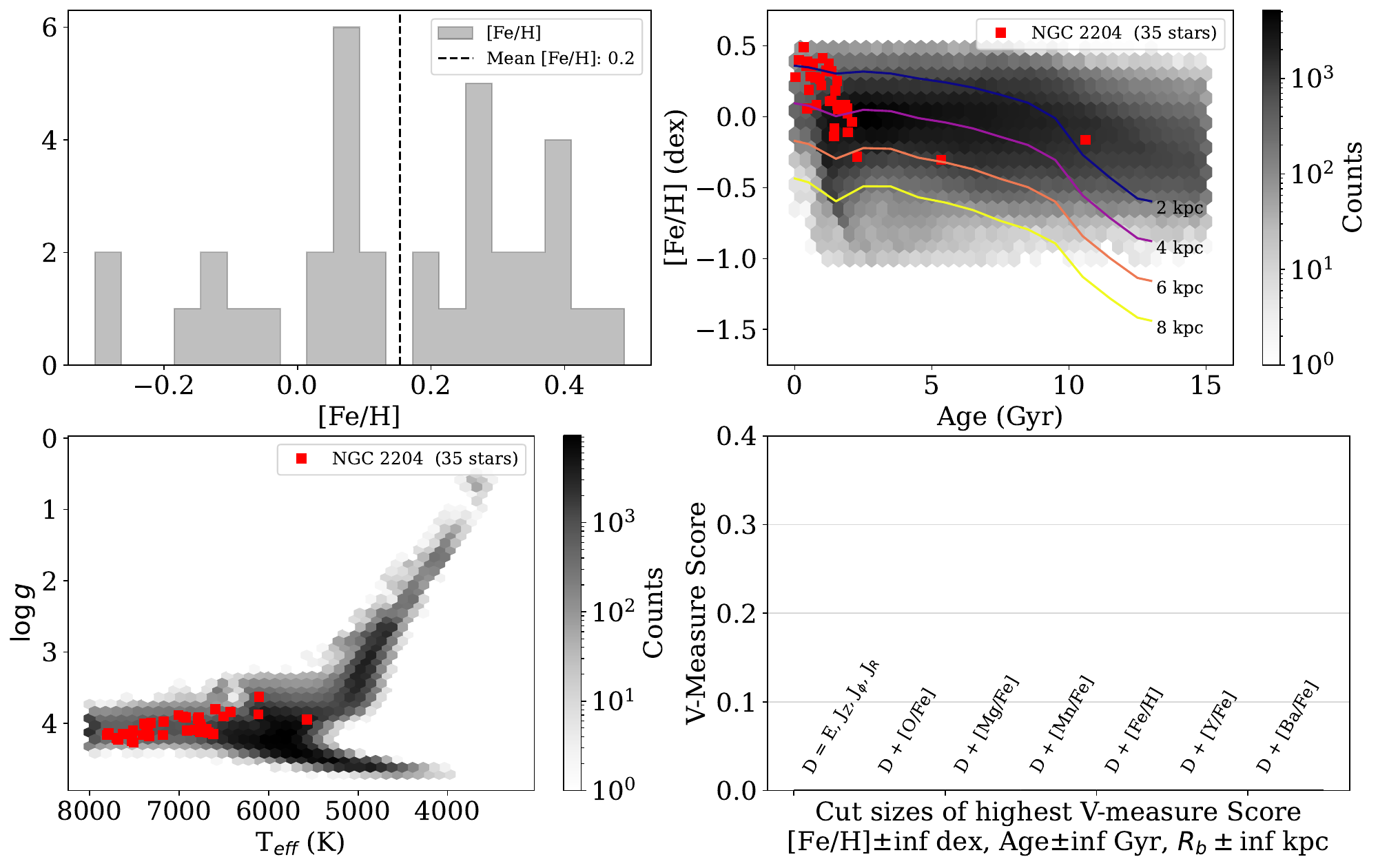}
        \caption{Same as Figure \ref{fig:sum_plot_Berkeley_32}, but for NGC 2204.}
    \end{figure}
    
    
    \begin{figure}
        \centering
        \includegraphics[width=0.9\linewidth]{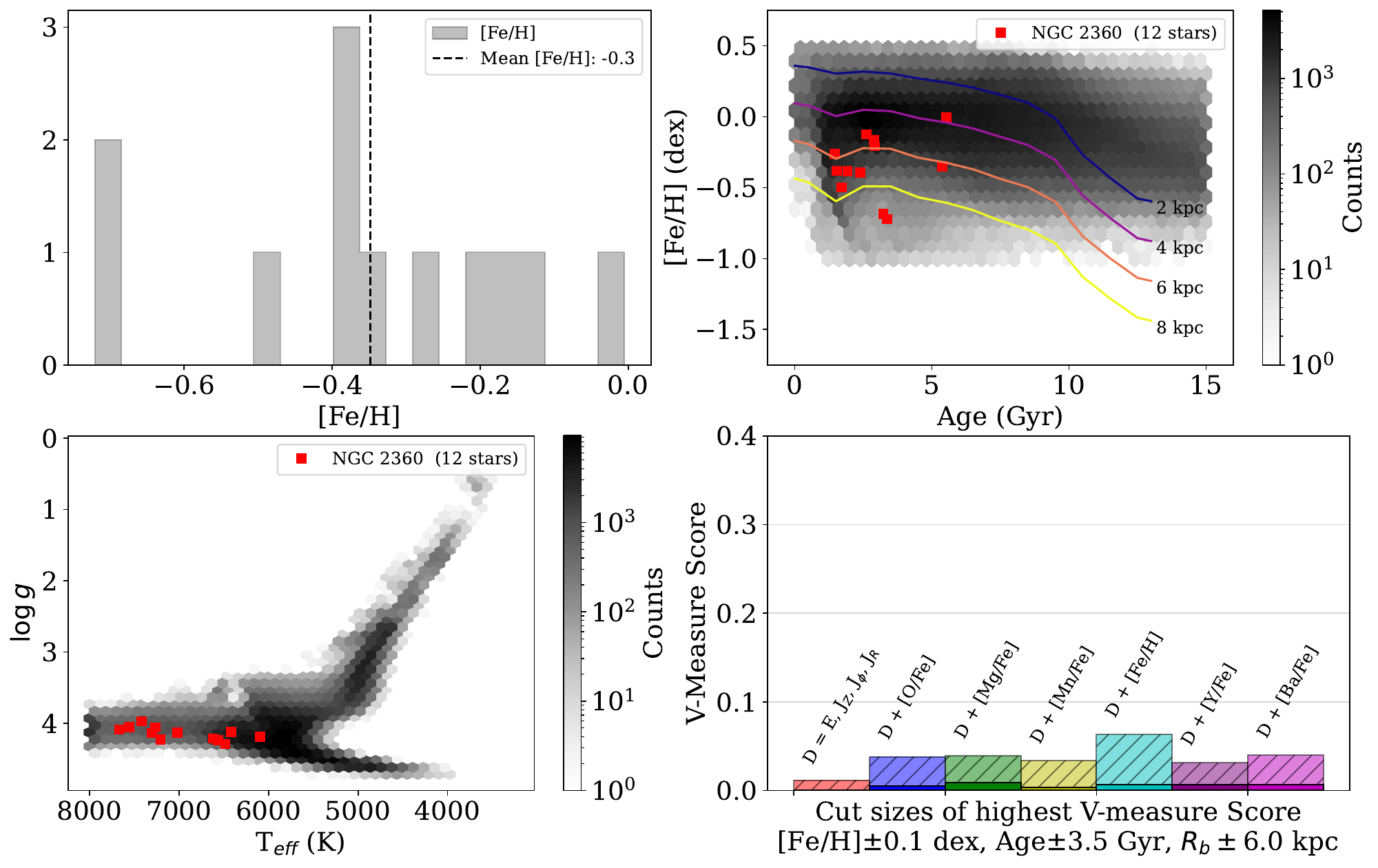}
        \caption{Same as Figure \ref{fig:sum_plot_Berkeley_32}, but for NGC 2360.}
    \end{figure}
    
    \begin{figure}
        \centering
        \includegraphics[width=0.9\linewidth]{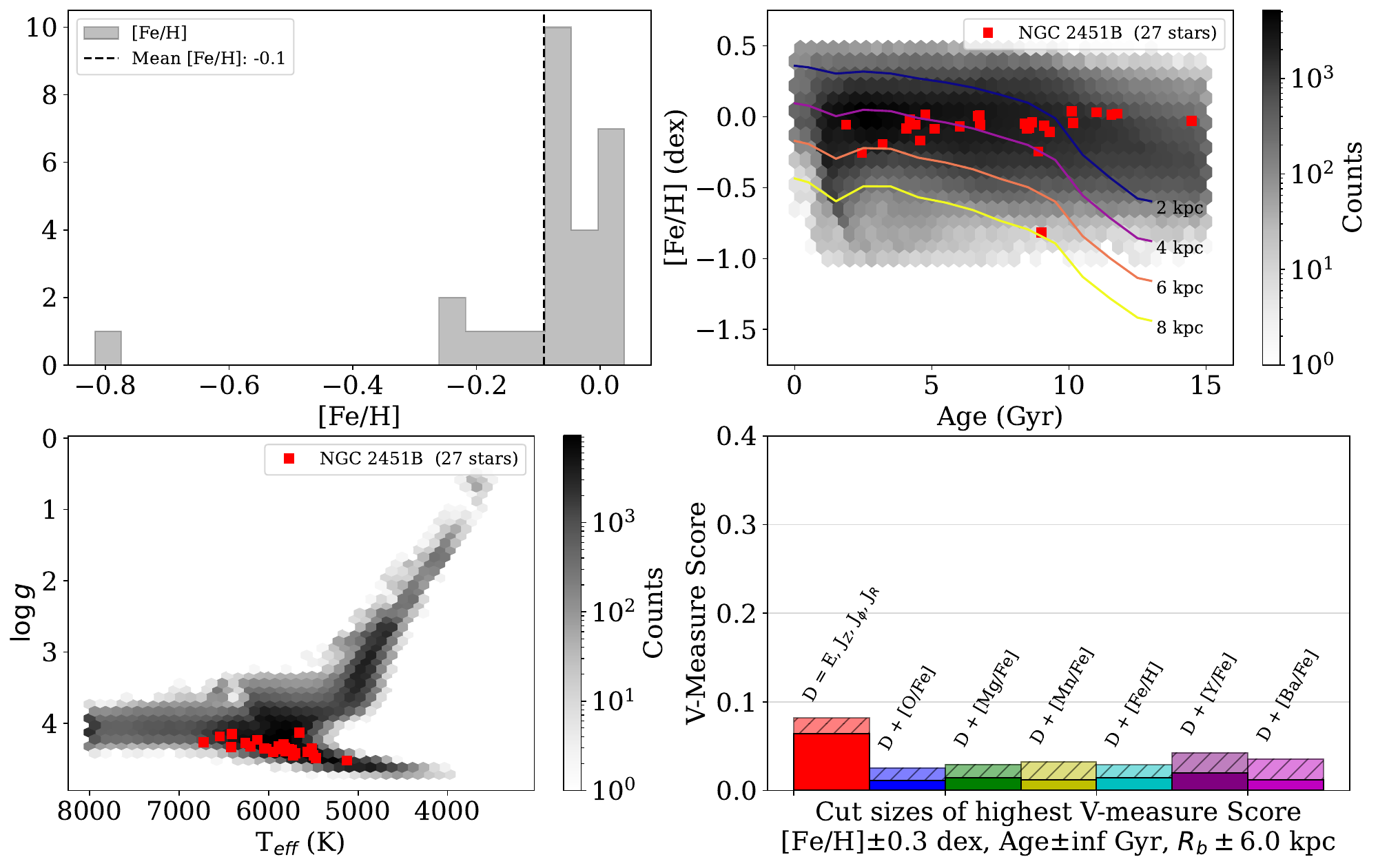}
        \caption{Same as Figure \ref{fig:sum_plot_Berkeley_32}, but for NGC 2451B.}
    \end{figure}
    
    \begin{figure}
        \centering
        \includegraphics[width=0.9\linewidth]{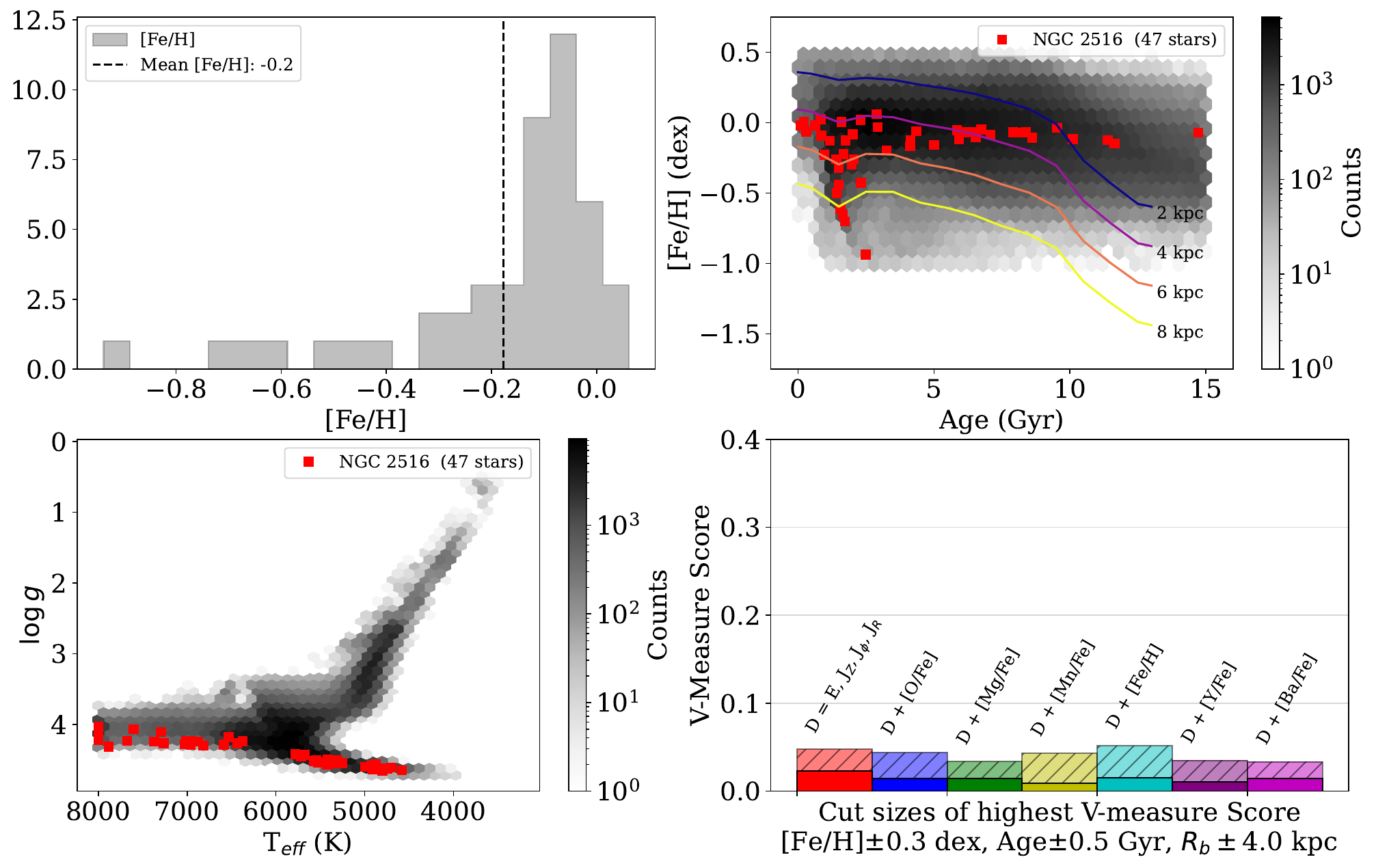}
        \caption{Same as Figure \ref{fig:sum_plot_Berkeley_32}, but for NGC 2516.}
    \end{figure}
    
    \begin{figure}
        \centering
        \includegraphics[width=0.9\linewidth]{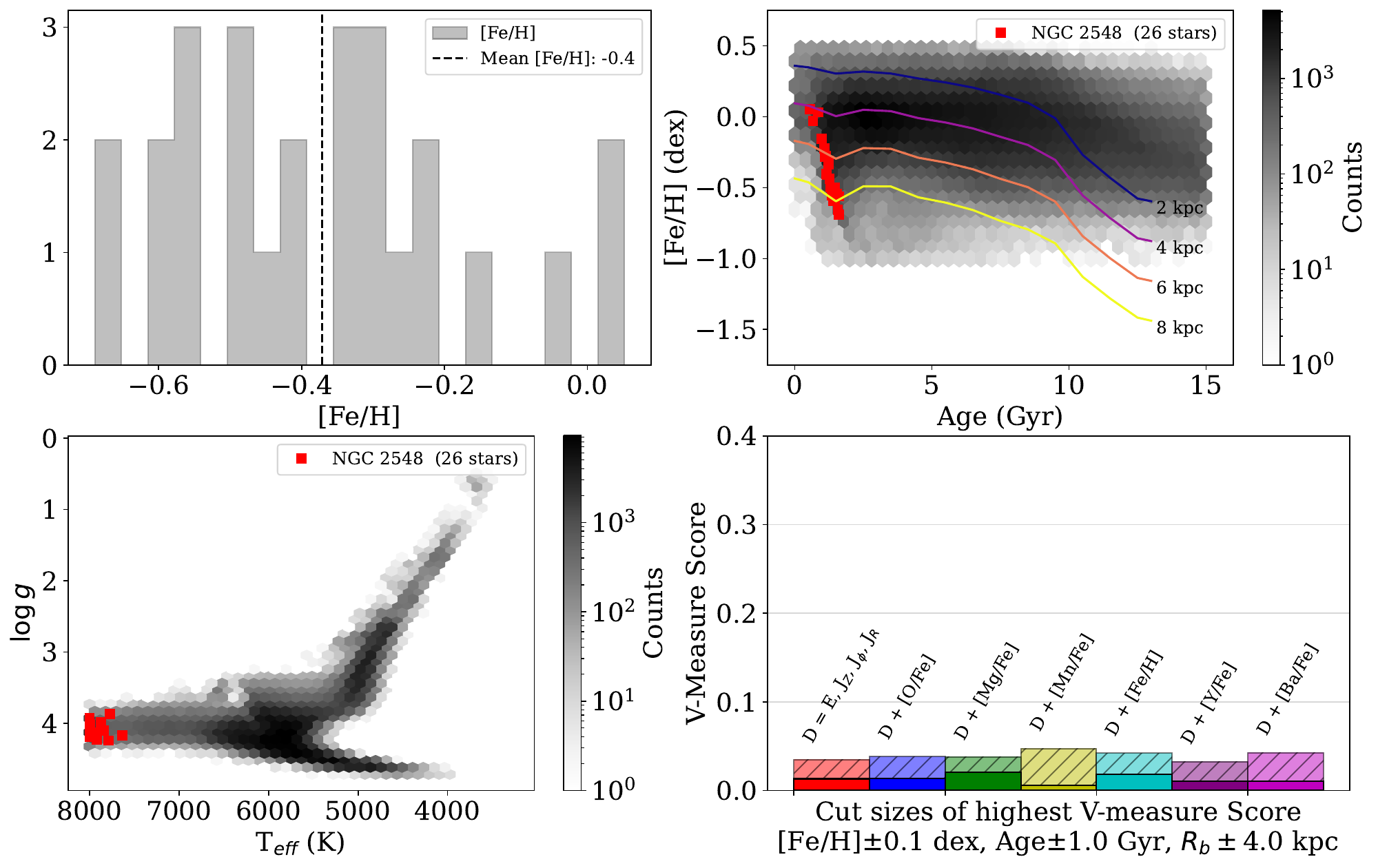}
        \caption{Same as Figure \ref{fig:sum_plot_Berkeley_32}, but for NGC 2548.}
    \end{figure}
    
    
    
    
    \begin{figure}
        \centering
        \includegraphics[width=0.9\linewidth]{summary_plot_NGC_2632.pdf}
        \caption{Same as Figure \ref{fig:sum_plot_Berkeley_32}, but for NGC 2632.}
    \end{figure}
    
    \begin{figure}
        \centering
        \includegraphics[width=0.9\linewidth]{summary_plot_NGC_2682.pdf}
        \caption{Same as Figure \ref{fig:sum_plot_Berkeley_32}, but for NGC 2682.}
    \end{figure}
    
    \begin{figure}
        \centering
        \includegraphics[width=0.9\linewidth]{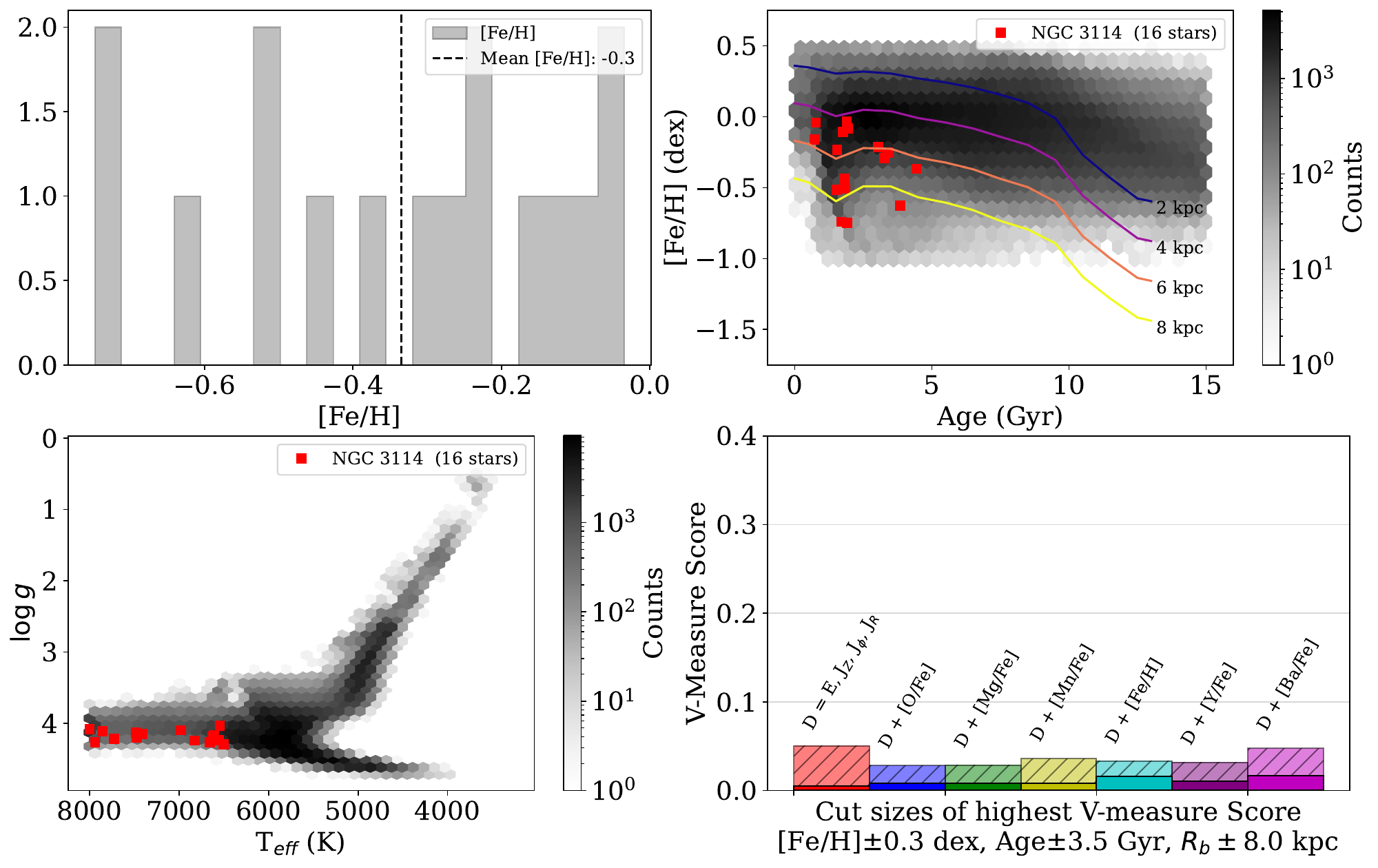}
        \caption{Same as Figure \ref{fig:sum_plot_Berkeley_32}, but for NGC 3114.}
    \end{figure}
    
    \begin{figure}
        \centering
        \includegraphics[width=0.9\linewidth]{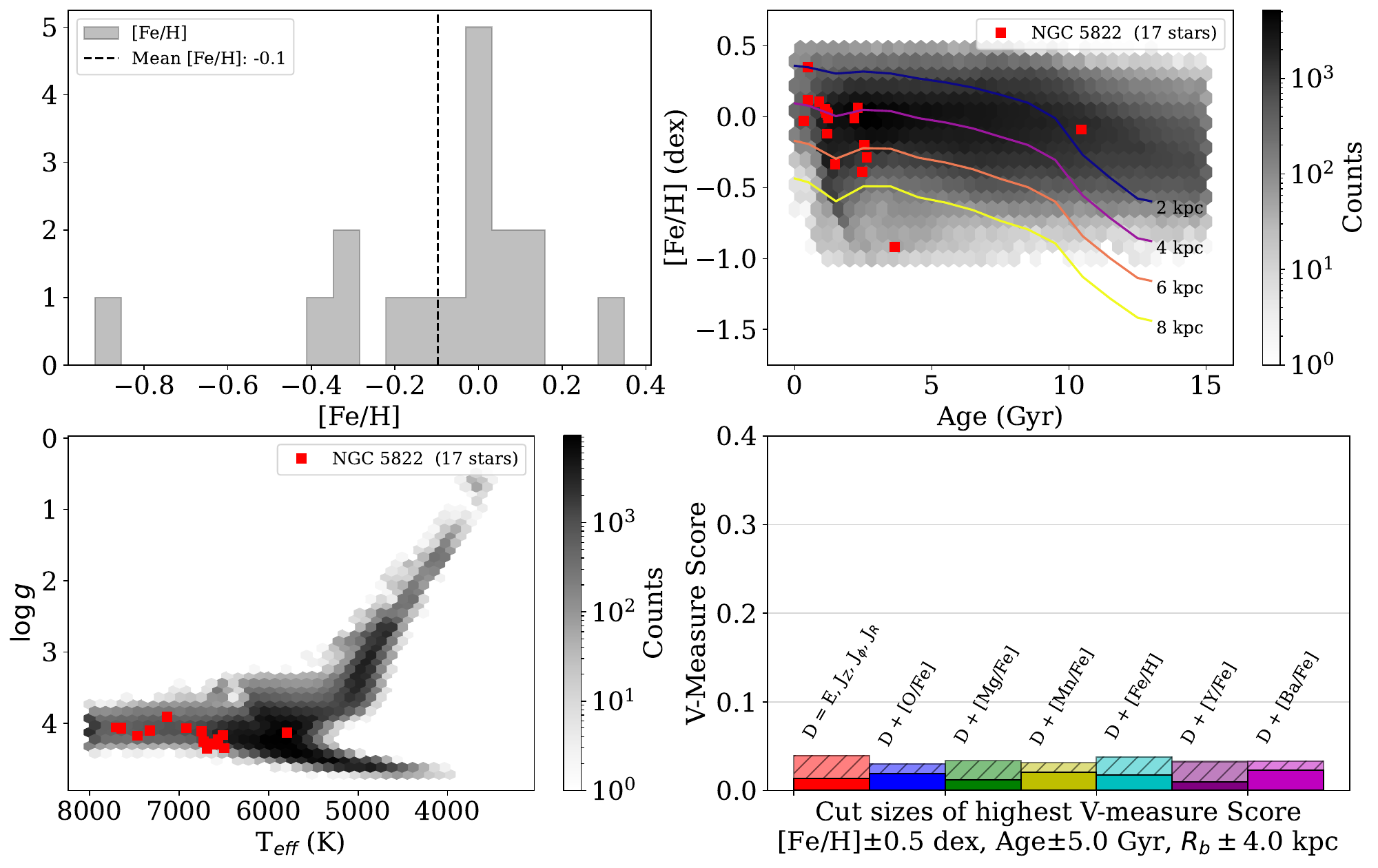}
        \caption{Same as Figure \ref{fig:sum_plot_Berkeley_32}, but for NGC 5822.}
    \end{figure}
    
    \begin{figure}
        \centering
        \includegraphics[width=0.9\linewidth]{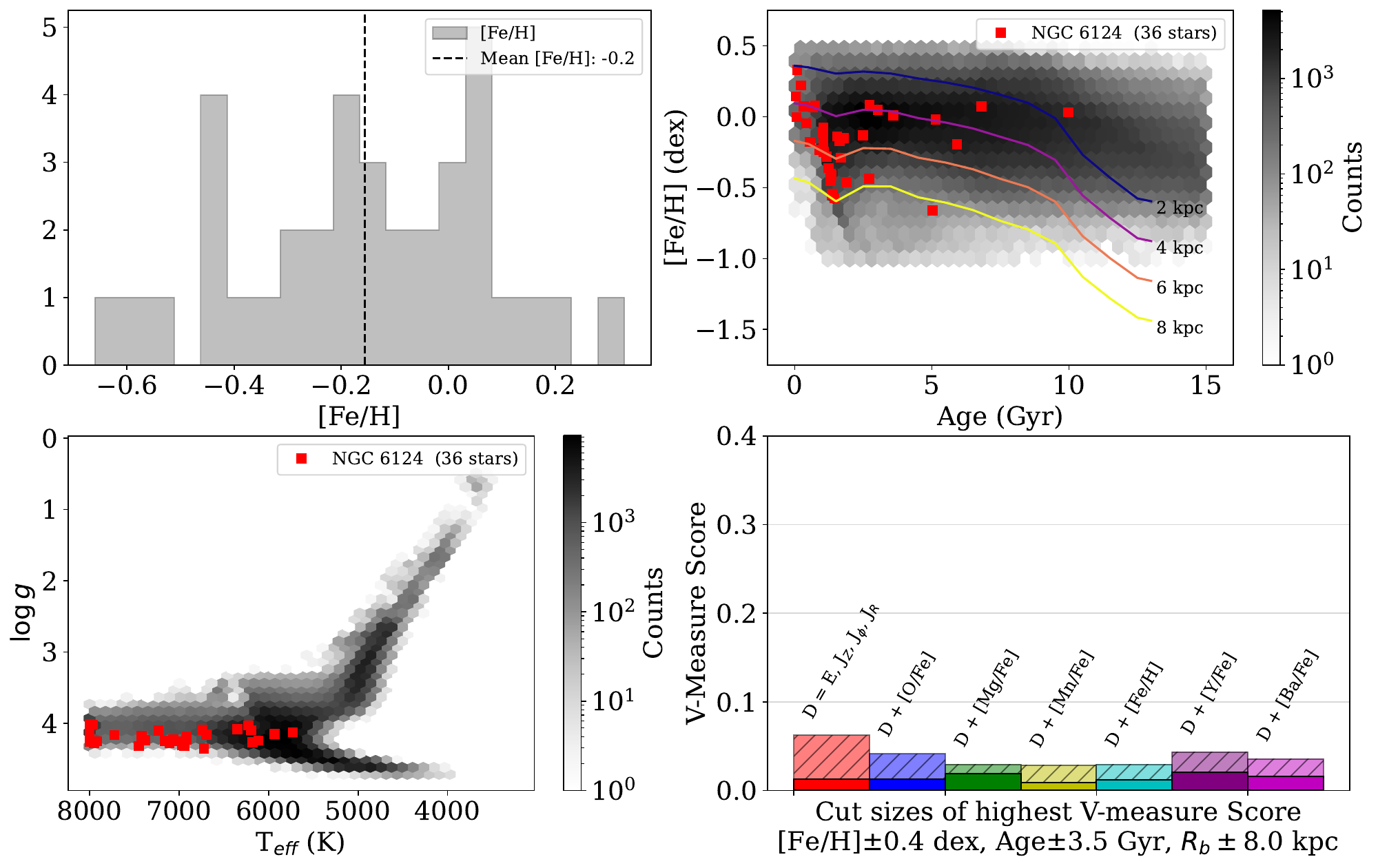}
        \caption{Same as Figure \ref{fig:sum_plot_Berkeley_32}, but for NGC 6124.}
    \end{figure}
    
    \begin{figure}
        \centering
        \includegraphics[width=0.9\linewidth]{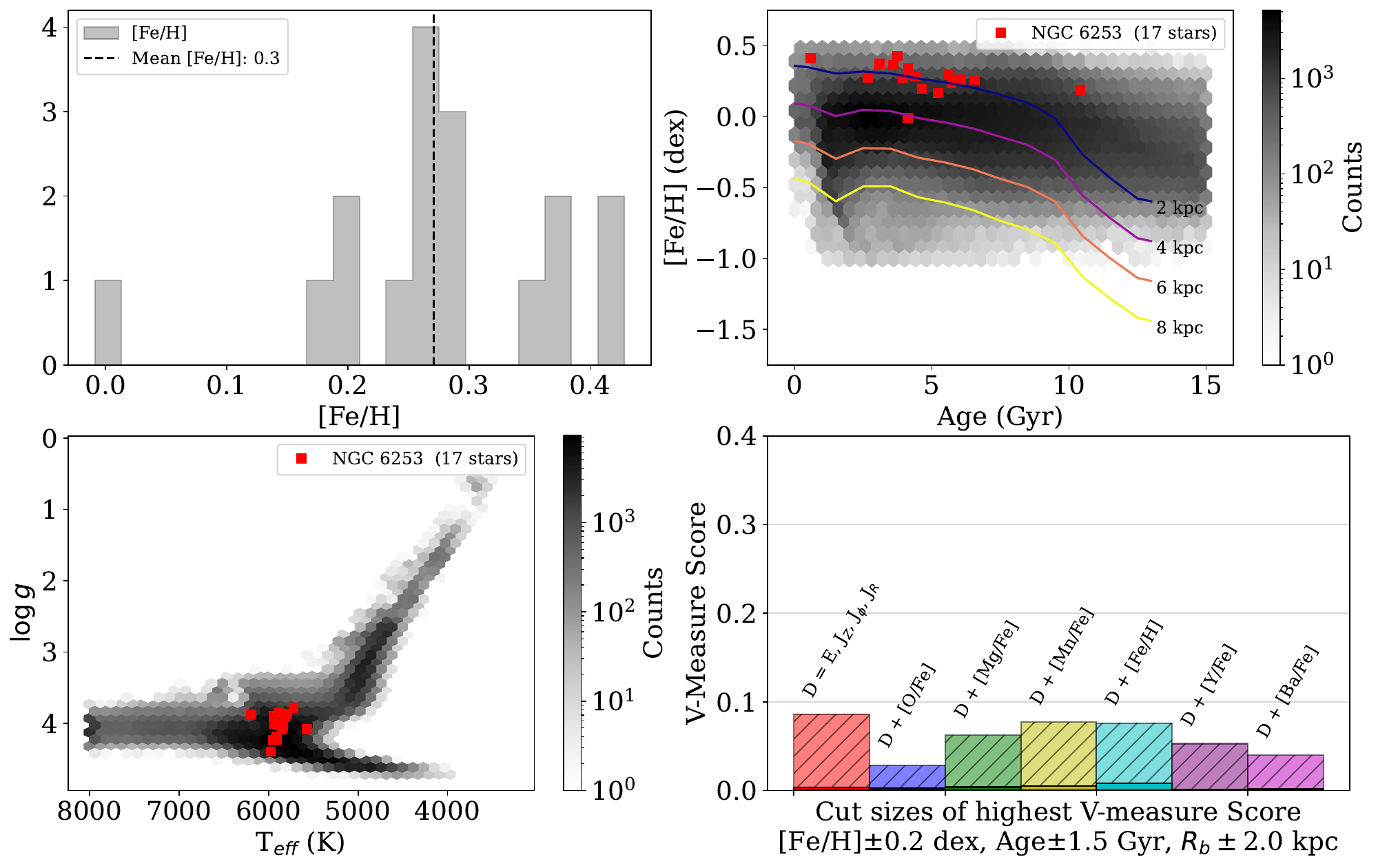}
        \caption{Same as Figure \ref{fig:sum_plot_Berkeley_32}, but for NGC 6253.}
    \end{figure}
    
    \begin{figure}
        \centering
        \includegraphics[width=0.9\linewidth]{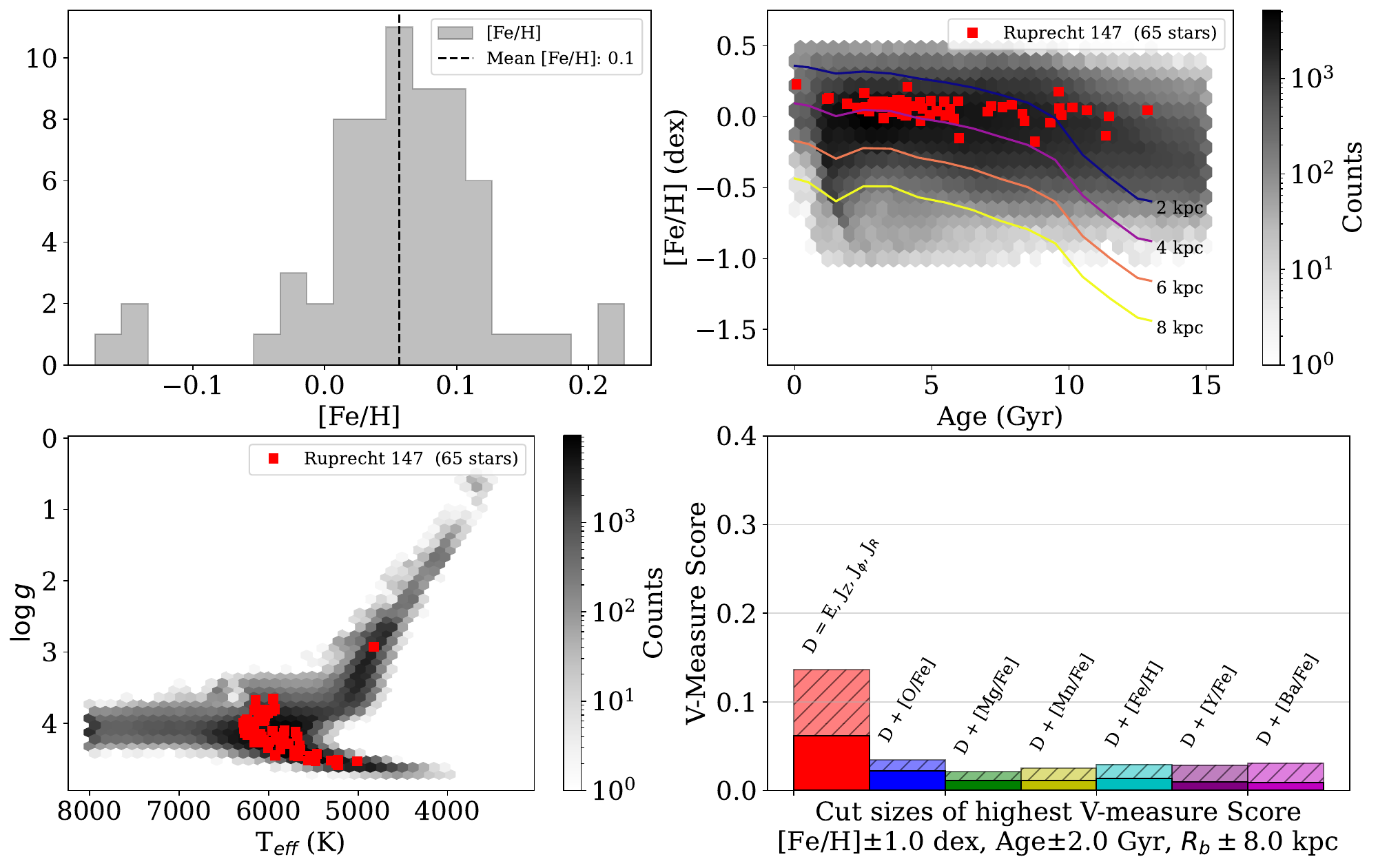}
        \caption{Same as Figure \ref{fig:sum_plot_Berkeley_32}, but for Ruprecht 147.}
    \end{figure}
    
    \begin{figure}
        \centering
        \includegraphics[width=0.9\linewidth]{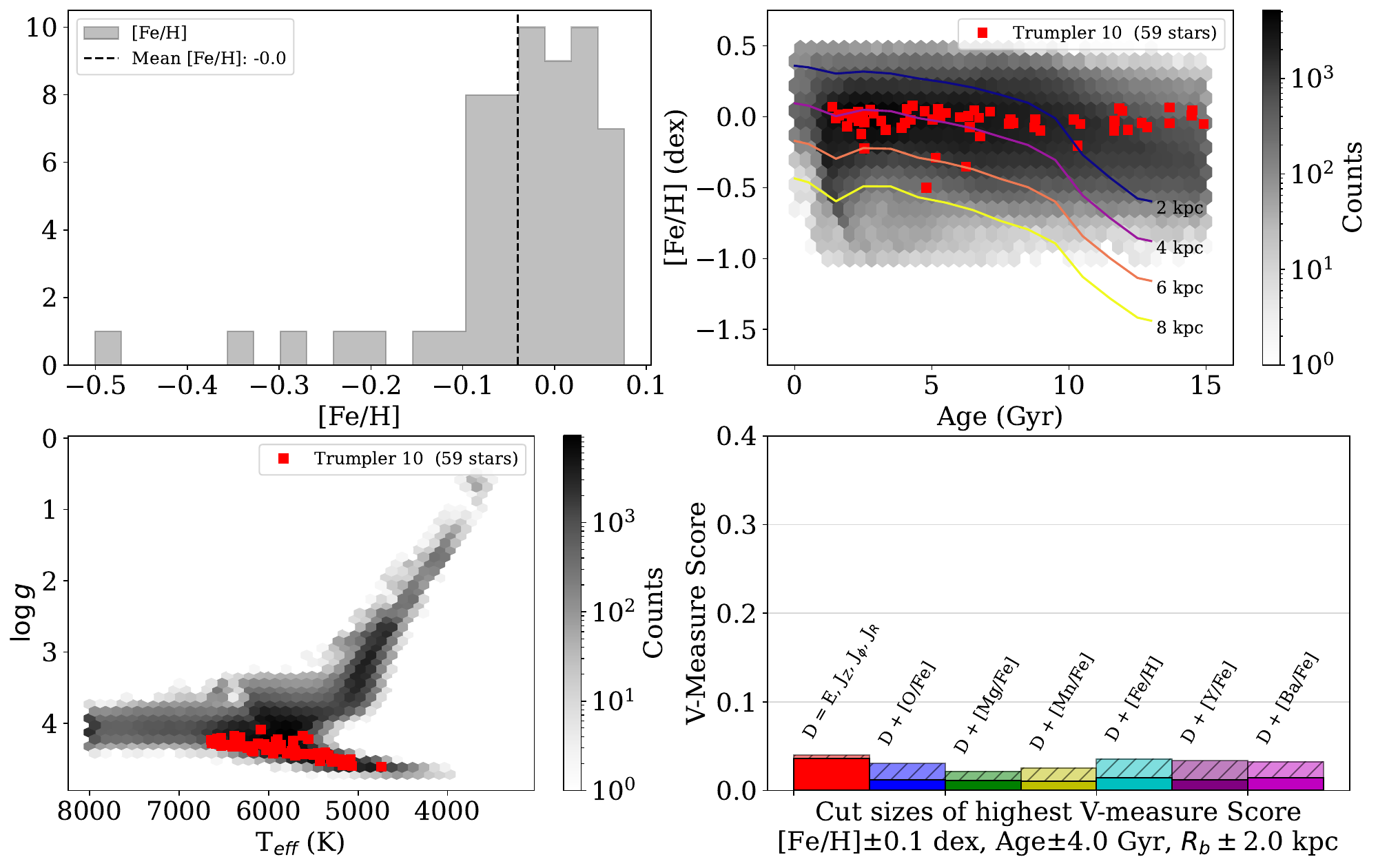}
        \caption{Same as Figure \ref{fig:sum_plot_Berkeley_32}, but for Trumpler 10.}
        \label{fig:trumpler10}
    \end{figure}
    
    \begin{figure}
        \centering
        \includegraphics[width=0.9\linewidth]{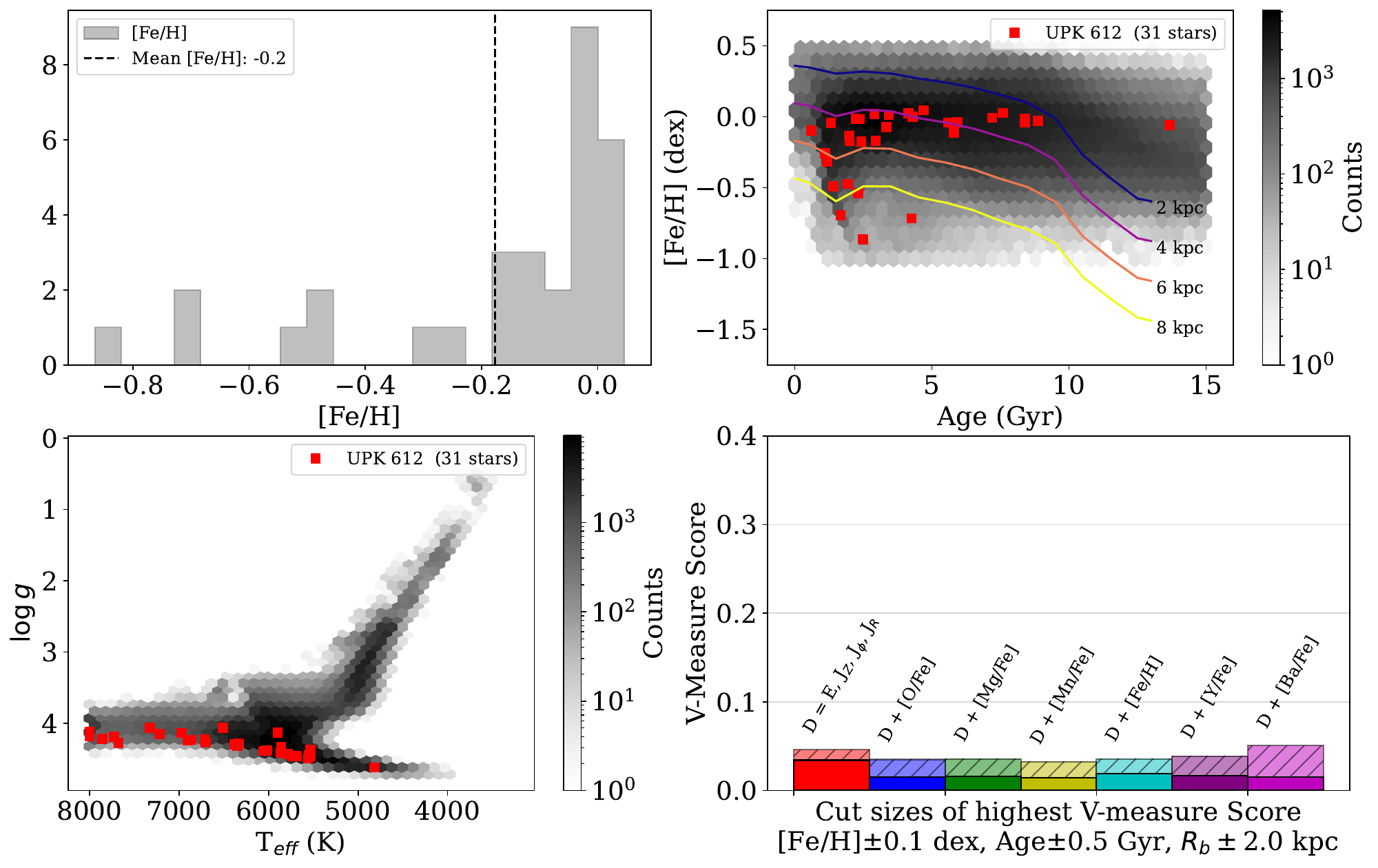}
        \caption{Same as Figure \ref{fig:sum_plot_Berkeley_32}, but for UPK 612.}
        \label{fig:UPK_612}
    \end{figure}

\end{subfigures}

\figsetstart

\figsetnum{A}

\figsettitle{Open Cluster Properties and Results}

\figsetgrpstart
\figsetgrpnum{A.1}
\figsetgrptitle{Berkeley 32}
\figsetplot{summary_plot_Berkeley_32} \label{fig:Berkeley_32}
\figsetgrpnote{Characteristics of the open cluster Berkeley 32. \textit{Top left}: Distribution of [Fe/H] abundances for member stars in Berkeley 32. The cluster's mean metallicity is shown as the vertical, dashed line. \textit{Top right}: 2D histogram of GALAH [Fe/H] abundances and ages derived in this work using stellar evolutionary track fitting, using the MIST isochrone models. Overlaid as red squares are the member stars of Berkeley 32. The horizontal lines correspond to mono-values of $R_b$ calculated from the GGC.  \textit{Bottom left}: Kiel diagram showing evolutionary stages of member stars of Berkeley 32, compared to the GGC. \textit{Bottom right}: Results of how well Berkeley 32 is recovered using different parameter combinations. The solid area of each bar represent the V-measure score of clustering the GGC, while the hatched section of the plot represents the best V-measure score achieved using a combination of [Fe/H], age, and birth radii cuts to reduce the size of the GGC.}
\figsetgrpend

\figsetgrpstart
\figsetgrpnum{A.2}
\figsetgrptitle{Blanco 1}
\figsetplot{summary_plot_Blanco_1} \label{fig:Blanco_1}
\figsetgrpnote{Characteristics of the open cluster Blanco 1. \textit{Top left}: Distribution of [Fe/H] abundances for member stars in Blanco 1. The cluster's mean metallicity is shown as the vertical, dashed line. \textit{Top right}: 2D histogram of GALAH [Fe/H] abundances and ages derived in this work using stellar evolutionary track fitting, using the MIST isochrone models. Overlaid as red squares are the member stars of Blanco 1. The horizontal lines correspond to mono-values of $R_b$ calculated from the GGC.  \textit{Bottom left}: Kiel diagram showing evolutionary stages of member stars of Blanco 1, compared to the GGC. \textit{Bottom right}: Results of how well Blanco 1 is recovered using different parameter combinations. The solid area of each bar represent the V-measure score of clustering the GGC, while the hatched section of the plot represents the best V-measure score achieved using a combination of [Fe/H], age, and birth radii cuts to reduce the size of the GGC.}
\figsetgrpend

\figsetgrpstart
\figsetgrpnum{A.3}
\figsetgrptitle{Collinder 135}
\figsetplot{summary_plot_Collinder_135} \label{fig:Collinder_135}
\figsetgrpnote{Characteristics of the open cluster Collinder 135. \textit{Top left}: Distribution of [Fe/H] abundances for member stars in Collinder 135. The cluster's mean metallicity is shown as the vertical, dashed line. \textit{Top right}: 2D histogram of GALAH [Fe/H] abundances and ages derived in this work using stellar evolutionary track fitting, using the MIST isochrone models. Overlaid as red squares are the member stars of Collinder 135. The horizontal lines correspond to mono-values of $R_b$ calculated from the GGC.  \textit{Bottom left}: Kiel diagram showing evolutionary stages of member stars of Collinder 135, compared to the GGC. \textit{Bottom right}: Results of how well Collinder 135 is recovered using different parameter combinations. The solid area of each bar represent the V-measure score of clustering the GGC, while the hatched section of the plot represents the best V-measure score achieved using a combination of [Fe/H], age, and birth radii cuts to reduce the size of the GGC.}
\figsetgrpend

\figsetgrpstart
\figsetgrpnum{A.4}
\figsetgrptitle{Collinder 140}
\figsetplot{summary_plot_Collinder_140} \label{fig:Collinder_140}
\figsetgrpnote{Characteristics of the open cluster Collinder 140. \textit{Top left}: Distribution of [Fe/H] abundances for member stars in Collinder 140. The cluster's mean metallicity is shown as the vertical, dashed line. \textit{Top right}: 2D histogram of GALAH [Fe/H] abundances and ages derived in this work using stellar evolutionary track fitting, using the MIST isochrone models. Overlaid as red squares are the member stars of Collinder 140. The horizontal lines correspond to mono-values of $R_b$ calculated from the GGC.  \textit{Bottom left}: Kiel diagram showing evolutionary stages of member stars of Collinder 140, compared to the GGC. \textit{Bottom right}: Results of how well Collinder 140 is recovered using different parameter combinations. The solid area of each bar represent the V-measure score of clustering the GGC, while the hatched section of the plot represents the best V-measure score achieved using a combination of [Fe/H], age, and birth radii cuts to reduce the size of the GGC.}
\figsetgrpend

\figsetgrpstart
\figsetgrpnum{A.5}
\figsetgrptitle{Collinder 261}
\figsetplot{summary_plot_Collinder_261} \label{fig:Collinder_261}
\figsetgrpnote{Characteristics of the open cluster Collinder 261. \textit{Top left}: Distribution of [Fe/H] abundances for member stars in Collinder 261. The cluster's mean metallicity is shown as the vertical, dashed line. \textit{Top right}: 2D histogram of GALAH [Fe/H] abundances and ages derived in this work using stellar evolutionary track fitting, using the MIST isochrone models. Overlaid as red squares are the member stars of Collinder 261. The horizontal lines correspond to mono-values of $R_b$ calculated from the GGC.  \textit{Bottom left}: Kiel diagram showing evolutionary stages of member stars of Collinder 261, compared to the GGC. \textit{Bottom right}: Results of how well Collinder 261 is recovered using different parameter combinations. The solid area of each bar represent the V-measure score of clustering the GGC, while the hatched section of the plot represents the best V-measure score achieved using a combination of [Fe/H], age, and birth radii cuts to reduce the size of the GGC.}
\figsetgrpend

\figsetgrpstart
\figsetgrpnum{A.6}
\figsetgrptitle{IC 4651}
\figsetplot{summary_plot_IC_4651} \label{fig:IC_4651}
\figsetgrpnote{Characteristics of the open cluster IC 4651. \textit{Top left}: Distribution of [Fe/H] abundances for member stars in IC 4651. The cluster's mean metallicity is shown as the vertical, dashed line. \textit{Top right}: 2D histogram of GALAH [Fe/H] abundances and ages derived in this work using stellar evolutionary track fitting, using the MIST isochrone models. Overlaid as red squares are the member stars of IC 4651. The horizontal lines correspond to mono-values of $R_b$ calculated from the GGC.  \textit{Bottom left}: Kiel diagram showing evolutionary stages of member stars of IC 4651, compared to the GGC. \textit{Bottom right}: Results of how well IC 4651 is recovered using different parameter combinations. The solid area of each bar represent the V-measure score of clustering the GGC, while the hatched section of the plot represents the best V-measure score achieved using a combination of [Fe/H], age, and birth radii cuts to reduce the size of the GGC.}
\figsetgrpend

\figsetgrpstart
\figsetgrpnum{A.7}
\figsetgrptitle{IC 4665}
\figsetplot{summary_plot_IC_4665} \label{fig:IC_4665}
\figsetgrpnote{Characteristics of the open cluster IC 4665. \textit{Top left}: Distribution of [Fe/H] abundances for member stars in IC 4665. The cluster's mean metallicity is shown as the vertical, dashed line. \textit{Top right}: 2D histogram of GALAH [Fe/H] abundances and ages derived in this work using stellar evolutionary track fitting, using the MIST isochrone models. Overlaid as red squares are the member stars of IC 4665. The horizontal lines correspond to mono-values of $R_b$ calculated from the GGC.  \textit{Bottom left}: Kiel diagram showing evolutionary stages of member stars of IC 4665, compared to the GGC. \textit{Bottom right}: Results of how well IC 4665 is recovered using different parameter combinations. The solid area of each bar represent the V-measure score of clustering the GGC, while the hatched section of the plot represents the best V-measure score achieved using a combination of [Fe/H], age, and birth radii cuts to reduce the size of the GGC.}
\figsetgrpend

\figsetgrpstart
\figsetgrpnum{A.8}
\figsetgrptitle{Mamajek 4}
\figsetplot{summary_plot_Mamajek_4} \label{fig:Mamajek_4}
\figsetgrpnote{Characteristics of the open cluster Mamajek 4. \textit{Top left}: Distribution of [Fe/H] abundances for member stars in Mamajek 4. The cluster's mean metallicity is shown as the vertical, dashed line. \textit{Top right}: 2D histogram of GALAH [Fe/H] abundances and ages derived in this work using stellar evolutionary track fitting, using the MIST isochrone models. Overlaid as red squares are the member stars of Mamajek 4. The horizontal lines correspond to mono-values of $R_b$ calculated from the GGC.  \textit{Bottom left}: Kiel diagram showing evolutionary stages of member stars of Mamajek 4, compared to the GGC. \textit{Bottom right}: Results of how well Mamajek 4 is recovered using different parameter combinations. The solid area of each bar represent the V-measure score of clustering the GGC, while the hatched section of the plot represents the best V-measure score achieved using a combination of [Fe/H], age, and birth radii cuts to reduce the size of the GGC.}
\figsetgrpend

\figsetgrpstart
\figsetgrpnum{A.9}
\figsetgrptitle{Melotte 22}
\figsetplot{summary_plot_Melotte_22} \label{fig:Melotte_22}
\figsetgrpnote{Characteristics of the open cluster Melotte 22. \textit{Top left}: Distribution of [Fe/H] abundances for member stars in Melotte 22. The cluster's mean metallicity is shown as the vertical, dashed line. \textit{Top right}: 2D histogram of GALAH [Fe/H] abundances and ages derived in this work using stellar evolutionary track fitting, using the MIST isochrone models. Overlaid as red squares are the member stars of Melotte 22. The horizontal lines correspond to mono-values of $R_b$ calculated from the GGC.  \textit{Bottom left}: Kiel diagram showing evolutionary stages of member stars of Melotte 22, compared to the GGC. \textit{Bottom right}: Results of how well Melotte 22 is recovered using different parameter combinations. The solid area of each bar represent the V-measure score of clustering the GGC, while the hatched section of the plot represents the best V-measure score achieved using a combination of [Fe/H], age, and birth radii cuts to reduce the size of the GGC.}
\figsetgrpend

\figsetgrpstart
\figsetgrpnum{A.10}
\figsetgrptitle{Melotte 25}
\figsetplot{summary_plot_Melotte_25} \label{fig:Melotte_25}
\figsetgrpnote{Characteristics of the open cluster Melotte 25. \textit{Top left}: Distribution of [Fe/H] abundances for member stars in Melotte 25. The cluster's mean metallicity is shown as the vertical, dashed line. \textit{Top right}: 2D histogram of GALAH [Fe/H] abundances and ages derived in this work using stellar evolutionary track fitting, using the MIST isochrone models. Overlaid as red squares are the member stars of Melotte 25. The horizontal lines correspond to mono-values of $R_b$ calculated from the GGC.  \textit{Bottom left}: Kiel diagram showing evolutionary stages of member stars of Melotte 25, compared to the GGC. \textit{Bottom right}: Results of how well Melotte 25 is recovered using different parameter combinations. The solid area of each bar represent the V-measure score of clustering the GGC, while the hatched section of the plot represents the best V-measure score achieved using a combination of [Fe/H], age, and birth radii cuts to reduce the size of the GGC.}
\figsetgrpend

\figsetgrpstart
\figsetgrpnum{A.11}
\figsetgrptitle{NGC 1901}
\figsetplot{summary_plot_NGC_1901} \label{fig:NGC_1901}
\figsetgrpnote{Characteristics of the open cluster NGC 1901. \textit{Top left}: Distribution of [Fe/H] abundances for member stars in NGC 1901. The cluster's mean metallicity is shown as the vertical, dashed line. \textit{Top right}: 2D histogram of GALAH [Fe/H] abundances and ages derived in this work using stellar evolutionary track fitting, using the MIST isochrone models. Overlaid as red squares are the member stars of NGC 1901. The horizontal lines correspond to mono-values of $R_b$ calculated from the GGC.  \textit{Bottom left}: Kiel diagram showing evolutionary stages of member stars of NGC 1901, compared to the GGC. \textit{Bottom right}: Results of how well NGC 1901 is recovered using different parameter combinations. The solid area of each bar represent the V-measure score of clustering the GGC, while the hatched section of the plot represents the best V-measure score achieved using a combination of [Fe/H], age, and birth radii cuts to reduce the size of the GGC.}
\figsetgrpend

\figsetgrpstart
\figsetgrpnum{A.12}
\figsetgrptitle{NGC 2112}
\figsetplot{summary_plot_NGC_2112} \label{fig:NGC_2112}
\figsetgrpnote{Characteristics of the open cluster NGC 2112. \textit{Top left}: Distribution of [Fe/H] abundances for member stars in NGC 2112. The cluster's mean metallicity is shown as the vertical, dashed line. \textit{Top right}: 2D histogram of GALAH [Fe/H] abundances and ages derived in this work using stellar evolutionary track fitting, using the MIST isochrone models. Overlaid as red squares are the member stars of NGC 2112. The horizontal lines correspond to mono-values of $R_b$ calculated from the GGC.  \textit{Bottom left}: Kiel diagram showing evolutionary stages of member stars of NGC 2112, compared to the GGC. \textit{Bottom right}: Results of how well NGC 2112 is recovered using different parameter combinations. The solid area of each bar represent the V-measure score of clustering the GGC, while the hatched section of the plot represents the best V-measure score achieved using a combination of [Fe/H], age, and birth radii cuts to reduce the size of the GGC.}
\figsetgrpend

\figsetgrpstart
\figsetgrpnum{A.13}
\figsetgrptitle{NGC 2204}
\figsetplot{summary_plot_NGC_2204} \label{fig:NGC_2204}
\figsetgrpnote{Characteristics of the open cluster NGC 2204. \textit{Top left}: Distribution of [Fe/H] abundances for member stars in NGC 2204. The cluster's mean metallicity is shown as the vertical, dashed line. \textit{Top right}: 2D histogram of GALAH [Fe/H] abundances and ages derived in this work using stellar evolutionary track fitting, using the MIST isochrone models. Overlaid as red squares are the member stars of NGC 2204. The horizontal lines correspond to mono-values of $R_b$ calculated from the GGC.  \textit{Bottom left}: Kiel diagram showing evolutionary stages of member stars of NGC 2204, compared to the GGC. \textit{Bottom right}: Results of how well NGC 2204 is recovered using different parameter combinations. The solid area of each bar represent the V-measure score of clustering the GGC, while the hatched section of the plot represents the best V-measure score achieved using a combination of [Fe/H], age, and birth radii cuts to reduce the size of the GGC.}
\figsetgrpend

\figsetgrpstart
\figsetgrpnum{A.14}
\figsetgrptitle{NGC 2360}
\figsetplot{summary_plot_NGC_2360} \label{fig:NGC_2360}
\figsetgrpnote{Characteristics of the open cluster NGC 2360. \textit{Top left}: Distribution of [Fe/H] abundances for member stars in NGC 2360. The cluster's mean metallicity is shown as the vertical, dashed line. \textit{Top right}: 2D histogram of GALAH [Fe/H] abundances and ages derived in this work using stellar evolutionary track fitting, using the MIST isochrone models. Overlaid as red squares are the member stars of NGC 2360. The horizontal lines correspond to mono-values of $R_b$ calculated from the GGC.  \textit{Bottom left}: Kiel diagram showing evolutionary stages of member stars of NGC 2360, compared to the GGC. \textit{Bottom right}: Results of how well NGC 2360 is recovered using different parameter combinations. The solid area of each bar represent the V-measure score of clustering the GGC, while the hatched section of the plot represents the best V-measure score achieved using a combination of [Fe/H], age, and birth radii cuts to reduce the size of the GGC.}
\figsetgrpend

\figsetgrpstart
\figsetgrpnum{A.15}
\figsetgrptitle{NGC 2451B}
\figsetplot{summary_plot_NGC_2451B} \label{fig:NGC_2451B}
\figsetgrpnote{Characteristics of the open cluster NGC 2451B. \textit{Top left}: Distribution of [Fe/H] abundances for member stars in NGC 2451B. The cluster's mean metallicity is shown as the vertical, dashed line. \textit{Top right}: 2D histogram of GALAH [Fe/H] abundances and ages derived in this work using stellar evolutionary track fitting, using the MIST isochrone models. Overlaid as red squares are the member stars of NGC 2451B. The horizontal lines correspond to mono-values of $R_b$ calculated from the GGC.  \textit{Bottom left}: Kiel diagram showing evolutionary stages of member stars of NGC 2451B, compared to the GGC. \textit{Bottom right}: Results of how well NGC 2451B is recovered using different parameter combinations. The solid area of each bar represent the V-measure score of clustering the GGC, while the hatched section of the plot represents the best V-measure score achieved using a combination of [Fe/H], age, and birth radii cuts to reduce the size of the GGC.}
\figsetgrpend

\figsetgrpstart
\figsetgrpnum{A.16}
\figsetgrptitle{NGC 2516}
\figsetplot{summary_plot_NGC_2516} \label{fig:NGC_2516}
\figsetgrpnote{Characteristics of the open cluster NGC 2516. \textit{Top left}: Distribution of [Fe/H] abundances for member stars in NGC 2516. The cluster's mean metallicity is shown as the vertical, dashed line. \textit{Top right}: 2D histogram of GALAH [Fe/H] abundances and ages derived in this work using stellar evolutionary track fitting, using the MIST isochrone models. Overlaid as red squares are the member stars of NGC 2516. The horizontal lines correspond to mono-values of $R_b$ calculated from the GGC.  \textit{Bottom left}: Kiel diagram showing evolutionary stages of member stars of NGC 2516, compared to the GGC. \textit{Bottom right}: Results of how well NGC 2516 is recovered using different parameter combinations. The solid area of each bar represent the V-measure score of clustering the GGC, while the hatched section of the plot represents the best V-measure score achieved using a combination of [Fe/H], age, and birth radii cuts to reduce the size of the GGC.}
\figsetgrpend

\figsetgrpstart
\figsetgrpnum{A.17}
\figsetgrptitle{NGC 2548}
\figsetplot{summary_plot_NGC_2548} \label{fig:NGC_2548}
\figsetgrpnote{Characteristics of the open cluster NGC 2548. \textit{Top left}: Distribution of [Fe/H] abundances for member stars in NGC 2548. The cluster's mean metallicity is shown as the vertical, dashed line. \textit{Top right}: 2D histogram of GALAH [Fe/H] abundances and ages derived in this work using stellar evolutionary track fitting, using the MIST isochrone models. Overlaid as red squares are the member stars of NGC 2548. The horizontal lines correspond to mono-values of $R_b$ calculated from the GGC.  \textit{Bottom left}: Kiel diagram showing evolutionary stages of member stars of NGC 2548, compared to the GGC. \textit{Bottom right}: Results of how well NGC 2548 is recovered using different parameter combinations. The solid area of each bar represent the V-measure score of clustering the GGC, while the hatched section of the plot represents the best V-measure score achieved using a combination of [Fe/H], age, and birth radii cuts to reduce the size of the GGC.}
\figsetgrpend

\figsetgrpstart
\figsetgrpnum{A.18}
\figsetgrptitle{NGC 2632}
\figsetplot{summary_plot_NGC_2632} \label{fig:NGC_2632}
\figsetgrpnote{Characteristics of the open cluster NGC 2632. \textit{Top left}: Distribution of [Fe/H] abundances for member stars in NGC 2632. The cluster's mean metallicity is shown as the vertical, dashed line. \textit{Top right}: 2D histogram of GALAH [Fe/H] abundances and ages derived in this work using stellar evolutionary track fitting, using the MIST isochrone models. Overlaid as red squares are the member stars of NGC 2632. The horizontal lines correspond to mono-values of $R_b$ calculated from the GGC.  \textit{Bottom left}: Kiel diagram showing evolutionary stages of member stars of NGC 2632, compared to the GGC. \textit{Bottom right}: Results of how well NGC 2632 is recovered using different parameter combinations. The solid area of each bar represent the V-measure score of clustering the GGC, while the hatched section of the plot represents the best V-measure score achieved using a combination of [Fe/H], age, and birth radii cuts to reduce the size of the GGC.}
\figsetgrpend

\figsetgrpstart
\figsetgrpnum{A.19}
\figsetgrptitle{NGC 2682}
\figsetplot{summary_plot_NGC_2682} \label{fig:NGC_2682}
\figsetgrpnote{Characteristics of the open cluster NGC 2682. \textit{Top left}: Distribution of [Fe/H] abundances for member stars in NGC 2682. The cluster's mean metallicity is shown as the vertical, dashed line. \textit{Top right}: 2D histogram of GALAH [Fe/H] abundances and ages derived in this work using stellar evolutionary track fitting, using the MIST isochrone models. Overlaid as red squares are the member stars of NGC 2682. The horizontal lines correspond to mono-values of $R_b$ calculated from the GGC.  \textit{Bottom left}: Kiel diagram showing evolutionary stages of member stars of NGC 2682, compared to the GGC. \textit{Bottom right}: Results of how well NGC 2682 is recovered using different parameter combinations. The solid area of each bar represent the V-measure score of clustering the GGC, while the hatched section of the plot represents the best V-measure score achieved using a combination of [Fe/H], age, and birth radii cuts to reduce the size of the GGC.}
\figsetgrpend

\figsetgrpstart
\figsetgrpnum{A.20}
\figsetgrptitle{NGC 3114}
\figsetplot{summary_plot_NGC_3114} \label{fig:NGC_3114}
\figsetgrpnote{Characteristics of the open cluster NGC 3114. \textit{Top left}: Distribution of [Fe/H] abundances for member stars in NGC 3114. The cluster's mean metallicity is shown as the vertical, dashed line. \textit{Top right}: 2D histogram of GALAH [Fe/H] abundances and ages derived in this work using stellar evolutionary track fitting, using the MIST isochrone models. Overlaid as red squares are the member stars of NGC 3114. The horizontal lines correspond to mono-values of $R_b$ calculated from the GGC.  \textit{Bottom left}: Kiel diagram showing evolutionary stages of member stars of NGC 3114, compared to the GGC. \textit{Bottom right}: Results of how well NGC 3114 is recovered using different parameter combinations. The solid area of each bar represent the V-measure score of clustering the GGC, while the hatched section of the plot represents the best V-measure score achieved using a combination of [Fe/H], age, and birth radii cuts to reduce the size of the GGC.}
\figsetgrpend

\figsetgrpstart
\figsetgrpnum{A.21}
\figsetgrptitle{NGC 5822}
\figsetplot{summary_plot_NGC_5822} \label{fig:NGC_5822}
\figsetgrpnote{Characteristics of the open cluster NGC 5822. \textit{Top left}: Distribution of [Fe/H] abundances for member stars in NGC 5822. The cluster's mean metallicity is shown as the vertical, dashed line. \textit{Top right}: 2D histogram of GALAH [Fe/H] abundances and ages derived in this work using stellar evolutionary track fitting, using the MIST isochrone models. Overlaid as red squares are the member stars of NGC 5822. The horizontal lines correspond to mono-values of $R_b$ calculated from the GGC.  \textit{Bottom left}: Kiel diagram showing evolutionary stages of member stars of NGC 5822, compared to the GGC. \textit{Bottom right}: Results of how well NGC 5822 is recovered using different parameter combinations. The solid area of each bar represent the V-measure score of clustering the GGC, while the hatched section of the plot represents the best V-measure score achieved using a combination of [Fe/H], age, and birth radii cuts to reduce the size of the GGC.}
\figsetgrpend

\figsetgrpstart
\figsetgrpnum{A.22}
\figsetgrptitle{NGC 6124}
\figsetplot{summary_plot_NGC_6124} \label{fig:NGC_6124}
\figsetgrpnote{Characteristics of the open cluster NGC 6124. \textit{Top left}: Distribution of [Fe/H] abundances for member stars in NGC 6124. The cluster's mean metallicity is shown as the vertical, dashed line. \textit{Top right}: 2D histogram of GALAH [Fe/H] abundances and ages derived in this work using stellar evolutionary track fitting, using the MIST isochrone models. Overlaid as red squares are the member stars of NGC 6124. The horizontal lines correspond to mono-values of $R_b$ calculated from the GGC.  \textit{Bottom left}: Kiel diagram showing evolutionary stages of member stars of NGC 6124, compared to the GGC. \textit{Bottom right}: Results of how well NGC 6124 is recovered using different parameter combinations. The solid area of each bar represent the V-measure score of clustering the GGC, while the hatched section of the plot represents the best V-measure score achieved using a combination of [Fe/H], age, and birth radii cuts to reduce the size of the GGC.}
\figsetgrpend

\figsetgrpstart
\figsetgrpnum{A.23}
\figsetgrptitle{NGC 6253}
\figsetplot{summary_plot_NGC_6253} \label{fig:NGC_6253}
\figsetgrpnote{Characteristics of the open cluster NGC 6253. \textit{Top left}: Distribution of [Fe/H] abundances for member stars in NGC 6253. The cluster's mean metallicity is shown as the vertical, dashed line. \textit{Top right}: 2D histogram of GALAH [Fe/H] abundances and ages derived in this work using stellar evolutionary track fitting, using the MIST isochrone models. Overlaid as red squares are the member stars of NGC 6253. The horizontal lines correspond to mono-values of $R_b$ calculated from the GGC.  \textit{Bottom left}: Kiel diagram showing evolutionary stages of member stars of NGC 6253, compared to the GGC. \textit{Bottom right}: Results of how well NGC 6253 is recovered using different parameter combinations. The solid area of each bar represent the V-measure score of clustering the GGC, while the hatched section of the plot represents the best V-measure score achieved using a combination of [Fe/H], age, and birth radii cuts to reduce the size of the GGC.}
\figsetgrpend

\figsetgrpstart
\figsetgrpnum{A.24}
\figsetgrptitle{Ruprecht 147}
\figsetplot{summary_plot_Ruprecht_147} \label{fig:Ruprecht_147}
\figsetgrpnote{Characteristics of the open cluster Ruprecht 147. \textit{Top left}: Distribution of [Fe/H] abundances for member stars in Ruprecht 147. The cluster's mean metallicity is shown as the vertical, dashed line. \textit{Top right}: 2D histogram of GALAH [Fe/H] abundances and ages derived in this work using stellar evolutionary track fitting, using the MIST isochrone models. Overlaid as red squares are the member stars of Ruprecht 147. The horizontal lines correspond to mono-values of $R_b$ calculated from the GGC.  \textit{Bottom left}: Kiel diagram showing evolutionary stages of member stars of Ruprecht 147, compared to the GGC. \textit{Bottom right}: Results of how well Ruprecht 147 is recovered using different parameter combinations. The solid area of each bar represent the V-measure score of clustering the GGC, while the hatched section of the plot represents the best V-measure score achieved using a combination of [Fe/H], age, and birth radii cuts to reduce the size of the GGC.}
\figsetgrpend

\figsetgrpstart
\figsetgrpnum{A.25}
\figsetgrptitle{Trumpler 10}
\figsetplot{summary_plot_Trumpler_10} \label{fig:Trumpler_10}
\figsetgrpnote{Characteristics of the open cluster Trumpler 10. \textit{Top left}: Distribution of [Fe/H] abundances for member stars in Trumpler 10. The cluster's mean metallicity is shown as the vertical, dashed line. \textit{Top right}: 2D histogram of GALAH [Fe/H] abundances and ages derived in this work using stellar evolutionary track fitting, using the MIST isochrone models. Overlaid as red squares are the member stars of Trumpler 10. The horizontal lines correspond to mono-values of $R_b$ calculated from the GGC.  \textit{Bottom left}: Kiel diagram showing evolutionary stages of member stars of Trumpler 10, compared to the GGC. \textit{Bottom right}: Results of how well Trumpler 10 is recovered using different parameter combinations. The solid area of each bar represent the V-measure score of clustering the GGC, while the hatched section of the plot represents the best V-measure score achieved using a combination of [Fe/H], age, and birth radii cuts to reduce the size of the GGC.}
\figsetgrpend

\figsetgrpstart
\figsetgrpnum{A.26}
\figsetgrptitle{UPK 612}
\figsetplot{summary_plot_UPK_612} \label{fig:UPK_612}
\figsetgrpnote{Characteristics of the open cluster UPK 612. \textit{Top left}: Distribution of [Fe/H] abundances for member stars in UPK 612. The cluster's mean metallicity is shown as the vertical, dashed line. \textit{Top right}: 2D histogram of GALAH [Fe/H] abundances and ages derived in this work using stellar evolutionary track fitting, using the MIST isochrone models. Overlaid as red squares are the member stars of UPK 612. The horizontal lines correspond to mono-values of $R_b$ calculated from the GGC.  \textit{Bottom left}: Kiel diagram showing evolutionary stages of member stars of UPK 612, compared to the GGC. \textit{Bottom right}: Results of how well UPK 612 is recovered using different parameter combinations. The solid area of each bar represent the V-measure score of clustering the GGC, while the hatched section of the plot represents the best V-measure score achieved using a combination of [Fe/H], age, and birth radii cuts to reduce the size of the GGC.}
\figsetgrpend

\figsetend








\end{document}